\documentclass{article}

\usepackage[preprint]{neurips_2022}

\usepackage[utf8]{inputenc} 
\usepackage[T1]{fontenc}    
\usepackage{hyperref}       
\usepackage{url}            
\usepackage{booktabs}       
\usepackage{amsfonts}       
\usepackage{nicefrac}       
\usepackage{microtype}      
\usepackage[dvipsnames]{xcolor} 
\usepackage{amsmath,amssymb,amsfonts,amsthm}
\usepackage{algorithm}
\usepackage{algorithmic}
\usepackage{graphicx, subcaption}
\usepackage[capitalise, noabbrev]{cleveref}
\usepackage{multicol}
\usepackage{changepage}
\usepackage{float}
\usepackage{ulem}
\usepackage{etoolbox}
\usepackage{xspace}

\newcommand{\geant}{\textsc{Geant4}\xspace}

\newtoggle{oldarch}   
\toggletrue{oldarch}   

\definecolor{dartmouthgreen}{rgb}{0.05, 0.5, 0.06}

\newcommand{\PR}[1]{\textcolor{NavyBlue}{\textbf{PR:} #1}}

\def\imgheight{5.5cm}
\def\figenvleftextend{0.5cm}
\def\figenvrightextend{0.5cm}

\title{A Generalisable Generative Model for Multi-Detector Calorimeter Simulation}

\author{%
   Piyush Raikwar \\
   CERN, Geneva, CH \\
  \texttt{piyush.raikwar@cern.ch} \\
   \And
   Mikołaj Piórczynski  \\
   CERN, Geneva, CH \\
  Warsaw University of Technology, Warsaw, PL \\
   \texttt{mikolaj.piorczynski.stud@pw.edu.pl} \\
   \And
   Anna Zaborowska \\
   CERN, Geneva, CH \\
   \texttt{anna.zaborowska@cern.ch} \\
   \And
   Peter McKeown \\
   CERN, Geneva, CH \\
   \texttt{peter.mckeown@cern.ch} \\
   \And
   Renato Cardoso \\
   CERN, Geneva, CH \\
   \texttt{renato.cardoso@cern.ch} \\
   \And
   Kyongmin Yeo \\
   Thomas J. Watson Research Center, Yorktown Heights, NY, USA \\
   \texttt{kyeo@us.ibm.com} \\
}

\begin{document}

\maketitle















\begin{abstract}
Collider experiments, such as those at the Large Hadron Collider, use the \textsc{Geant4} toolkit to simulate particle–detector interactions with high accuracy. However, these experiments increasingly require larger amounts of simulated data, leading to huge computing cost. Generative machine learning methods could offer much faster calorimeter shower simulations by directly emulating detector responses.

In this work, we present CaloDiT-2, a diffusion model which uses transformer blocks.  As is the case for other models explored for this task, it can be applied to specific geometries, however its true strength lies in its generalisation capabilities. Our approach allows pre-training on multiple detectors and rapid adaptation to new ones, which we demonstrate on the LEMURS dataset. It reduces the effort required to develop accurate models for novel detectors or detectors which are under development and have geometries that are changed frequently, requiring up to $25\times$ less data and $20\times$ less training time. To the best of our knowledge, this is the first pre-trained model to be published that allows adaptation in the context of particle shower simulations, with the model also included in the Geant4 toolkit. We also present results on benchmarks on Dataset-2 from the community-hosted CaloChallenge, showing that our models provide one of the best trade-offs between accuracy and speed from the published models. Our contributions include a mechanism for the creation of detector-agnostic data representations, architectural modifications suitable for the data modality, a pre-training and adaptation strategy, and publicly released datasets and pre-trained models for broad use.  

\end{abstract}


\section{Introduction}
In high energy physics (HEP), detector simulation plays an essential role, from the initial design and optimisation of a detector, through to the operation phase of an experiment, where simulation is required for the estimation of systematic errors. Consequently, large numbers of events must be simulated for a given experimental context. \geant \cite{agostinelli2003geant4} is a state-of-the-art Monte Carlo (MC) toolkit for simulating the passage of particles through matter, with applications in HEP, medical physics, astrophysics, and material science. The simulation involves a step--by-step transportation of individual particles through the volume of a detector, providing a highly accurate, yet computationally expensive response of the detector. This is particularly true for the calorimeter subsystems of a detector, which aim to stop incident particles by producing a shower (cascade) of secondary particles. These subsystems are typically separated into the electromagnetic calorimeter (ECal) and the hadronic calorimeter (HCal). At the upcoming high-luminosity phase of the LHC and at future colliders, the amount of simulated data required by experiments will increase\cite{CERN-LHCC-2022-005, Software:2815292}, requiring the development of fast yet accurate simulation alternatives.  

To this end, numerous approaches have been developed to accelerate calorimeter shower simulations. These fast simulation (FastSim) approaches seek to directly emulate the detector response, skipping a multitude of steps. Recently, significant progress has been made in the development of surrogate simulators based on generative models. Numerous approaches have been explored, including Generative Adversarial Networks (GANs) \cite{paganini2018calogan, Musella:2018rdi, Erdmann:2018jxd, Khattak:2021ndw, ATLAS:2020quw}, Variational Autoencoders (VAEs) and their variants \cite{ATLAS:2018wpe, gettinghigh, Cresswell:2022tof, Diefenbacher:2023prl, Salamani:2023ttx, Raikwar:2024peb, liu2024calo}, Normalising Flows \cite{Krause:2021ilc, Krause:2021wez, Diefenbacher:2023vsw, Ernst:2023qvn, Buss:2024orz} and Diffusion Models \cite{Mikuni:2022xry, Buhmann:2023bwk, Amram:2023onf, Mikuni:2023tqg, Buhmann:2023kdg, Favaro:2024rle, Buss:2025cyw}. A recent community challenge compared a large number of models on a standard set of datasets and benchmarks \cite{krause2024calochallenge}. These approaches are now sufficiently mature that they have also started to be deployed as part of the production ecosystems of major HEP experiments \cite{ATLAS:2021pzo}. The majority of these works have focused on simulating showers represented in the form of regular grids, typically at the level of the detector readout. However, recent work has proposed creating shower representations by recording energy depositions in a manner not attached to the detector readout, either with a scoring mesh \cite{Par04, Salamani:2023ttx} or by directly clustering energy deposits into a point cloud \cite{Buhmann:2023bwk, Buhmann:2023kdg, Buss:2025cyw}.

While significant progress has been made, the development of a fast simulation tool requires significant computational resources. This includes not only the design and the development of the model itself, but also the production of sufficiently large simulation datasets for training. These challenges are compounded when detector designs are frequently updated during development, as is the case with future colliders such as the Future Circular Collider (FCC-ee) \cite{FCC:2025lpp}. Motivated by these challenges, our goal is to develop a generalisable approach to fast simulation that is both easy to use and resource-efficient. In practice, the user should only need to collect shower data from the desired detector, train the model using ready-to-use scripts, and reintegrate the generated showers into the simulation pipeline.

In the literature, a few prior works have explored the development of generalisable models for HEP. MetaHEP \cite{Salamani:2023ttx} was probably the first to introduce a meta-learning approach, training across multiple detectors. This work demonstrated that a VAE-based model, similar to the one proposed in \cite{salamanivae}, could achieve accurate shower profiles on a new detector using approximately 400 meta-adaptation steps. However, MetaHEP relies on meta-learning, which involves complex training procedures and can be challenging to apply effectively to more complex models. Meta-learning is typically used in settings where rapid adaptation to new tasks is required, such as zero- or few-shot learning. We believe this to be an unnecessary constraint for a fast simulation application, which forms part of typical offline computing workflows. This means that \geant can be used to simulate sufficient quantities of events for an adaptation step. While MetaHEP also reports results using fine-tuning, these are based on pre-training with only a single detector, which we consider insufficient for a generic tool.  

More recently, OmniJet-$\alpha$ \cite{birk2024omnijet} advanced the idea of a general-purpose HEP model capable of both classification and jet generation. However, OmniJet-$\alpha$ addresses multiple tasks within a single detector, whereas our focus lies on a single task across different detectors. It employs an autoregressive GPT-based model for sampling \cite{radford2018improving}, which requires sequential token generation and therefore leads to slower inference. VQVAE, in contrast, is known to be incredibly difficult to train \cite{lancucki2020robust} and has been shown to struggle with capturing the voxel energy distribution in the context of showers \cite{Raikwar:2024peb}. This issue becomes more pronounced when the decoder must be adapted to a new detector that differs from the one used during initial training. OmniJet-$\alpha_C$ \cite{birk2025omnijet} further extended this framework to generate single-particle showers. However, this extension is trained on a single detector and does not involve any pre-training or adaptation mechanisms.    

For these reasons, we follow the foundation model paradigm \cite{bommasani2021opportunities, reed2022generalist, brown2020language}, where a sufficiently large model is trained on a diverse dataset to learn robust representations of the data modality. Such models can then be efficiently fine-tuned for new tasks, or even used as a part of a different task altogether. In the context of FastSim, our approach is to train on data from multiple detectors to capture generalizable shower characteristics, enabling the reuse of the knowledge learned to quickly adapt to unseen detectors.

To develop such an adaptable FastSim tool requires two key components. The first is an expressive generative model that can be adapted effectively. The second is a means of representing showers across disparate detectors, with different materials, geometries, and readouts. This limits the possible approaches to either the aforementioned point cloud approach or the use of a detector-agnostic scoring mesh. In order to properly exploit the flexibility of a point cloud, the clustering of many thousands of \geant steps (individual energy deposits) into a more tractable set of points for model training would be required \cite{Buhmann:2023bwk}. This would currently require significant hand-crafting in order to create appropriate representations for each of the different geometries used, which feature drastically different readout granularities. The larger resulting differences in data distributions may also present challenges for model development. For this reason, we adopt a physics-inspired virtual cylindrical mesh, originally introduced in the \geant Par04 example \cite{Par04}. This mesh allows us to transfer the difficulty from data representation to the reintegration step, enabling a simpler and more uniform representation of showers. As a result, the model’s expressive capacity can be focused primarily on capturing how particle showers develop across different detector geometries.

Even without the added complexity of training a generalisable model, simulating particle shower data with a generative model is a challenging problem. Showers are inherently stochastic and sparse, with their dynamic range spanning several orders of magnitude. To address these challenges we introduce CaloDiT-2, a diffusion model based on transformers \cite{vaswani2017attention} which operates at the voxel-level. Diffusion models have recently established themselves as state-of-the-art for image and video generation \cite{rombach2022high, blattmann2023stable}, with various approaches proposed to mitigate their slow inference \cite{song2020denoising, DBLP:conf/iclr/SalimansH22}. Transformers, in contrast, are powerful and generalisable models. Their low architectural inductive bias and attention mechanism for modeling long-range dependencies make them particularly well-suited for the complex, non-trivial structure of shower data.  

Our results indicate that CaloDiT-2 achieves a significantly improved trade-off between accuracy and inference speed compared to prior works when trained on a single detector. Furthermore, by pre-training on multiple detectors, our model generalizes to produce a FastSim model that can be quickly adapted to new detectors. This marks the first step towards a potential foundation model for FastSim and beyond \cite{radford2018improving, kirillov2023segment}. The contributions of this paper are as follows: 
\begin{enumerate}
    \item We present CaloDiT-2, a diffusion-based model that achieves good agreement with \textsc{Geant4} on widely used physics observables. Its advantages are reflected in the community-hosted CaloChallenge \cite{krause2024calochallenge}, where it demonstrates the best trade-off between accuracy and speed (Section \ref{sec:results_single_detector}).  
    \item We propose and validate a multi-detector training framework, demonstrating the benefits of cross-detector adaptation (Section~\ref{sec:generic_model}).  
    \item Our pre-trained models are released as part of \textsc{Geant4} with ready-to-use scripts for either adapting the pre-trained model or for training a new one from scratch.  
    \item The LEMURS (Large-scale multi-detector ElectroMagnetic Universal Representation of Showers) dataset, consisting of ECal showers from multiple detectors, is also made publicly available on Zenodo~\cite{zaborowska_2025_17045562} and is described in detail in ~\cite{mckeown2025lemursdatasetlargescalemultidetector}.
\end{enumerate} 

The paper is structured as follows. In Section \ref{sec:shower_dataset}, we introduce the problem setting, describe the dataset, and present its validation. Section \ref{sec:methodology} outlines the preliminaries and then details our proposed model, CaloDiT-2, along with its training setup. In Section \ref{sec:results_single_detector}, we evaluate the model’s performance when trained on a single detector and provide additional results on Dataset-2 from the CaloChallenge. Section \ref{sec:generic_model} extends CaloDiT-2 into a generalisable model, discusses key observations, and offers guidance for users interested in such a setup. Finally, Section \ref{sec:discussion} and Section \ref{sec:conclusion} present the discussion of current limitations and future work, and the conclusions, respectively.

\section{Dataset Description and Validation}
\label{sec:shower_dataset}

\begin{figure}[!ht]
    \centering
    \begin{subfigure}{.33\textwidth}
    \centering
    \includegraphics[width=\textwidth,trim={30cm 11.5cm 20cm 8cm},clip]{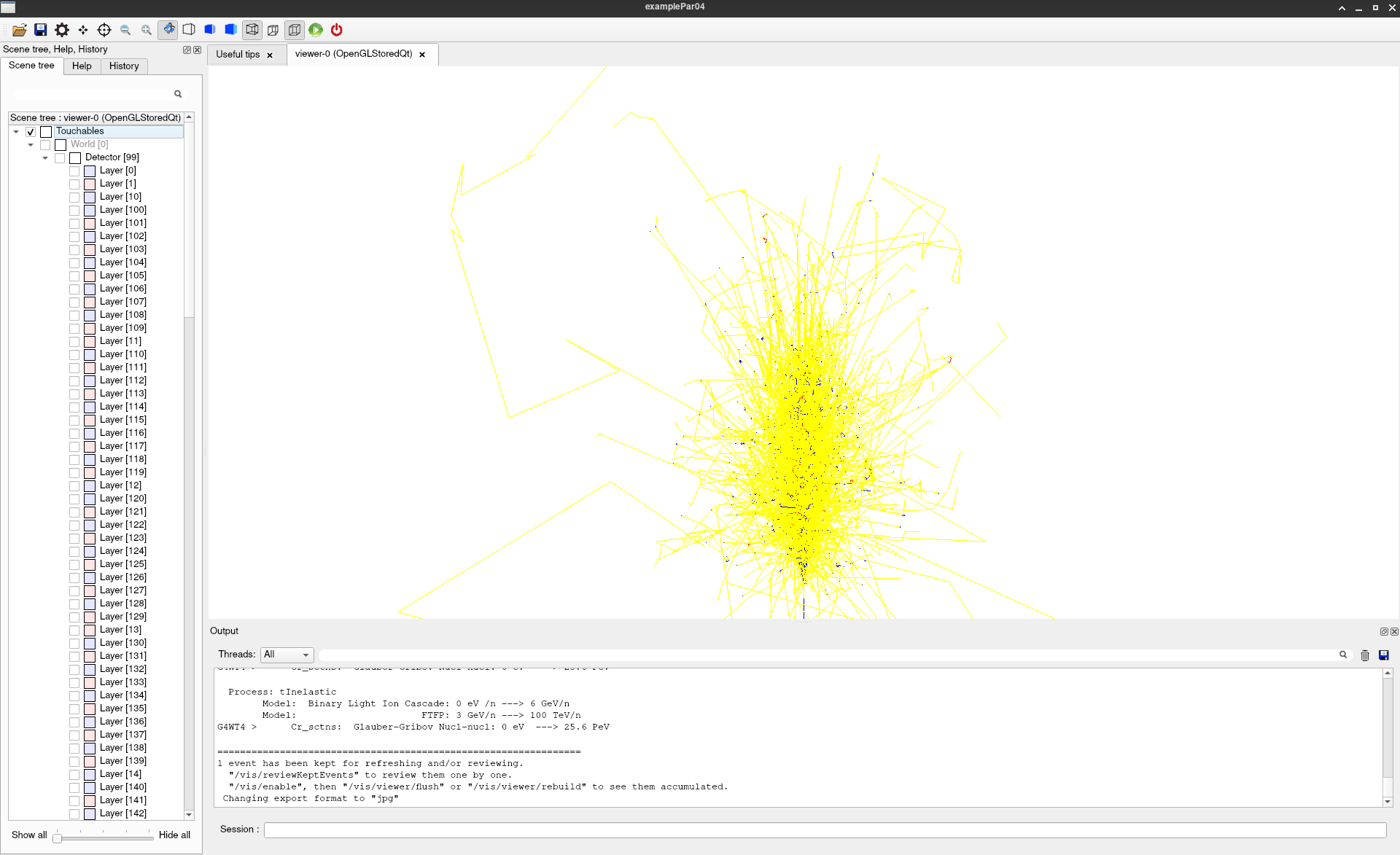}
    \caption{Particles above 0.1\,keV.}\label{visFull}
    \end{subfigure}~
    \begin{subfigure}{.33\textwidth}
    \centering
    \includegraphics[width=\textwidth,trim={30cm 11.5cm 20cm 8cm},clip]{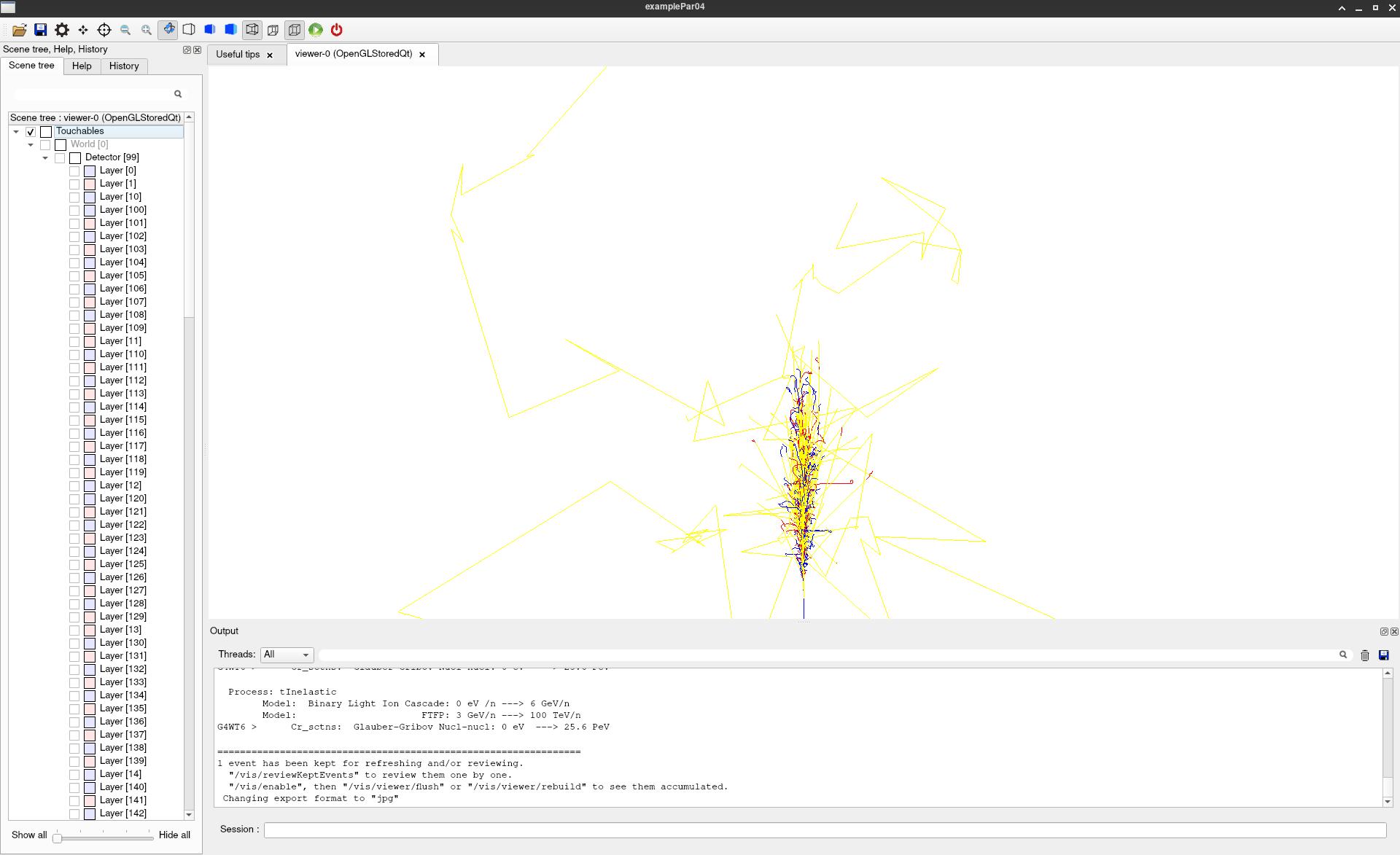}
    \caption{Particles above 10\,MeV.}\label{visMid}
    \end{subfigure}~
    \begin{subfigure}{.33\textwidth}
    \centering
    \includegraphics[width=\textwidth,trim={30cm 11.5cm 20cm 8cm},clip]{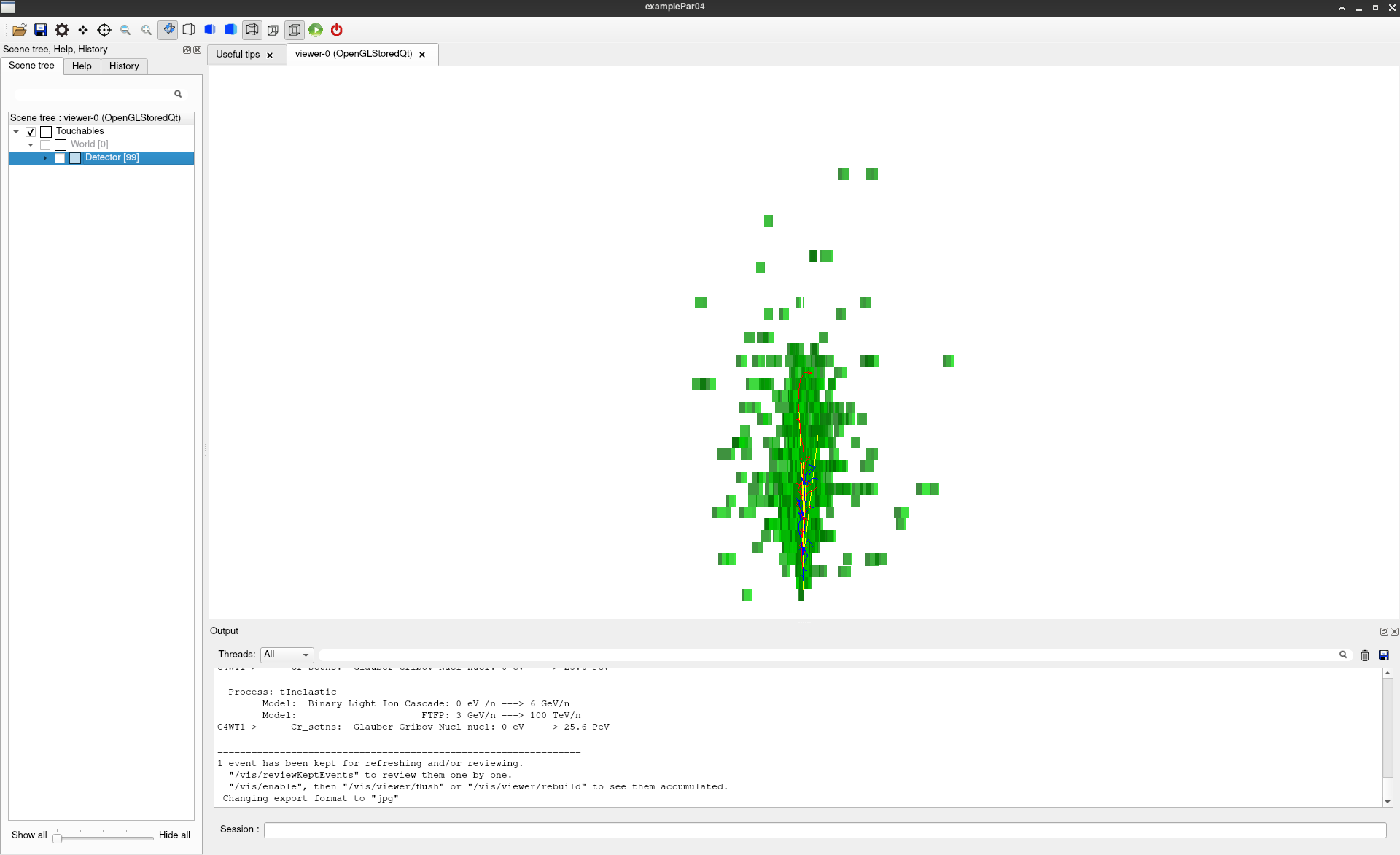}
    \caption{Particles above 100\,MeV and all the energy deposits.}\label{visDep}
    \end{subfigure}%
    \caption{Particle cascade simulation originated from a 10\,GeV electron traversing a Par04 detector~\cite{Par04}. Photons are depicted in yellow, electrons in blue, and positrons in red. Deposited energy in scored in voxels indicated in green.}
  \label{fig:shower_steps}
\end{figure}

\subsection{Shower Representation}
Our objective is to model the properties of individual particle showers in the ECal, as simulated by \textsc{Geant4} \cite{agostinelli2003geant4}. Specifically, the focus is on the spatial pattern of energy deposits left in the detector as particles interact with its material, an example of which is illustrated in Figure \ref{fig:shower_steps}. Instead of focusing on the signal from the entire volume of the calorimeter, only a region around the incident particle is considered, as it contains the produced shower.

\begin{figure}[htbp]
  \centering
  \begin{subfigure}[b]{0.56\linewidth}
    \centering
    \includegraphics[width=\linewidth]{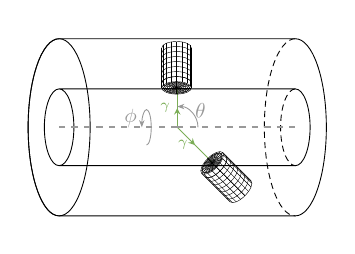}
    \caption{}
    \label{fig:mesh_within_detector}
  \end{subfigure}
  \hfill
  \begin{subfigure}[b]{0.43\linewidth}
    \centering
    \includegraphics[width=\linewidth]{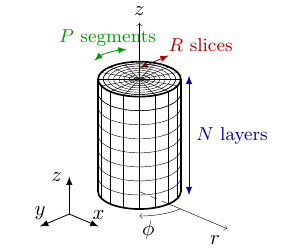}
    \caption{}
    \label{fig:mesh}
  \end{subfigure}
  \caption{Universal Representation of Showers is a cylindrical virtual (dynamic) mesh: (a) Placement of the mesh inside the detector's calorimeter. (b) Regular division of the mesh into voxels \cite{Par04}.}
\end{figure}

We adopt the 3D cylindrical mesh introduced in the Par04 example \cite{Par04} and called the Universal Representation of Showers~\cite{mckeown2025lemursdatasetlargescalemultidetector}. This physics-inspired mesh is aligned along the incoming particle direction, as shown in Figure \ref{fig:mesh_within_detector}. It also exploits the symmetry of the shower; in other words, this representation has an approximate azimuthal symmetry, and captures the physical intuition of shower evolution being decomposed into longitudinal and radial components, as depicted in Figure \ref{fig:mesh}. Moreover, the detector geometry-agnostic nature of this scoring (voxels have finer granularity than the detector readout) allows it to be used for complex detector geometries while providing a consistent representation for training a model. The size of the mesh is chosen to contain the most energetic showers in our dataset. The granularity of the mesh is chosen to preserve relevant shower observables whilst being efficient for the generative model. For now, we adopt the same configuration as used in the CaloChallenge Dataset-2~\cite{faucci_giannelli_2022_6366271}, with a resolution of $9 \times 16 \times 45$ in the $r \times \phi \times z$ directions, respectively.

\subsection{Dataset}
The development of a shower depends on the properties of the incident particle and the detector structure and material composition. In our setup, we restrict our focus to showers generated by photons ($\gamma$) with a given energy ($E$), and we represent the local region indirectly through two angles: $\phi$ and $\theta$. Consequently, the generative model is conditioned on $(E, \phi, \theta)$.  

The dataset produced for this study is now published under the name LEMURS (Large-scale multi-detector ElectroMagnetic Universal Representation of Showers)~\cite{zaborowska_2025_17045562} and described in more detail in~\cite{mckeown2025lemursdatasetlargescalemultidetector}. The dataset comprises 1 million simulated showers per detector. The photon energy $E$ spans a wide dynamic range from 1 GeV to either 100 GeV or 1 TeV (depends on the detector), while $\phi$ ranges from $0$ to $2\pi$, and $\theta$ from $0.87$ to $2.27$ radians. All three parameters are sampled uniformly, allowing us to uniformly cover most of the barrel region of each ECal. We emphasize the importance of sampling energies from a uniform spectrum. For machine learning models, the sampling distribution is generally less critical than ensuring the training data is unbiased across the conditioning variables. Unless otherwise noted, we use 900,000 showers for training and reserve 100,000 for validation for each detector type.  

In this paper, we consider the following detectors and samples from the LEMURS dataset:  

\textbf{Par04 Detectors:}
Based on the \textsc{Geant4} Par04 example \cite{Par04}, we include two simplified calorimeter setups, namely, \textit{Par04-SiW} and \textit{Par04-SciPb}. These detectors serve as proof-of-concept models and as a dataset for model development. In particular, Par04-SiW has been widely adopted in the CaloChallenge \cite{krause2024calochallenge} to define Dataset-2 and Dataset-3. Both detectors consist of concentric cylindrical layers alternating between sensitive and absorber materials. Par04-SiW uses silicon as the sensitive material and tungsten as the absorber, while Par04-SciPb uses a scintillator and lead. The differing material compositions and layer thicknesses lead to distinct shower development patterns in each detector.  

\textbf{Open Data Detector (ODD):}
The \textit{Open Data Detector} (ODD) \cite{Gessinger-Befurt:2023snx} is a realistic benchmark detector developed for algorithmic research in areas including tracking, simulation, and reconstruction. It features a detailed silicon-based calorimeter and includes elements typically present in actual detectors, such as air gaps, support structures, and electronics. This makes its simulation more representative of real detectors.  

\textbf{Detectors for Future Colliders:}
We also consider two proposed detector concepts for future lepton colliders such as FCC-ee, namely, \textit{CLD} \cite{CLDBacchetta:2019fmz} and \textit{ALLEGRO} \cite{ALLEGRO}. We refer to these as \textit{FCCeeCLD} and \textit{FCCeeALLEGRO}, respectively. Note that for both the incident energy is limited to the range of 1 GeV to 100 GeV as those detectors are proposed for an electron--positron collider. These detectors feature realistic geometry descriptions that account for non-calorimeter components such as support structures, air, cables, electronics, and cooling elements. FCCeeCLD employs a highly granular silicon-tungsten calorimeter, with planar calorimeter layers arranged in a polyhedral barrel. In contrast, FCCeeALLEGRO uses a liquid argon lead calorimeter, with a turbine-like orientation of layers in a cylindrical barrel. Since both detectors are still in development, their designs undergo frequent changes.  

\begin{figure}[htbp]
  \centering
  \begin{subfigure}[b]{0.49\linewidth}
    \centering
    \includegraphics[width=\linewidth]{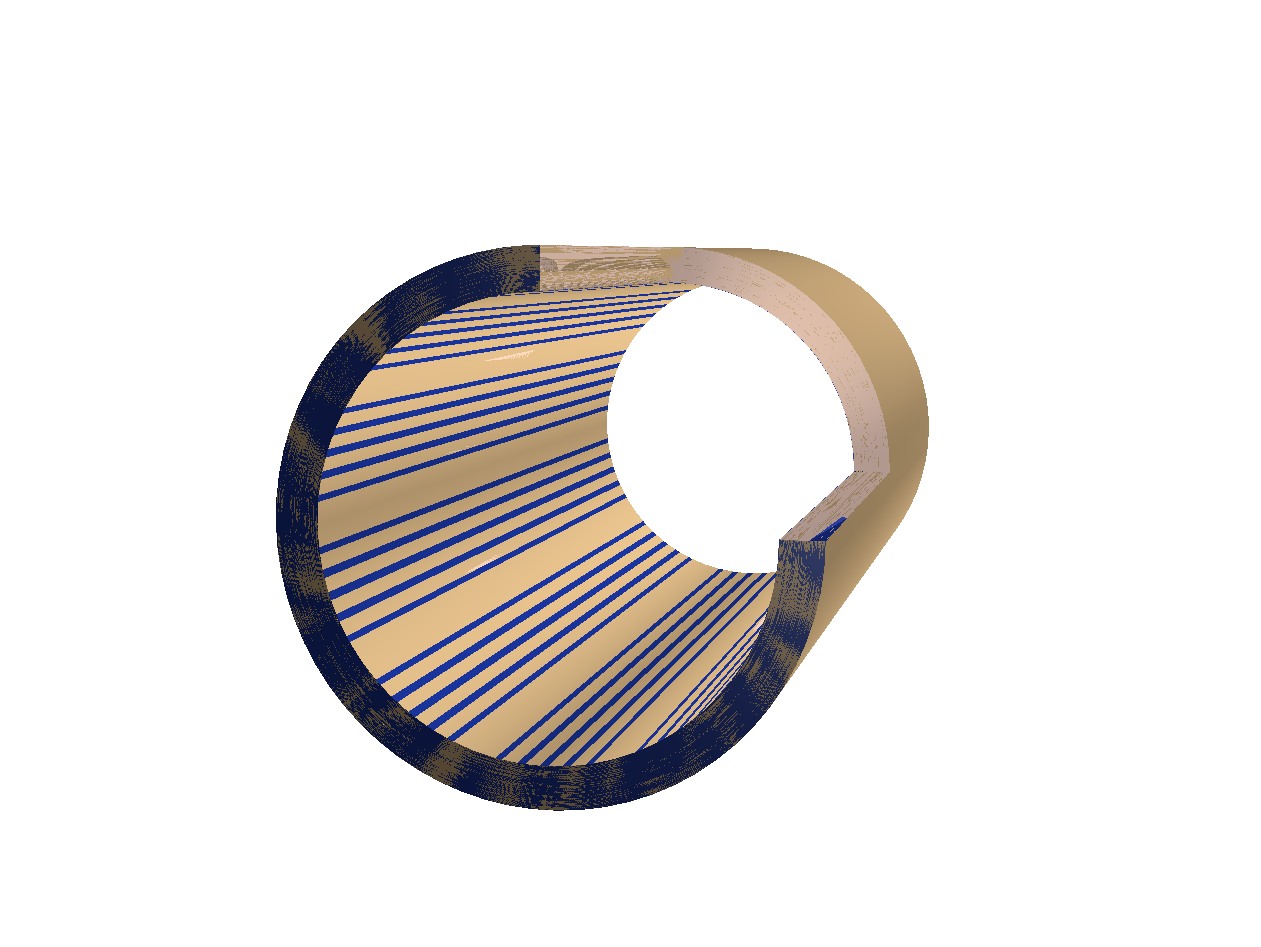}
    \caption{Par04 detectors}
    \label{fig:par04_detectors}
  \end{subfigure}
  \hfill
  \begin{subfigure}[b]{0.49\linewidth}
    \centering
    \includegraphics[width=\linewidth]{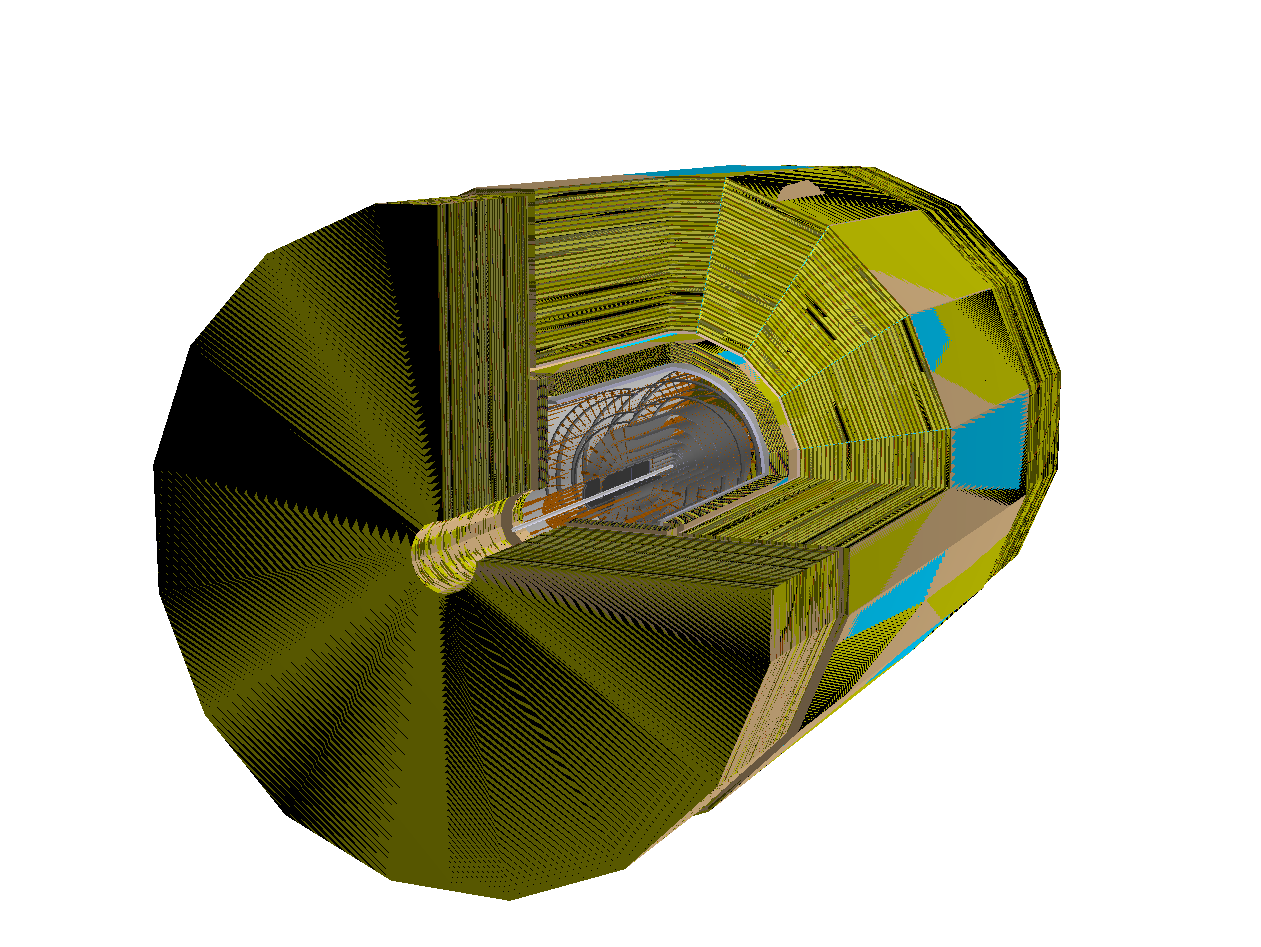}
    \caption{ODD}
    \label{fig:odd}
  \end{subfigure}
  \hfill
  \begin{subfigure}[b]{0.49\linewidth}
    \centering
    \includegraphics[width=\linewidth]{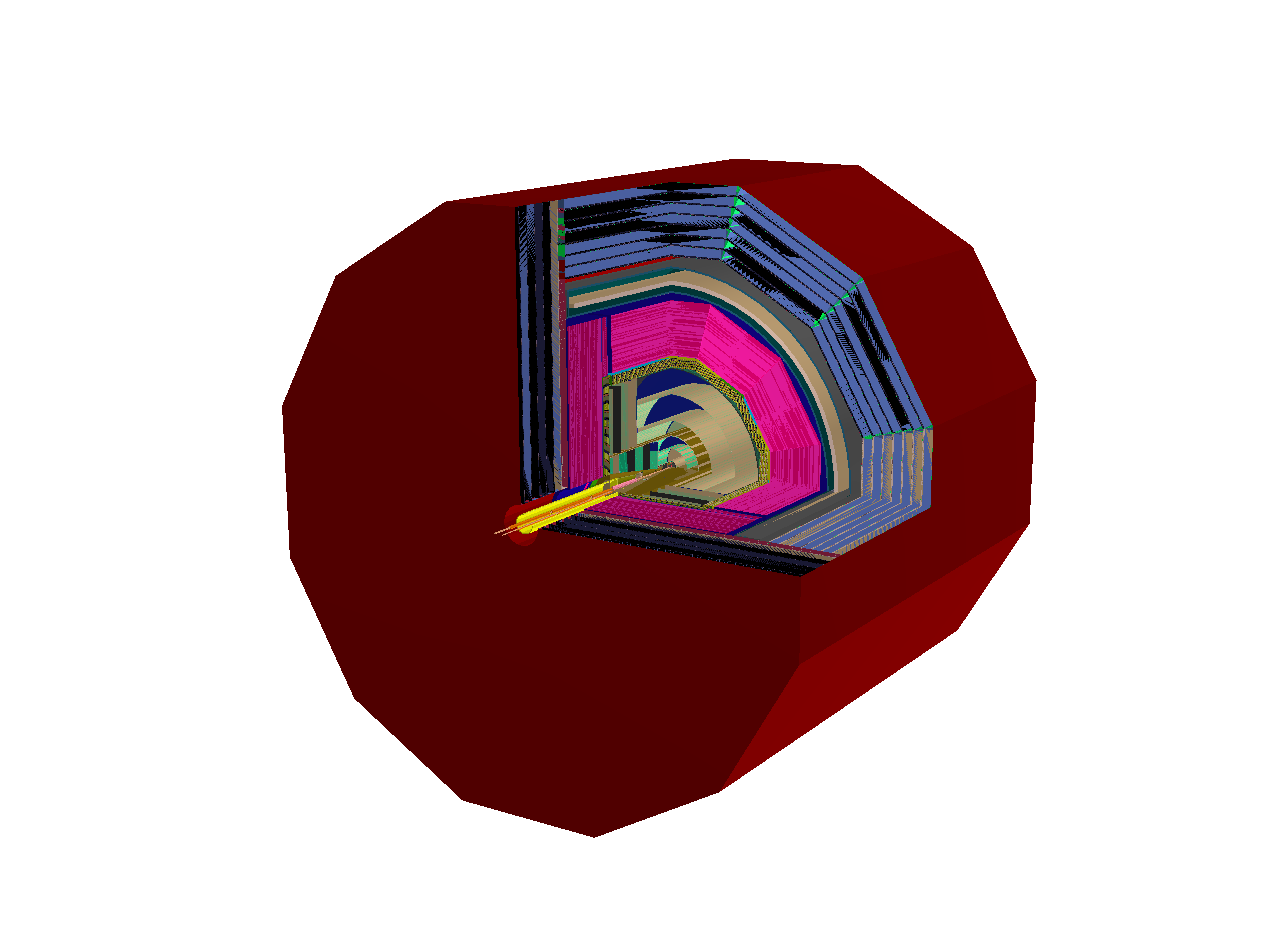}
    \caption{FCCeeCLD}
    \label{fig:cld}
  \end{subfigure}
  \hfill
  \begin{subfigure}[b]{0.49\linewidth}
    \centering
    \includegraphics[width=\linewidth]{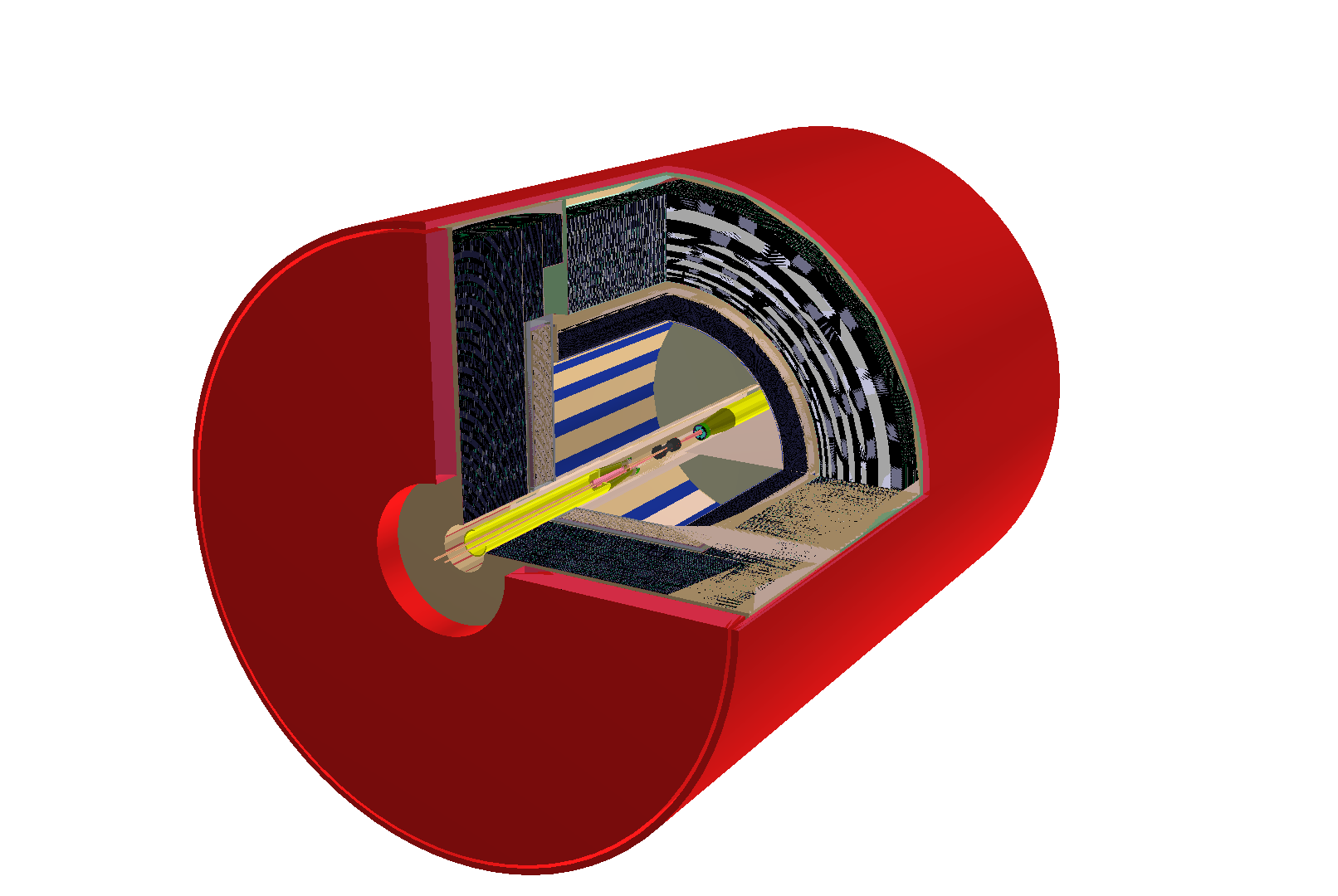}
    \caption{FCCeeALLEGRO}
    \label{fig:allegro}
  \end{subfigure}
  \caption{Geometry displays of the various detectors used in this study. While Par04-SiW \& Par04-SciPb follow a simple geometry (a), others consist of complex systems (b, c, d). All detectors are visualised using DD4hep \cite{frank_markus_2018_1464634} GeoDisplays}
  \label{fig:detector_sketches}
\end{figure}

Figure \ref{fig:detector_sketches} illustrates the varying geometric complexity of calorimeter systems across different detectors with varied levels of realism in their geometry. The nuances introduced by the realistic alignment of detector modules, combined with the sophisticated design necessary for precise particle measurements, can introduce significant irregularities in the spatial structure of the detector geometry. As a result, a model trained on data from a specific local region cannot necessarily be directly applied across the entire calorimeter.  

In this paper, we first use Par04-SiW sample of LEMURS dataset to develop and evaluate a single-detector baseline model (see Section \ref{sec:results_single_detector}). For pre-training our generalisable model, we leverage four detectors - Par04-SiW, Par04-SciPb, ODD, and FCCeeCLD. To assess the generalization capability of the generalisable model, we use FCCeeALLEGRO for adaptation, which is discussed in Section \ref{sec:generic_model}. FCCeeALLEGRO's more complex geometry and distinct calorimeter make it a valuable but challenging testbed for evaluating the model's ability to adapt to new detectors.   

\subsection{Validation Metrics}
\label{sec:valmetrics}
Since particle showers are inherently stochastic, comparisons between \textsc{Geant4}-simulated and model-generated data are typically performed using distributions of higher-level quantities, referred to as shower observables. Throughout this paper, we use the term \textit{accuracy} qualitatively to denote the degree of agreement between these distributions.  

Following standard practices in HEP, the comparisons are carried out under fixed-point conditions. Specifically, we validate on a test dataset consisting of 1,000 showers, each simulated by firing a particle with fixed energy and direction. Ideally, generated showers should be placed back into the detector geometry for evaluation. However, due to differences in each detector readout, we operate on the intermediate step that allows us to compare all shower observables directly on our virtual mesh representation. This validation would need to be taken one step further into the detector-specific readouts for each of the new adaptations.

The key shower observables considered in this paper include the following:  
\begin{enumerate}
    \item \textit{Voxel energy distribution} - is the distribution of voxel energies. An entry corresponds to the $i^{th}$ voxel $x_i$.
    \item \textit{Total energy distribution} - is the distribution of total energy of all voxels in a shower. An entry corresponds to $\sum_{r,\varphi,z} x_{i}$. Total energy is a crucial observable, useful in determining the type and energy of a particle in the reconstruction chain.  
    \item Marginal profiles - are the distributions along each dimension of the mesh. The \textit{longitudinal profile} shows how energies are distributed along the z-axis, where an entry corresponds to $\sum_{r, \varphi} x_{i}$. Similarly, the \textit{transverse profile} corresponds to an entry being $\sum_{\varphi, z} x_{i}$, and the \textit{phi profile} corresponds to an entry being $\sum_{r, z} x_{i}$.  
\end{enumerate}

In general, longitudinal, transverse, and phi profiles are relatively easy for machine learning models to learn and are typically the first observables examined to assess whether the model is functioning correctly. In contrast, the voxel energy distribution poses a greater challenge due to the wide dynamic range of energy values, spanning up to several orders of magnitude. Similarly, accurately predicting the total energy, obtained by summing over all voxels, is non-trivial for the model. We discuss these challenges in more detail in Section \ref{sec:training_dynamics}.  

In addition to these core shower observables, we also evaluate our models using quantitative metrics introduced by the CaloChallenge~\cite{krause2024calochallenge}, as detailed in the later sections of the paper.  

\section{Methodology}
\label{sec:methodology} 

\subsection{Preliminaries}
In this section, we describe two key components of the generative process underlying CaloDiT-2: the diffusion process and diffusion distillation. The diffusion process provides a principled framework for iterative generative modeling, while diffusion distillation enables the conversion of a multi-step diffusion process into a single-step generative model without significant loss in performance.  

\subsubsection{EDM Diffusion}
\label{subsec:edm}
In this work, we adopt the continuous-time diffusion framework (EDM) introduced in \cite{karras2022elucidating_EDM_diffusion}. One key advantage of using a continuous rather than a discrete time diffusion process is that it decouples the procedure used to train the model from the sampling. This provides access to a broader suite of sampling tools, including both ordinary differential equation (ODE) and stochastic differential equation (SDE) solvers, as well as procedures for reducing the number of diffusion steps during sampling. Details on the approach used for reducing the number of sampling steps will be given in Section \ref{subsec:CD}.

Following \cite{song2021scorebased_generativemodelling}, a score-based approach to diffusion is as follows. The forward diffusion occurs across continuous time variable $t \in [0,T], T>0$, and the process maps the initial data distribution $p_{t=0}(x)$ at time $t=0$ onto some tractable distribution $p_{t=T}(x)$ at time $t=T$, where $p_{t=T}(x)$ is typically chosen to be a fixed mean and variance Gaussian. The forward process is therefore described by the SDE
\begin{equation}
dx = \mu(x,t) \ dt + \sigma(t) \ dW,
\end{equation}
where $\mu$ and $\sigma$ are the drift and diffusion coefficients respectively and $W$ is a standard Wiener process. 

For such an SDE, there exists also a reverse diffusion process, given by the reverse-time SDE
\begin{equation} \label{eq:RevDiffSDE}
dx = \big[ \mu(x,t) - \sigma(t)^2 \ \nabla_{x} \log p_{t}(x) \big] \ dt + \sigma(t) \ d\overline{W},
\end{equation}
where the time variable now runs backwards from $T$ to $0$, and $\overline{W}$ is a standard Wiener process. $\nabla_{x} \log p_{t}(x)$ is known as the score function, the analytical form of which is in general unknown, but which we aim to approximate with a neural network $s_{\phi}(x,t)$ with weights $\phi$. Since for the reverse-time SDE given in equation \ref{eq:RevDiffSDE}, there exists a probability flow ODE \cite{Maoutsa_2020} from noise to data given by 
\begin{equation} \label{eq:PFODE}
dx = \big[ \mu(x,t) -\frac{1}{2} \ \sigma(t)^2 \ \nabla_{x} \log p_{t}(x) \big] \ dt \ \cite{song2021scorebased_generativemodelling},
\end{equation}
it is then possible to sample from $p_{t=0}(x)$. We set $\mu(x,t) = 0$ and $\sigma(t)=\sqrt{2}t$, following \cite{karras2022elucidating_EDM_diffusion}.

Our actual implementation follows the approach described in \cite{karras2022elucidating_EDM_diffusion}. Following these recommendations, rather than directly training the neural network $s_{\phi}(x,t)$, a separate neural network $v_{\theta}$ is trained. The objective can then be decomposed as

\begin{equation}
s_{\phi}(x,t) = c_{\text{skip}}(t) \ x + c_{\text{out}}(t) \cdot v_{\theta}(c_{\text{in}}(t) \ x, c_{\text{noise}}(\sigma)),
\end{equation}

    where the coefficient $c_{\text{skip}}=\frac{\sigma_{\text{data}}^{2}}{\sigma_{\text{data}}^{2} + t^{2}}$ controls the skip connection, the coefficient $c_{\text{in}}= t \cdot \frac{\sigma_{\text{data}}}{\sqrt{\sigma_{\text{data}}^2 +t^{2}}}$ controls the input scaling and $c_{\text{out}} =\frac{1}{\sqrt{\sigma_{\text{data}}^2 +t^{2}}}$ controls the output scaling and $c_{\text{noise}}=\frac{\kappa}{4}\ln{t}$ controls a scaling from noise level to conditional input to $v_{\theta}$, where $\kappa=1\times10^{4}$ is a factor chosen empirically \cite{karras2022elucidating_EDM_diffusion}. The parameter $\sigma_{\text{data}}=0.5$ is chosen to approximately match the standard deviation of the data distribution, while $\sigma_{\text{min}}=0.002$ and $\sigma_{\text{max}}=80$ represent the minimum and maximum noise levels. During training, a timestep is drawn from the distribution $\ln{t}\sim\mathcal{N}(P_{\text{mean}}, P_{\text{std}}^{2})$, with $P_{\text{mean}}=-1.2$ and  $P_{\text{std}}=1.2$. In each case, the default values for the parameters from \cite{karras2022elucidating_EDM_diffusion} have been used. The loss is then calculated via

\begin{equation}
\mathcal{L_{\text{EDM}}} = \mathbb{E}\big[ \lambda(t)|| s_{\phi}(x, t) - x_{0}||_{2}^{2}\big],
\end{equation}

where $x_{0}$ is a sample drawn from $p_{t=0}(x)$ and $\lambda$ is a time dependent loss weighting, given by $\lambda(t) = \frac{t^2+\sigma_{\text{data}}}{t \ \cdot \ \sigma_{\text{data}} }$. Sampling is performed using the $2^{\text{nd}}$ order Heun ODE solver from \cite{karras2022elucidating_EDM_diffusion}, using $32$ steps.

\subsubsection{Consistency Distillation} \label{subsec:CD}
Consistency distillation \cite{song2023consistencymodels} is a means of reducing the number of diffusion steps in a diffusion model. In this work we employ consistency distillation to reduce the number of steps of an EDM diffusion model. This approach relies on enforcing \textit{self-consistency} of a learnt function $f \ : \ (x_{t},t) \longmapsto x_{\epsilon}$; i.e. requiring that at any two points on the probability flow ODE trajectory (E.q. \ref{eq:PFODE}) return an identical result from $f$, i.e. $f(x_{t}, t) = f(x_{t'},t') \ \forall \ t,t' \in [\epsilon, T]$.

A diffusion model can be distilled to a consistency model from a trained (continuous-time) diffusion model as follows. The training involves a \textit{teacher model} $s_{\phi}(x,t)$, which is the initial, pre-trained diffusion model, a \textit{student model} $f_{\zeta}(x,t)$, which is actively trained during the procedure, and a \textit{target model} $f_{\zeta^{-}}$, which acts as the final target of distillation. The space of continuous time $[\epsilon, T]$ is discretized into $N-1$ sub-intervals- in this study we employ $32$ discretization steps. In each training step, a random discretized time step $t_{n}$ where $n \ \in \ [1, N]$ is selected. Enforcing the self-consistency condition gives a training loss for the student model of 

\begin{equation}
\mathcal{L_{\text{CD}}} = \mathbb{E}\big [ \lambda(t_{n})|| f_{\zeta}(x_{t_{n+1}}, t_{n+1}) - f_{\zeta^{-}}(x_{t_{n}, t_{n}}) ||^{2}_{2} \big],
\end{equation}
where $\lambda(t_{n})$ is a time-dependent loss scaling, set to unity \cite{song2023consistencymodels}.

The student model can then be trained via stochastic gradient descent. The weights of the target model $\zeta^{-}$ are updated after every iteration as an exponential moving average of the student models' weights $\zeta$. Our sampling is performed in the same manner as described in Section \ref{subsec:edm}, but with a single solver step.

\subsection{CaloDiT-2}

Here, we introduce CaloDiT-2, starting with the architectural modifications designed to accommodate our shower data representation, followed by details of the preprocessing and training setup. We note that a preliminary version of our model, CaloDiT-1 \cite{krause2024calochallenge}, was previously submitted to the CaloChallenge. That version relied on DDPM \cite{ho2020denoising} for its diffusion process, required 400 reverse diffusion steps per shower, and lacked key architectural modifications. We compare them in Section \ref{subsec:cc_results} via CaloChallenge Dataset-2 setup.    

\subsubsection{The Model Architecture}

\begin{figure}[htbp]
  \centering
  \includegraphics[width=0.8\linewidth]{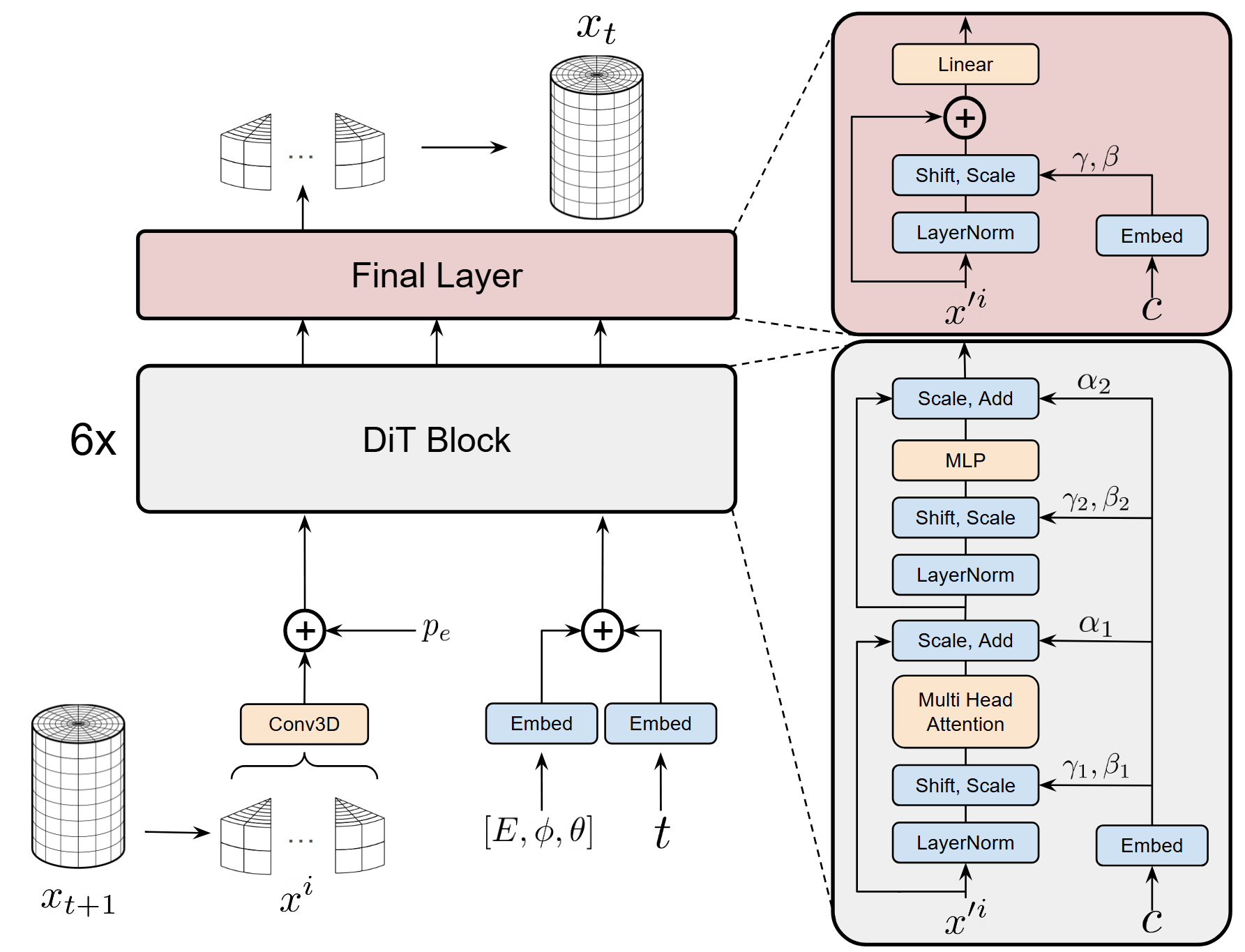}
  \caption{Diagram illustrating the architecture of CaloDiT-2.}
  \label{fig:arch}
\end{figure}

\iftoggle{oldarch}
  {
    CaloDiT-2’s architecture is based on the diffusion transformer (DiT) model \cite{peebles2023scalable}, with modifications that enable it to handle a full 3D \textit{image} of a calorimeter shower. Unlike DiT’s hybrid latent-diffusion design, CaloDiT-2 is an end-to-end diffusion model built entirely on a transformer backbone. We deliberately avoid VAE-like components, as they often produce overly smoothed or blurry showers \cite{Raikwar:2024peb}.    

    As described in Section \ref{subsec:edm}, the diffusion process is iterative. The model’s input has the same dimensions as the required output, i.e. $9 \times 16 \times 45$. A noisy shower ($x_{t+1}$) is passed to the network together with the conditioning variables of the diffusion process. At the end, the model outputs the denoised shower ($x_t$). The complete architecture is shown in Figure \ref{fig:arch}.  

    The conditions include the properties of the incident particle, $E$, $\phi$ \& $\theta$, and the diffusion timestep, $t$. The incident particle's properties are concatenated and processed through a nonlinear projection, while $t$ passes through a separate projection layer. The resulting vectors are concatenated into a condition embedding $c$, which is provided to every DiT block. The shower itself is split into patches, which are then fed through the stack of DiT blocks and recombined at the output stage.      

    Because the model is transformer-based, showers must be tokenized. We extend ViT-style patching \cite{dosovitskiy2020image} to three dimensions. Although patches of size $1 \times 1 \times 1$ would maximize the use of the attention mechanism, they are prohibitively expensive. In practice, we found that $3 \times 2 \times 3$ patches offer a good trade-off between accuracy and computational efficiency, resulting in 360 tokens. Overall, the shower ($x$) is first projected with a shared Conv3D layer to form patches, to which the positional embeddings ($p_e$) are added, yielding $x'$. These patches are then passed to the stack of DiT blocks. Positional embeddings are also extended to 3D. The $r$, $\varphi$, and $z$ directions each occupy one-third of the embedding space. As a baseline, we also tested 1D fixed and learned positional embeddings, but both underperformed compared to the 3D sinusoidal embeddings.   

    The model comprises 6 DiT blocks with an embedding dimension of 384. Each block uses 6 attention heads of dimension 64, and the MLP expands from 384 to 1536 before projecting back. After the final DiT block, the patches are modulated and linearly projected back into the original patch of shape $3 \times 2 \times 3$. These are then unpatchified to reconstruct the full shower of shape $9 \times 16 \times 45$. The size of the model is the result of having the necessity to allow for faster inference, while still allowing for enough capacity.    
  }
  {
    CaloDiT-2's architecture is based on the DiT model \cite{peebles2023scalable}, along with modifications that helped us better learn the shower representation. Unlike DiT’s hybrid latent-diffusion design, CaloDiT-2 is an end-to-end diffusion model built entirely on a transformer backbone. We deliberately avoid VAE-like components, as they often produce overly smoothed or blurry showers \cite{Raikwar:2024peb}.   

    \PR{Restructure}

    As described in Section \ref{subsec:edm}, the diffusion process works iteratively. Hence, the input to CaloDiT-2 is of the same dimensions as the required output shower, i.e. $9 \times 16 \times 45$ dimensional shower. A noisy shower ($x_{t+1}$) is fed to the model, along with the conditions for the diffusion process. At the end, the model outputs a denoised shower ($x_t$). The architecture is shown in Figure \ref{fig:arch}. The conditions include the properties of the incident particle, $E$, $\phi$ \& $\theta$, and the diffusion timestep, $t$. The properties, $E$, $\phi$ \& $\theta$ are concatenated and are passed through a non-linear projection, whereas $t$ goes through a separate projection layer. These two projections are concatenated to form a condition vector, $c$, which is then passed to each of the CaloDiT-2 blocks. As for the shower, it is first projected onto patches via a linear projection, which are then fed to the CaloDiT-2 blocks. The output from the model is obtained in the form of several patches. These are then stitched together to get the output shower. We now describe the patching and CaloDiT-2 block in detail.   
    
    \subsubsection{Patching}
    Since the architecture of CaloDiT-2 is transformer-based, the showers need to be split into tokens. We extend the ViT-like \cite{dosovitskiy2020image} patching to 3-dimensions. Ideally, the patches could have been $1 \times 1 \times 1$, thus utilizing the attention mechanism to the lowest detail. This is, however, computationally expensive. Empirically, we found that patches of size $3 \times 2 \times 3$ from $9 \times 16 \times 45$, respectively, offer a good trade-off between accuracy and speed. This results in the number of patches or the sequence length being 360. Similar to patches, the sinusoidal positional embeddings ($p_e$) are also extended to 3-dimensions by representing each direction, i.e. $r$, $\varphi$ \& $z$ in one-third of the embedding space. As a baseline, we explored 1D fixed positional embeddings and learned positional embeddings, both of which performed a bit worse than 3D positions. The patches are linearly projected using a shared layer, and the positional embeddings are added to them. These patches are then passed to the stack of CaloDiT-2 blocks.    

    \PR{Write better the factorized adaLN.}  
    
    \subsubsection{CaloDiT-2 Block}
    The CaloDiT-2 block proposes small, but important, modifications to the DiT block to suit our shower data. We start from the adaLN-zero DiT block. We replace the point-wise feedforward network with SwiGLU activation \cite{shazeer2020glu}. This helps in modelling a low-level physics observable better (will be discussed in the results). We replace the scale, $\alpha_1$ \& $\alpha_2$, by a gating mechanism. This is done by bounding the main branch and the residual branch via $\sigma(\alpha_i)$, where $\sigma$ is the sigmoid function. This helps in reducing the outliers and results in improved training stability. Moreover, as opposed to natural images, the showers can have very different representations depending on the position. With LayerNorm, patches are normalized irrespective of the position, and the affine transformation in adaLN-zero does not depend on the position of the patch, hence leading to a loss of information. That is why, along with the condition vector, $c$, we condition the calculation of affine parameters on positional embeddings, $p_e$. This allows the model to learn more expressive modulation based on the combination of both the position and condition for a single patch. Since this projection can be expensive to compute, we factorize it. That is, the final affine parameters are the result of the addition of the affine parameters calculated separately for positions and conditions.  

    The model consists of 4 such blocks with an embedding dimension of 144. There are 8 attention heads, with each head's dimension being 18. The projection in the SwiGLU block goes to and fro from the embedding dimension to a hidden layer's dimension of 576. After the final CaloDiT-2 block, the patches undergo modulation and are linearly projected to their original dimensions of $3 \times 2 \times 3$ via a linear layer, and then unpatchified to get the shower's dimensions of $9 \times 16 \times 45$. The size of the model is the result of a trade-off between faster inference and sufficient model capacity.    
  }

\subsubsection{Preprocessing}

\iftoggle{oldarch}{}{\PR{Add scale by factor? Exp running}}%

First, we clip the voxels below 15.15 KeV to 0. These are low-energy voxels and are usually within the bounds of the noise from the detector readout. Then, we employ a simple log transformation followed by normalization. The parameters for the normalization, $\mu$ \& $\sigma$, are scalars and are shared between voxels. 
\iftoggle{oldarch}
{}
{To keep the formulations consistent with the diffusion process, we scale $\sigma$ by a factor of 2. This makes the standard deviation of the preprocessed dataset 0.5, which is equal to the $\sigma_{data}$ used in EDM \cite{karras2022elucidating_EDM_diffusion}.}%
That is for a voxel $x_i$,
\begin{align}
\hat{x_i} &= \frac{\log(x_i + \epsilon) - \mu}{%
  \iftoggle{oldarch}{\;\sigma\;}{\;2\sigma\;}%
} 
\; \; \text{where}%
\iftoggle{oldarch}{}{%
  \label{eq:norm}%
} \\
\mu &= \mathbb{E}[\log(x_i + \epsilon)] \nonumber\\
\sigma &= \sqrt{\mathbb{E}[(\log(x_i + \epsilon)-\mu)^2]}, \nonumber \; \; \text{and}\\
\epsilon &= 10^{-6} \nonumber
\end{align}

As for the conditions, $E$ \& $\theta$ are scaled by a fixed constant and $\phi$ is made to preserve the cyclic nature as follows:
\begin{align*}
    \hat{E} &= \frac{E}{E_{max}} \\
    \hat{\theta} &= \frac{\theta}{\pi} \\
    \mathbf{\hat{\phi}} &= [sin(\phi), cos(\phi)]
\end{align*}

We also investigated the effect of removing the log transformation in preprocessing. Without the log transformation, learning the total shower energy distribution became easier since it remained a linear operation. However, the agreement of the voxel energy distribution, particularly at lower energies, degraded due to the large number of very low-energy voxels present in the data. For this reason, we decided to retain the log transformation in preprocessing. In addition, unlike some prior works \cite{Krause:2021ilc}, we do not normalize voxel energies on a per-layer basis by dividing by the total energy of that layer. While this scaling can simplify training, it disrupts correlations between adjacent layers and may introduce inconsistencies when adapting to new detectors; thus, we chose to preserve the raw correlations.    

\subsubsection{Training Setup}
\label{sec:training_dynamics}

The training of CaloDiT-2 occurs in two parts. At first, the diffusion process (EDM) is trained to generate the showers. The accuracy of the EDM model is validated by sampling using 32 diffusion steps. Once this process has been determined to generate sufficiently accurate showers, it is then followed by the consistency distillation process (CD). Here, the EDM model trained earlier is used as the teacher model. This teacher model is then frozen and distilled into the target model, allowing for a single-step diffusion process. Note that, unless otherwise stated, both EDM and CD use the default hyperparameters ($\sigma_{min}$, $\sigma_{max}$, $\rho$, etc.), and can be referred to in the original papers \cite{karras2022elucidating_EDM_diffusion, song2023consistency}. \iftoggle{oldarch}
  {}
  {We keep $\sigma_{data}$ unchanged, hence the factor of 2 in the equation \ref{eq:norm}.}%

We train both the EDM and the CD model for \iftoggle{oldarch}{100K}{400K} steps with the AdamW \cite{loshchilov2019decoupledweightdecayregularization} optimizer, using a maximum learning rate of $0.0005$, weight decay of \iftoggle{oldarch}{$0.0001$}{$0.00001$}, and batch size of $256$. We utilize the \textit{wsd} scheduler \cite{hu2024minicpm, wen2025understanding} for the learning rate, where the first 10\% of steps are warm-up steps following a linear schedule. $50\%$ of the steps employ a stable learning rate equal to the maximum. For the remaining $40\%$ of the steps, we use a square root decay.  

We utilize the AdamW optimizer with a low weight decay as a conservative measure. Given the small size of the model, the huge size of the dataset, and the complexity of the shower modality, we are not in an overparameterized regime, and thus a small amount of weight decay should be sufficient. We also experimented with decaying non-bias parameters, but the results were mostly similar; hence, we go for a simpler formulation. In contrast, in diffusion processes, the learning signal from the gradients is not easily interpretable for the model. Hence, we avoid learning rate schedulers like \textit{ReduceLROnPlateau} \cite{paszke2019pytorch} or step-wise learning rate. \iftoggle{oldarch}{}{Further, we observed that reducing $\beta_2$ of AdamW helped in stabilizing the gradients. This was more apparent in wider or deeper networks; however, as a conservative measure, we use a value of $0.95$ for $\beta_2$.}  

Furthermore, we utilize the EMA (exponential moving average) to stabilize the EDM training, which is a standard technique in the literature \cite{morales2024exponential, grill2020bootstrap}. We use a scheduler to vary the EMA decay throughout the training. The decay increases from 0 and caps at 0.999. For CD, we do not use the EMA for stability, but rather for updating the student model to the target model (eq. 8 in \cite{song2023consistency}). This decay, $\mu=0.95$, is taken from the paper, and we keep it as it is. Note that it is possible to incorporate the standard EMA for stability, but the memory requirements are already quite high due to maintaining a teacher model (EDM), a student model, and a target model for CD training.  

Furthermore, while training the distillation process, we initialize the student model from the teacher model. Hence, they share the same architecture. We also tried random initialization of the student model weights, but the results were slightly worse. We use 32 discretization steps for the CD training (which is assigned the variable $N$ in the original paper \cite{song2023consistency}). This value is chosen heuristically.

\section{Results}
\label{sec:results_single_detector}
In this section, we present the results of CaloDiT-2 for the standard literature case of training on a single detector. This simplified scenario was also used to explore the model architecture and the training setup. The conclusions from here were used for the subsequent pre-training of the model on multiple detectors, the results of which are presented in the later sections of the paper.  

\subsection{Single Detector Performance}
\label{sec:results_par04}
Here, we show the results for the simplistic Par04-SiW detector with CaloDiT-2 for both the diffusion process (EDM) and the distillation process (CD). The sampling for EDM is done with 32 steps. For CD, single-step sampling is used.  
\begin{figure}[htbp]
\begin{adjustwidth}{-\figenvleftextend}{-\figenvrightextend}
  \centering
  \begin{subfigure}[b]{0.49\linewidth}
    \centering
    \includegraphics[height=\imgheight]{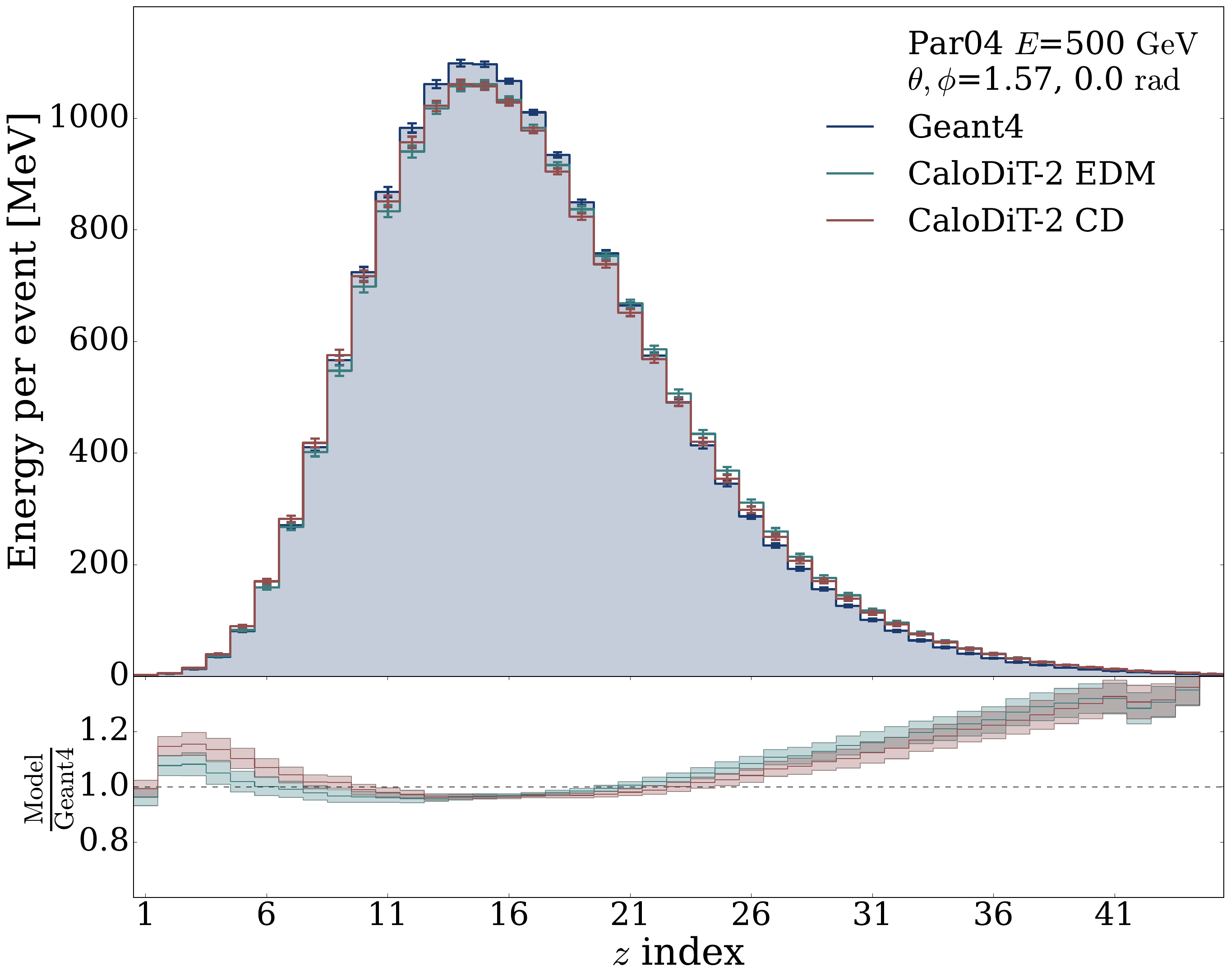}
    \caption{Longitudinal profile}
  \end{subfigure}
  \hfill
  \begin{subfigure}[b]{0.49\linewidth}
    \centering
    \includegraphics[height=\imgheight]{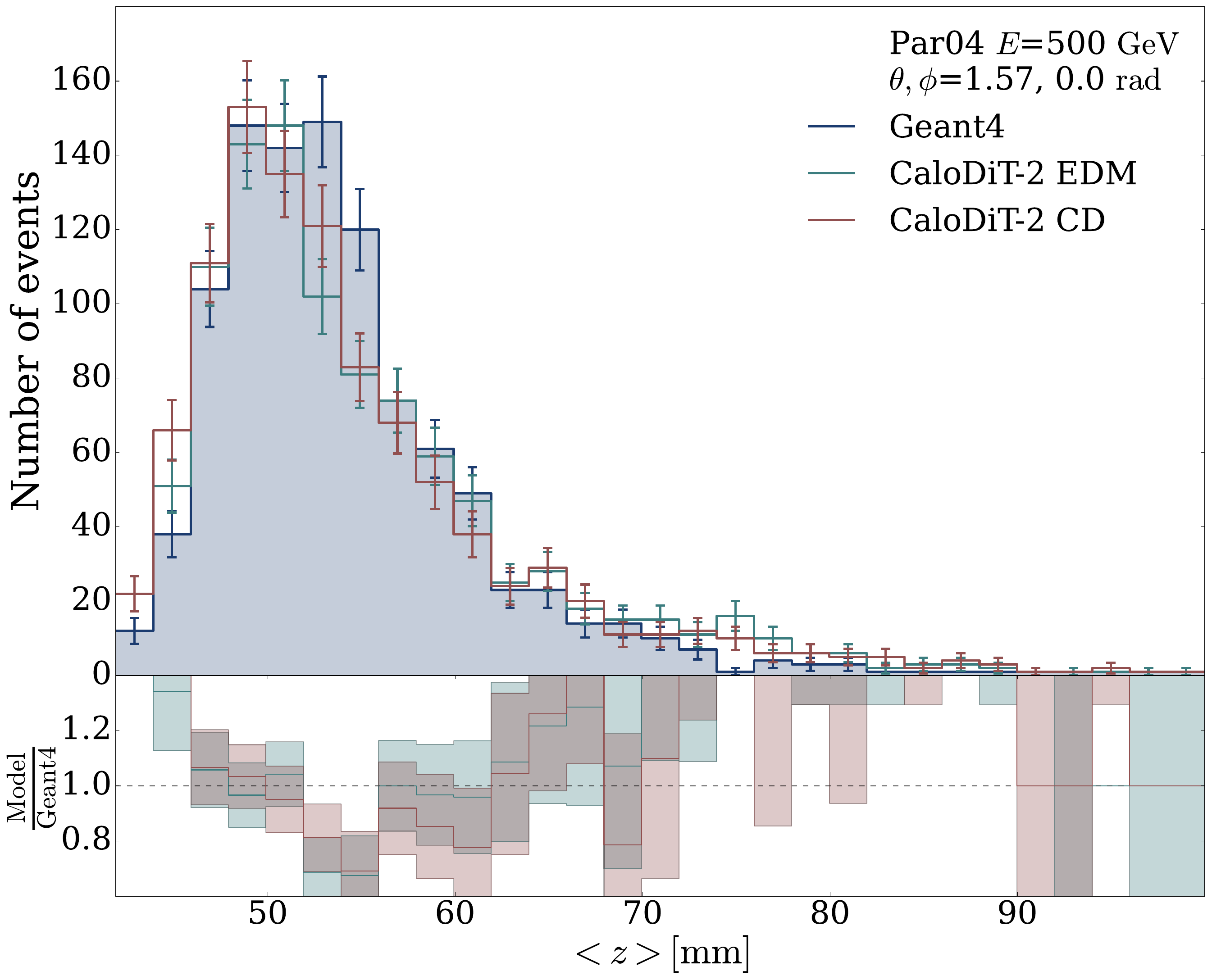}
    \caption{First moment of longitudinal profile}
  \end{subfigure}
  \hfill
  \begin{subfigure}[b]{0.49\linewidth}
    \centering
    \includegraphics[height=\imgheight]{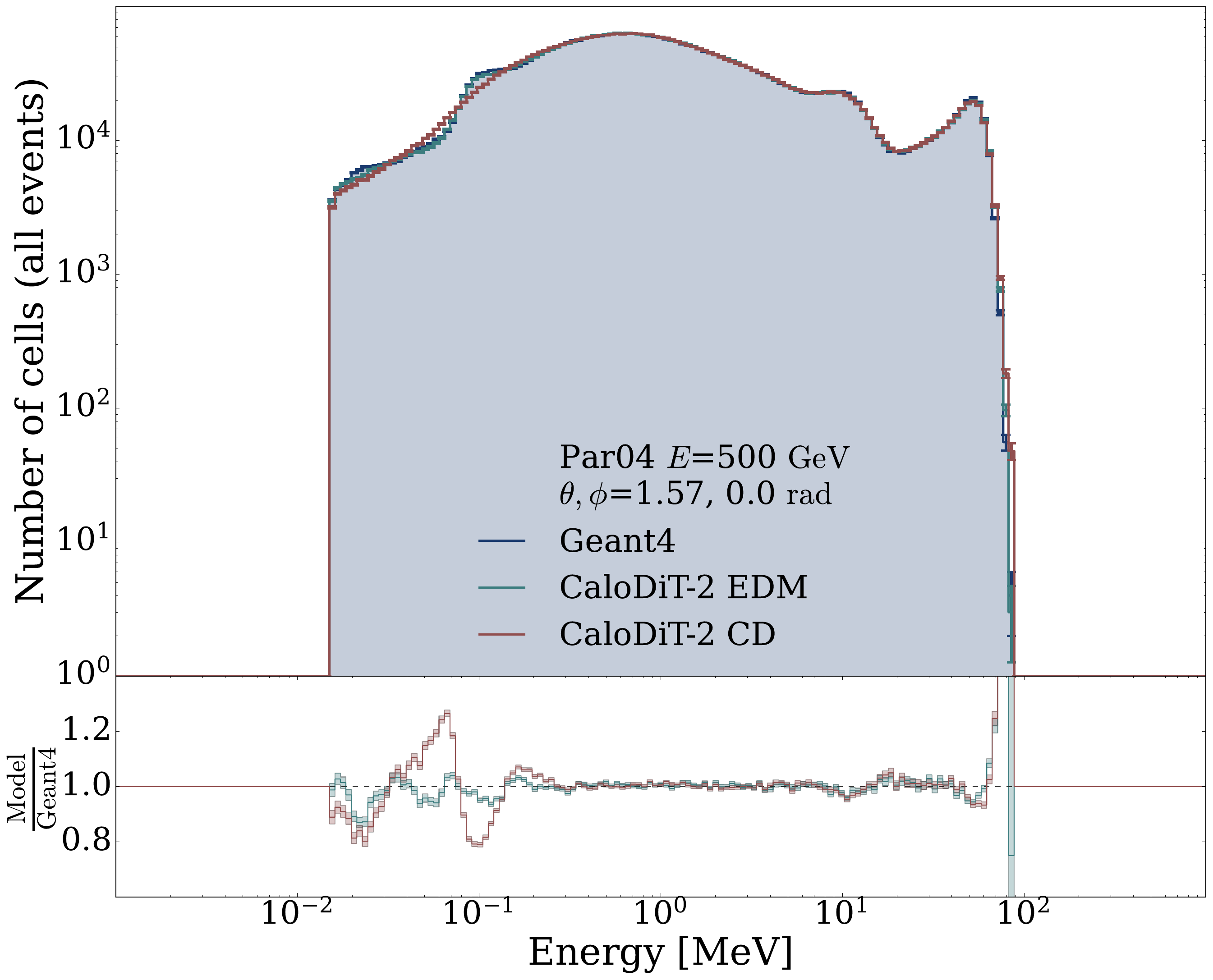}
    \caption{Voxel energy distribution}
    \label{fig:cell_500GeV}
  \end{subfigure}
  \hfill
  \begin{subfigure}[b]{0.49\linewidth}
    \centering
    \includegraphics[height=\imgheight]{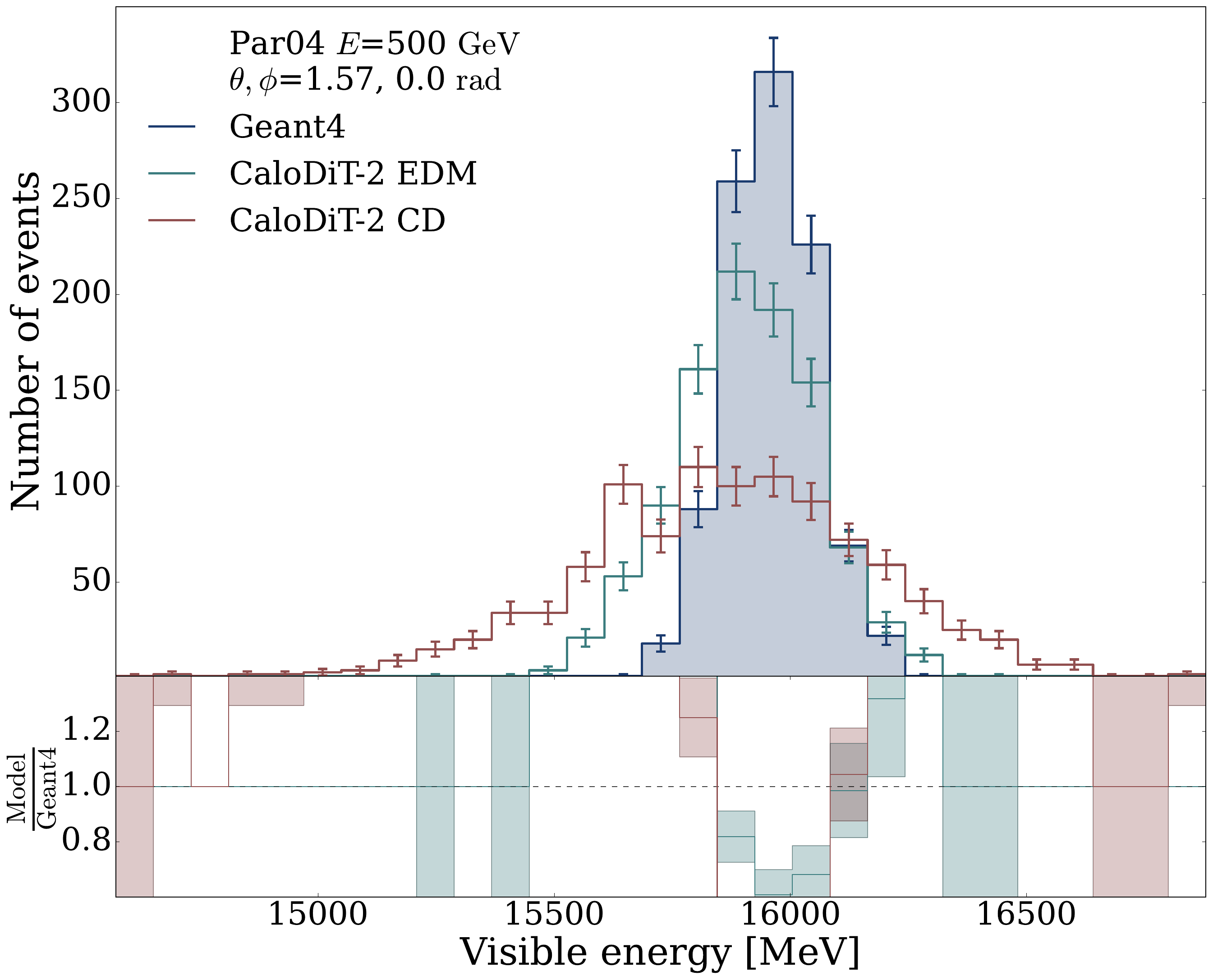}
    \caption{Total visible energy distribution}
    \label{fig:tot_500GeV}
  \end{subfigure}
  \caption{Shower observables for $500$ GeV $\gamma$ generated for Par04-SiW detector. $1,000$ showers were generated for each of \geant and the CaloDiT-2 models. The error bars represent the Poissonian error on each bin.}
  \label{fig:single_detector_results_500GeV}
\end{adjustwidth}
\end{figure}

\begin{figure}[htbp]
\begin{adjustwidth}{-\figenvleftextend}{-\figenvrightextend}
  \centering
  \begin{subfigure}[b]{0.49\linewidth}
    \centering
    \includegraphics[height=\imgheight]{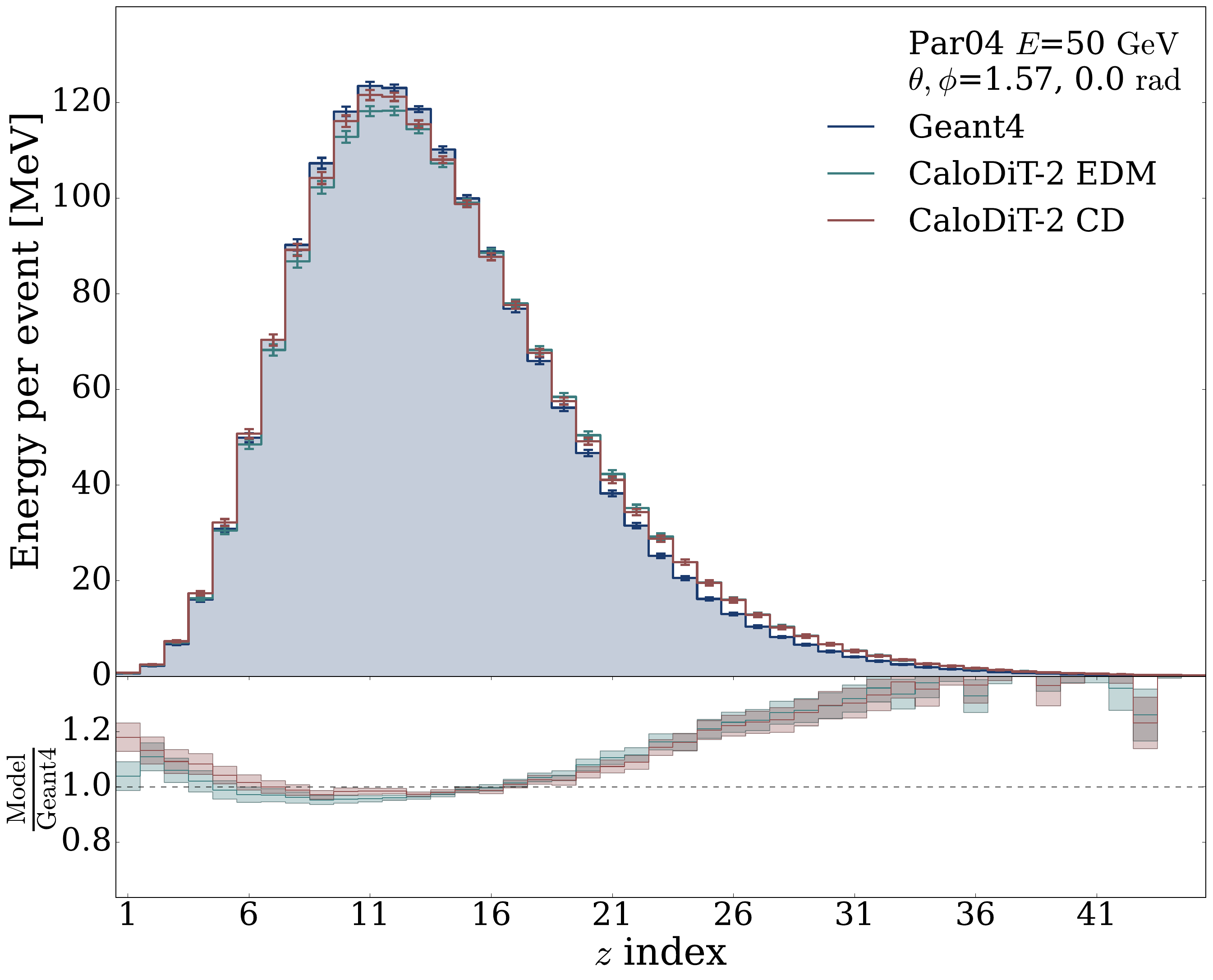}
    \caption{Longitudinal profile}
  \end{subfigure}
  \hfill
  \begin{subfigure}[b]{0.49\linewidth}
    \centering
    \includegraphics[height=\imgheight]{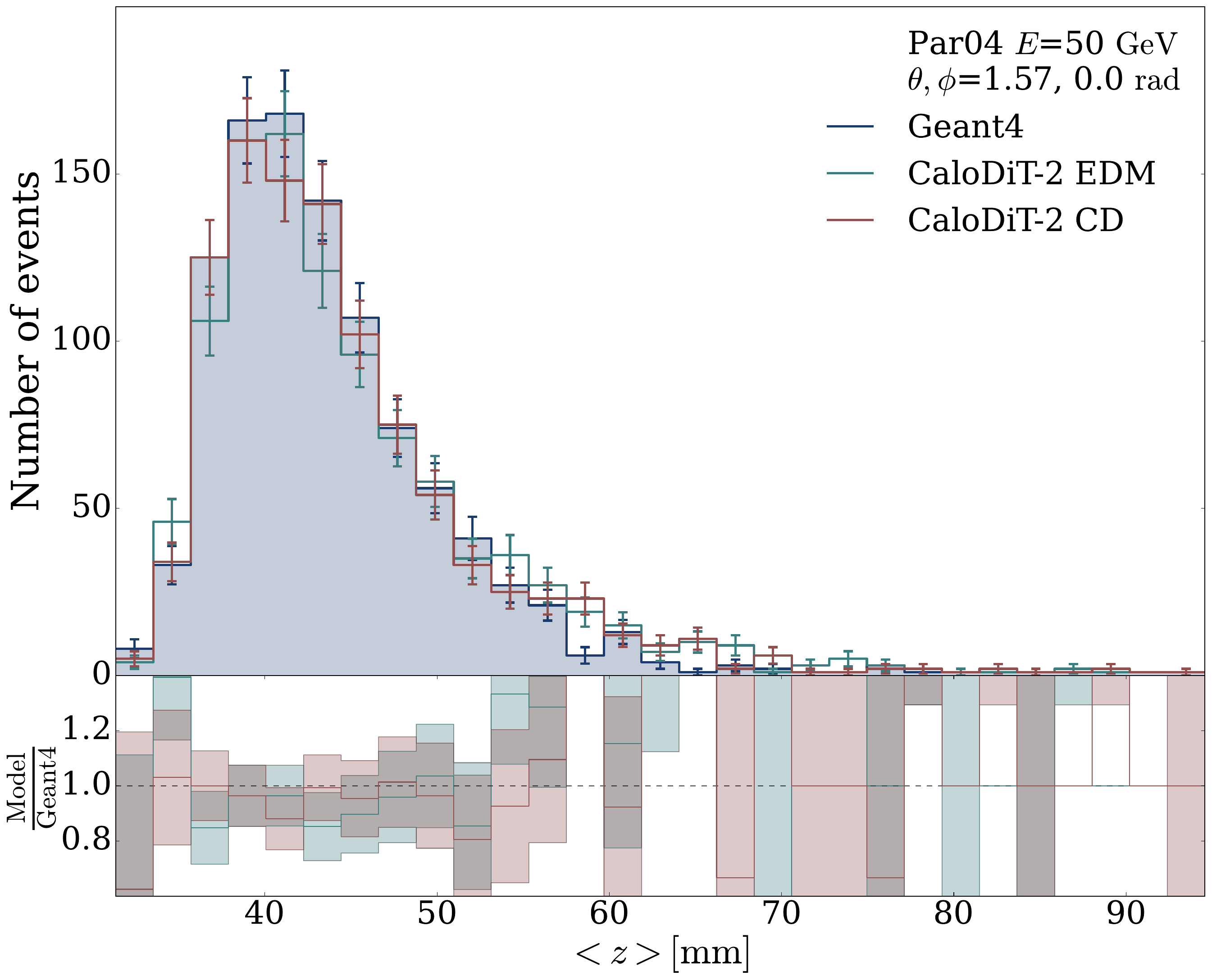}
    \caption{First moment of longitudinal profile}
  \end{subfigure}
  \hfill
  \begin{subfigure}[b]{0.49\linewidth}
    \centering
    \includegraphics[height=\imgheight]{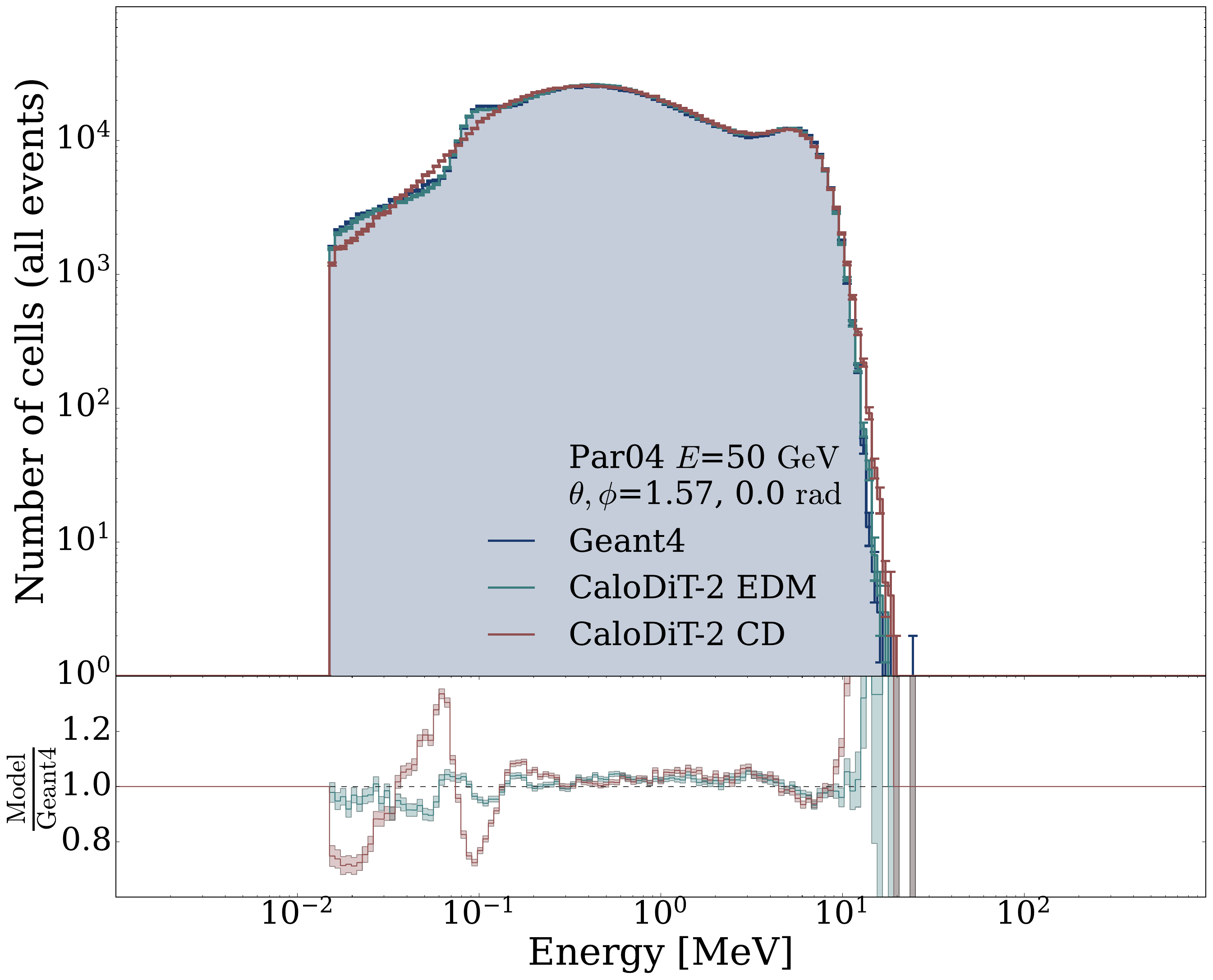}
    \caption{Voxel energy distribution}
  \end{subfigure}
  \hfill
  \begin{subfigure}[b]{0.49\linewidth}
    \centering
    \includegraphics[height=\imgheight]{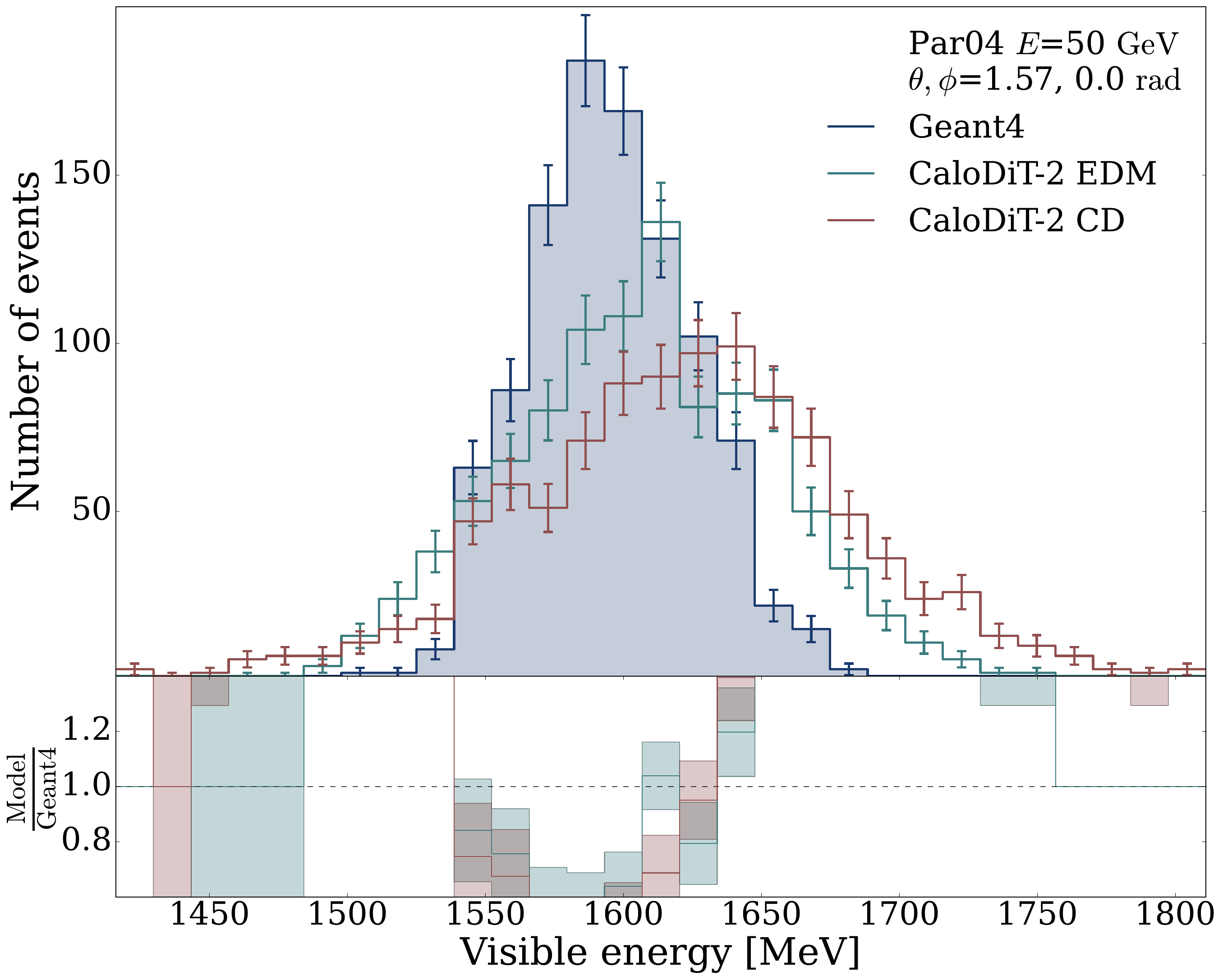}
    \caption{Total visible energy distribution}
  \end{subfigure}
  \caption{Shower observables for $50$ GeV $\gamma$ generated for Par04-SiW detector. $1,000$ showers were generated for each of \geant and the CaloDiT-2 models. The error bars represent the Poissonian error on each bin.}
  \label{fig:single_detector_results_50GeV}
\end{adjustwidth}
\end{figure}

Figure \ref{fig:single_detector_results_500GeV} shows various shower observables for a $500$ GeV $e^-$ fired at $\theta=1.57$ and $\phi=0.0$ rad. The histograms are computed over the test dataset consisting of 1000 such events. Most of the shower observables show good agreement with the \textsc{Geant4} ground-truth, except for the total visible energy (Figure \ref{fig:tot_500GeV}). In particular, the voxel energy distribution (Figure \ref{fig:cell_500GeV}) is quite well modeled due to the diffusion process being better at learning finer details compared to other generative processes. There is a slight degradation in the quality of CD over EDM, which is expected due to the CD model having to generate the showers in a single pass. Since, $500$ GeV is at the centre of the training distribution, we also show the same shower observables for $50$ GeV $e^-$ in Figure \ref{fig:single_detector_results_50GeV}.    

As for the training dynamics, during the initial phase of the training, the longitudinal profile, the transverse profile, and their moments are learned well. These and the voxel energy distribution get refined during the middle phase. The total energy distribution only starts to be learned towards the end of the training; otherwise, it remains rather inconsistent. In parallel, the model first learns to generate showers belonging to the centre of the data, i.e. close to $E=500$ GeV, and showers belonging to extreme $E$ are modeled at the end. It is worth noting that the decay was particularly helpful in modeling the total energy distribution, hence a longer-than-usual decay phase. We believe this is due to the difficulty of performing the summation of non-log data in the log space, which requires much finer adjustments to the weights.  

\begin{figure}[htbp]
  \begin{adjustwidth}{-\figenvleftextend}{-\figenvrightextend}
  \centering
  \begin{subfigure}[b]{0.49\linewidth}
    \centering
    \includegraphics[height=\imgheight]{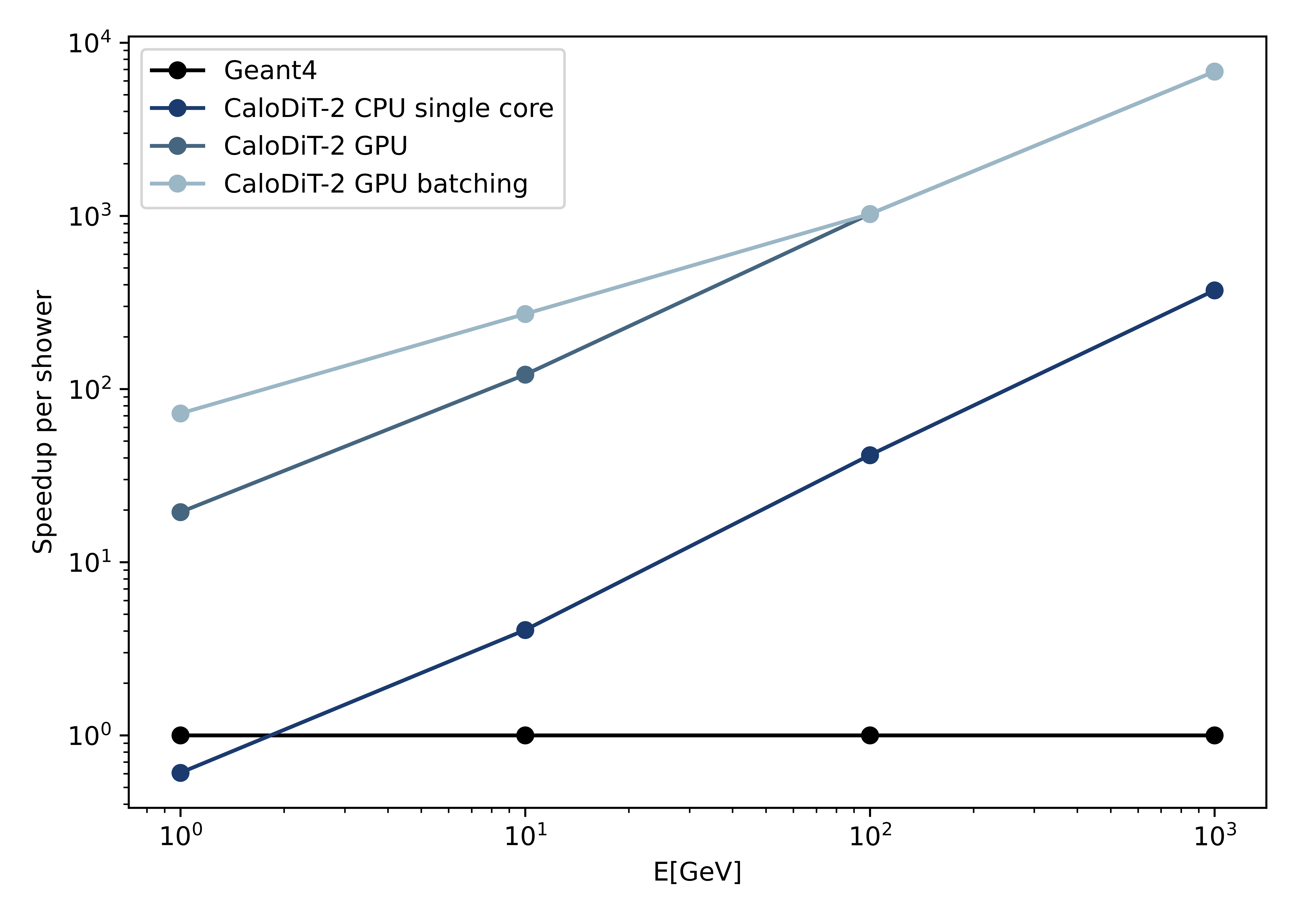}
    \caption{Timings}
    \label{fig:speedup}
  \end{subfigure}
  \hfill
  \begin{subfigure}[b]{0.49\linewidth}
    \centering
    \includegraphics[height=\imgheight]{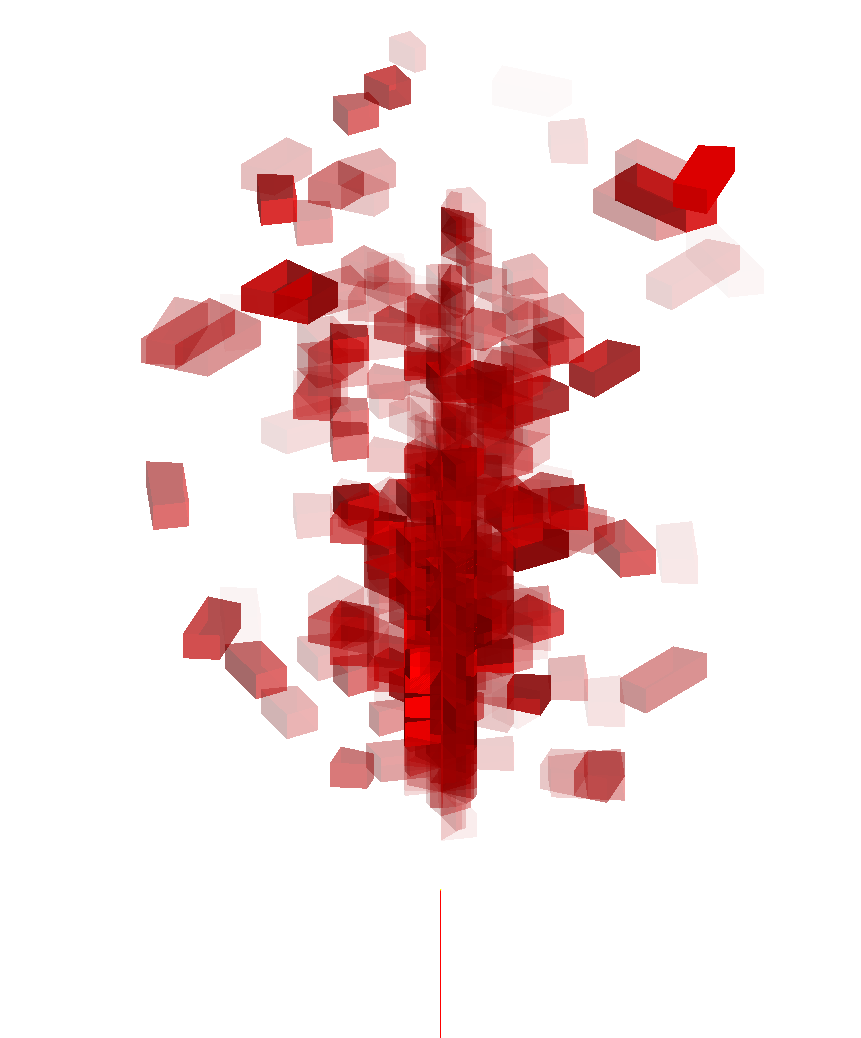}
    \caption{Generated shower}
    \label{fig:calodit_g4_display}
  \end{subfigure}
  \end{adjustwidth}
  \caption{(a) Speedup compared to \textsc{Geant4}. (b) \textsc{Geant4} display of CaloDiT-2 generated shower.}
  \label{fig:display_and_speedup}
\end{figure}

We also measure the time required to generate a shower using the CD model. These measurements are done on an AMD EPYC 9334 CPU and an NVIDIA RTX 6000 Ada GPU. We calculate these measurements in PyTorch, as numerous tweaks can be applied to C++ environments depending on the deployment requirements. The time is calculated by taking an average over 5 rounds, where each round generates 1000 showers of a given batch size. For CPU inference, we restrict the number of threads and the batch size to 1. For GPU, we assume that \textsc{Geant4} can offload as many showers as it can to the GPU, thus allowing for a batch size larger than 1. The exact batch size would depend on the number of shower events available to be run in parallel. Following \cite{batchingAnna}, this batching and the time to place the shower back into \textsc{Geant4} is taken into account to calculate the final expected speedup per shower, which is shown in Figure \ref{fig:speedup}. The speedup offered is significant, even for CPU, for incident particles whose energy is above $2$ GeV.   



The CD model is used for inference in \textsc{Geant4} via first converting it C++ compatible format.  Figure \ref{fig:calodit_g4_display} shows the shower generated by CaloDiT-2 and visualized within \textsc{Geant4}. More details in Section \ref{sec:for_users}.

\subsection{CaloChallenge Dataset-2 Performance}
\label{subsec:cc_results}
\begin{figure}[htbp]
  \begin{adjustwidth}{-1cm}{-1cm}
  \centering
  \begin{subfigure}[b]{0.325\linewidth}
    \centering
    \includegraphics[height=\imgheight]{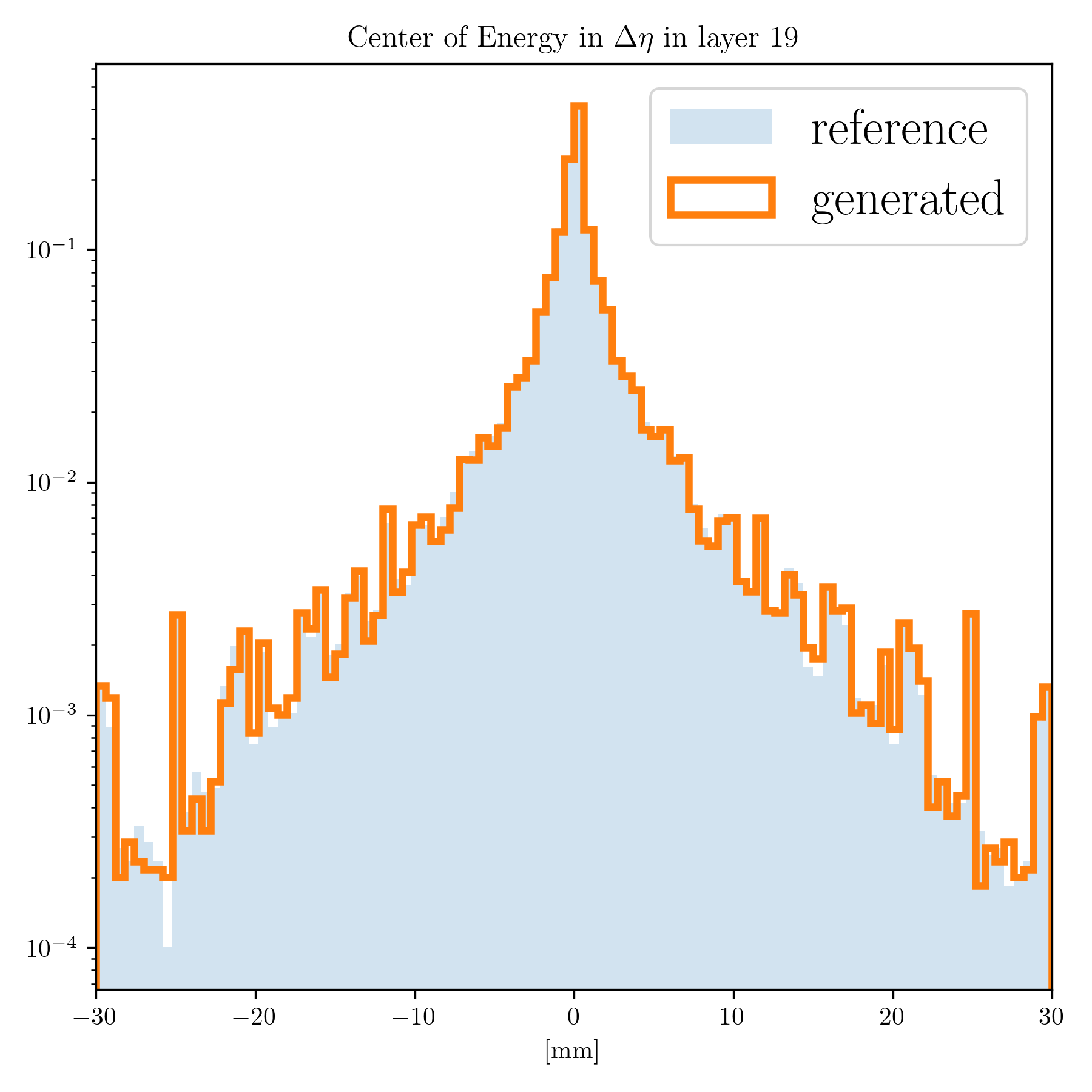}
    \caption{}
  \end{subfigure}
  \hfill
  \begin{subfigure}[b]{0.325\linewidth}
    \centering
    \includegraphics[height=\imgheight]{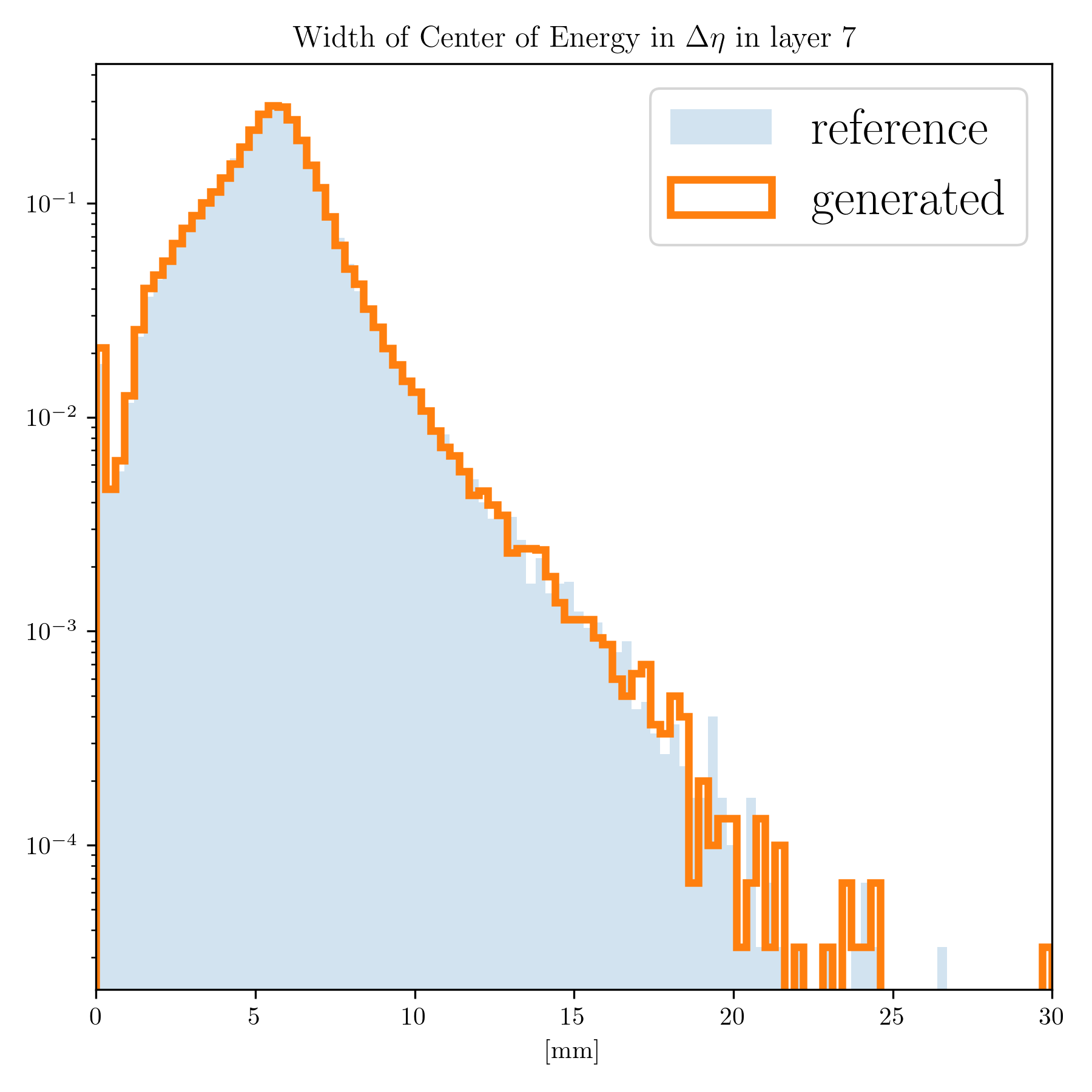}
    \caption{}
  \end{subfigure}
  \hfill
  \begin{subfigure}[b]{0.325\linewidth}
    \centering
    \includegraphics[height=\imgheight]{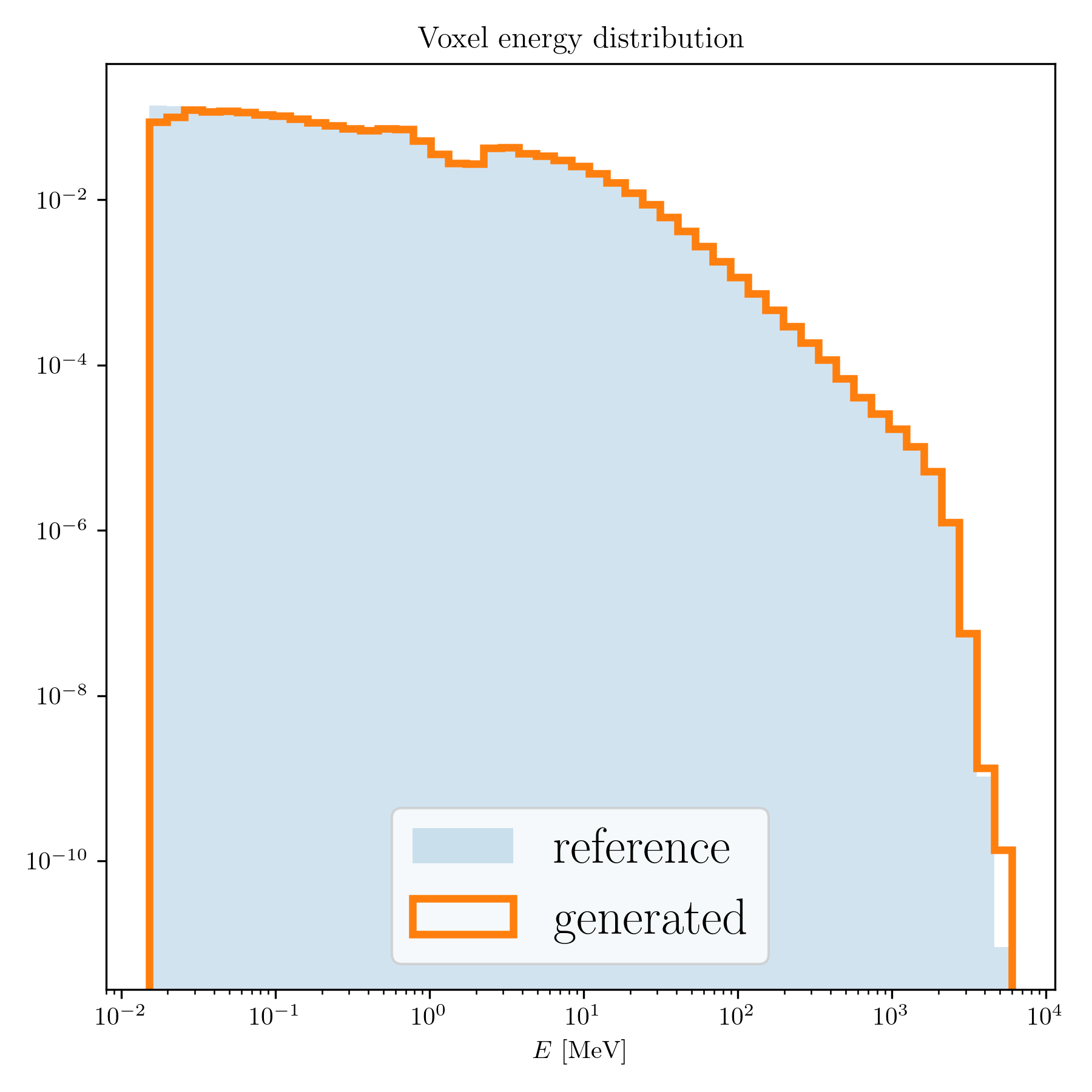}
    \caption{}
  \end{subfigure}
  \end{adjustwidth}
  \caption{(a) Center of energy for layer 19. (b) Width of the center of energy for layer 7. (c) Voxel energy distribution. Reference is \textsc{Geant4} and generated refers to showers generated by CaloDiT-2.}
  \label{fig:ccd2_results}
\end{figure}

We also briefly report results on CaloChallenge Dataset-2. Figure \ref{fig:ccd2_results} shows a selection of randomly chosen plots, indicating that CaloDiT-2 models nearly all shower observables defined in CaloChallenge with high accuracy.

\begin{table}[htbp]
\begin{adjustwidth}{0cm}{0cm}
\centering
\caption{Quantitative metrics from the CaloChallenge \cite{krause2024calochallenge} on Dataset-2. For all metrics, lower values indicate better performance. All values are obtained using the default hyperparameters provided in the CaloChallenge repository.}
\label{table:ccd2}
\begin{small}
\begin{tabular}{p{3cm}cccccc}
\toprule
\raisebox{-0.8ex}{Models} \textbackslash \raisebox{0.8ex}{ Metrics} & \parbox[t]{1.5cm}{AUC\\low-level} & \parbox[t]{1.5cm}{AUC\\high-level} & FPD ($\times10^3$) & KPD ($\times10^3$) \\ [0.1cm]
\midrule
CaloDiT-2 EDM & 0.5939$\pm$ 0.002 & 0.5598$\pm$ 0.005 & 20.06$\pm$ 0.74 & 0.089$\pm$ 0.09 \\ [0.1cm]
CaloDiT-2 CD & 0.5975$\pm$ 0.002 & 0.6947$\pm$ 0.004 & 68.83$\pm$ 3.37 & 0.39$\pm$ 0.15 \\ [0.1cm]
\midrule
CaloDiT-1 DDPM \cite{krause2024calochallenge} & 0.984$\pm$ 0.001 & 0.912$\pm$ 0.001 & 1690.98$\pm$ 6.77 & 11.03$\pm$ 0.43 \\ [0.1cm]
\midrule
CaloDiffusion \cite{Amram:2023onf} & 0.577$\pm$ 0.004 & 0.591$\pm$ 0.009 & 146.933$\pm$ 0.87 & 0.174$\pm$ 0.04 \\ [0.1cm]
CaloDREAM \cite{Favaro:2024rle} & 0.531$\pm$ 0.003 & 0.521$\pm$ 0.002 & 24.65$\pm$ 1.035 & 0.023$\pm$ 0.036 \\ [0.1cm]
CaloINN \cite{Ernst:2023qvn} & 0.743$\pm$ 0.002 & 0.865$\pm$ 0.003 & 732.83$\pm$ 5.33 & 2.82$\pm$ 0.42 \\ [0.1cm]
Calo-VQ \cite{liu2024calo} & 0.986$\pm$ 0.001 & 0.994$\pm$ 0.0 & 1315.72$\pm$ 7.03 & 8.52$\pm$ 0.5 \\ [0.1cm]
\bottomrule
\end{tabular}
\end{small}
\end{adjustwidth}
\end{table}

In Table \ref{table:ccd2}, we present a comparison of CaloDiT-2 with CaloDiT-1 and several prior generative models across a range of CaloChallenge quantitative metrics. The baseline models include iterative approaches, such as flow-matching-based CaloDREAM \cite{Favaro:2024rle} and diffusion-based CaloDiffusion \cite{Amram:2023onf}, as well as single-shot methods, including VAE-based Calo-VQ \cite{liu2024calo} and normalizing-flow-based CaloINN \cite{Ernst:2023qvn}. Both variants of CaloDiT-2, namely EDM and CD, significantly outperform single-shot models, as well as the preliminary CaloDiT-1. When compared to the iterative models, CaloDiT-2 EDM demonstrates performance that is on par or, in some metrics (e.g., FPD), better. The CD variant, in contrast, achieves substantial speed gains with a slight drop in accuracy. Despite this, CaloDiT-2 CD still outperforms other single-shot models by a huge margin. These results indicate that CaloDiT-2 provides a compelling trade-off between accuracy and speed.  

\begin{table}[htbp]
\begin{adjustwidth}{0cm}{0cm}
\centering
\caption{Measure of CPU and GPU times using batch size of 1. $^*$ indicate CaloChallenge timings.}
\label{table:timings}
\begin{small}
\begin{tabular}{p{3cm}cc}
\toprule
\raisebox{-0.8ex}{Models} \textbackslash \raisebox{0.8ex}{ Timings} & CPU (ms) & GPU (ms) \\ [0.1cm]
\midrule
CaloDiT-2 EDM & 6349.0$\pm$ 7.6 & 171.8$\pm$ 0.13 \\ [0.1cm]
CaloDiT-2 CD & 101.2$\pm$ 0.04 & 2.9$\pm$ 0.02 \\ [0.1cm]
\midrule
CaloDiT-1 DDPM \cite{krause2024calochallenge} & \parbox[c]{2cm}{17322.9$\pm$ 33.9 \\ 24642$\pm$ 1883$^*$} & \parbox[c]{2cm}{639.96$\pm$ 2.4 \\ 1036$\pm$ 18$^*$} \\ [0.1cm]
\bottomrule
\end{tabular}
\end{small}
\end{adjustwidth}
\end{table}

Table \ref{table:timings} shows a comparison of inference times. Compared to CaloDiT-1, CaloDiT-2 is both significantly faster and more accurate on all metrics. CPU (single-thread) and GPU times correspond to per-shower generation with batch size 1 using PyTorch, measured on the machine mentioned in Section \ref{sec:results_par04}. Note that the timings differ from those reported in the CaloChallenge paper for CaloDiT-1, as we measure only the generation process, whereas CaloChallenge timings include end-to-end overhead, from container initialization to saving generated showers. To highlight the relative differences, we also include the CaloChallenge timings, marked with $^*$.  

\section{A Generalisable Model}
\label{sec:generic_model}
Now that we have established a working model for a single detector, we explore the possibility of having a generalisable model. A generalisable model is expected to learn rich detector-agnostic representations and generalize across different detectors. Such a model would be exposed to multiple detectors, so the model can learn how the shower develops across different detectors. The same model can then be quickly adapted for use with an unseen detector. This is desired for several reasons, such as the changes in the full simulation pipeline for experiments at CERN; for detectors under development, whose geometries change frequently, e.g. future collider detectors; and for smaller experiments that lack the expertise and manpower to design these models from scratch. The motivation is to make an open-source FastSim foundation model available, which is affordable and accessible to everyone, so that every user can benefit from the advantages these models provide.   

\subsection{Extending to Multiple Detectors}

Showers in different detectors evolve differently depending on the material composition and structure, and the manner in which hit information is recorded depends strongly on the detector readout. This makes it difficult to unify the differing shower structures into a single representation or to design a model capable of handling multiple geometries. However, the virtual cylindrical mesh mentioned earlier provides a convenient means of scoring energies irrespective of the underlying detector geometry. This transfers the difficulty from model design to placing the generated shower back into the geometry.  

Compared to the single-detector CaloDiT-2, the difference now lies in the number of detectors which the model is trained on. It should be noted that a typical description of a realistic detector geometry is extremely complicated, with many variables governing any given geometry description. Even if these are restricted to a feasible set, the range needed to cover the phase space would likely require an enormous amount of data. Hence, we believe that the simplest and most viable approach is to use a categorical encoding of the detector. To this end, we introduce a geometry condition ($G$) to the model, along with the already existing $E$, $\phi$ \& $\theta$ conditions. $G$ is a one-hot vector denoting the detector for the given sample. The size of $G$ is $K + 1$, $K$ being the number of detectors used in the training of the model. The additional element is to denote any new detector, which need not be known beforehand. This means that the model can first be pre-trained to learn generalizable representations of showers using a given set of detectors, and later adapted to any desired detector. Such adaptation is expected to require less training data and optimization steps compared to training from scratch, while achieving a comparable level of performance. The pre-training phase refers to the extensive training on multiple detectors, so the model can learn how the shower develops across different detectors. The adaptation phase refers to the fine-tuning or retraining of the model on the desired detector using a small dataset. Thus, we replace the difficulty of describing the geometry of the detector to the model with a practical adaptation approach.   

Another difference is in the preprocessing of the data. A log transformation is applied as before, followed by the normalization. However, now normalization is performed over the data from all $K$ detectors available for pre-training. We also explored normalizing data belonging to each detector separately, but the former performed well in practice. We now discuss the results for adaptation in the next subsection. Unless otherwise stated, the experiments are executed using the hyperparameters mentioned in Section \ref{sec:training_dynamics}.  

In our setting, we first pre-train CaloDiT-2 on the aforementioned Par04-SiW detector, its variant Par04-SciPb, ODD, and FCCeeCLD. Later, the adaptation is performed on FCCeeALLEGRO. Practically, this is achieved by finetuning a diffusion process (EDM) using the pre-trained EDM process. Subsequently, the distillation process (CD) is trained using the pre-trained CD as the student model and the newly finetuned EDM as the teacher model.  

\subsection{Results}
\label{subsec:adaptation_results}

To illustrate the benefits of the adaptation, we vary the number of training steps and the amount of FCCeeALLEGRO data for both the EDM and CD training and compare with training these from scratch. As a metric to assess the quality of shower generation, we use the FPD (Fréchet physics distance) \cite{kansal2023evaluating}. We calculate FPD on shower observables calculated on a set of 50,000 test samples belonging to the entire distribution, i.e. covering the entire manifold of $E$, $\phi$, and $\theta$. It is important to note that FPD is a relative measure. For instance, a reduction from 100 to 50 is significantly more impactful than a reduction from 500 to 300, despite the latter having a larger absolute difference.   

\begin{figure}[htbp]
  \centering
  \includegraphics[width=0.8\linewidth]{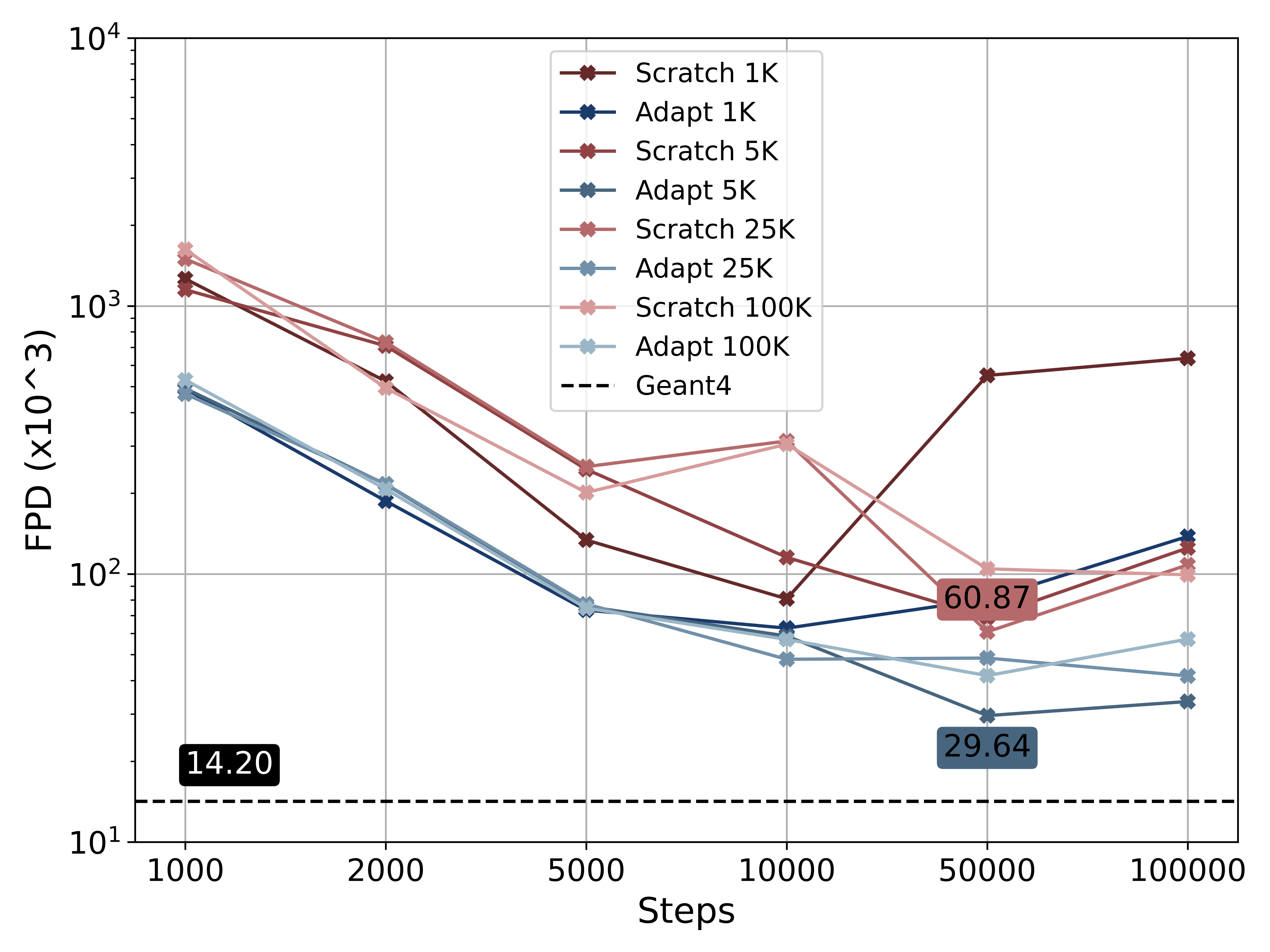}
  \caption{FPD against numbers of training steps for various amounts of FCCeeALLEGRO training data. The FPD value for \textsc{Geant4} represents the baseline, drawn by a black dashed line. Red hues represent training from scratch, and blue hues represent adaptation. Darker hues represent less data used for training from scratch or adaptation, whereas lighter hues represent more data. Values indicate the minimum FPD achieved in each case.}
  \label{fig:adaptation_fpd}
\end{figure}

Figure \ref{fig:adaptation_fpd} shows the FPD across a varying number of training steps and data in a logarithmic scale. The x-axis represents the number of training steps taken, and the legend indicates the amount of training data. Note that each point on the plot is a different training run and each one of them follows a \textit{wsd} schedule, as described in Section \ref{sec:training_dynamics}. We empirically choose a learning rate of $0.001$ for training from scratch and a learning rate of $0.0001$ for adaptation. This is based on the intuition that training from scratch would benefit from bigger steps in scenarios where the number of training steps is limited, while adaptation would benefit from shorter steps to not destroy the previously learnt representations. It should be noted that the points on the plot can be significantly influenced by hyperparameters affecting convergence speed, such as the learning rate, the EMA decay used in CD, or the degree of distributional shift introduced by the new detector, i.e. how different the new detector is. The optimal values for these hyperparameters depend on factors like whether the model is trained from scratch, the size of the dataset, and the total number of training steps. However, systematically tuning these parameters is computationally expensive and beyond the scope of this work. Nevertheless, the overall trends presented in the plot should remain valid under any reasonable choice of hyperparameters.  

From Figure \ref{fig:adaptation_fpd}, we can infer some of the expected trends. First, in general, the more training steps, the better. Second, in general, the more data, the better. However, if the number of training steps saturates the amount of data, the generation quality decreases (Scratch 1K after 10,000 steps, Adapt 1K after 10,000 steps). Similarly, if the amount of data saturates the number of training steps, the generation quality decreases (e.g. Scratch at 10,000 steps, Adapt at 100,000 steps). Third, given the same amount of data and the same number of training steps, adaptation consistently outperforms training from scratch. There are some exceptions to these trends, e.g. the FPD increases for 10,000 steps when using more data, but then goes down with more training steps. We believe this is tied to the \textit{wsd} schedule and the coverage of data during the decay phase.  

\begin{figure}[htbp]
  \begin{adjustwidth}{-\figenvleftextend}{-\figenvrightextend}
  \centering
  \begin{subfigure}[b]{0.49\linewidth}
    \centering
    \includegraphics[height=\imgheight]{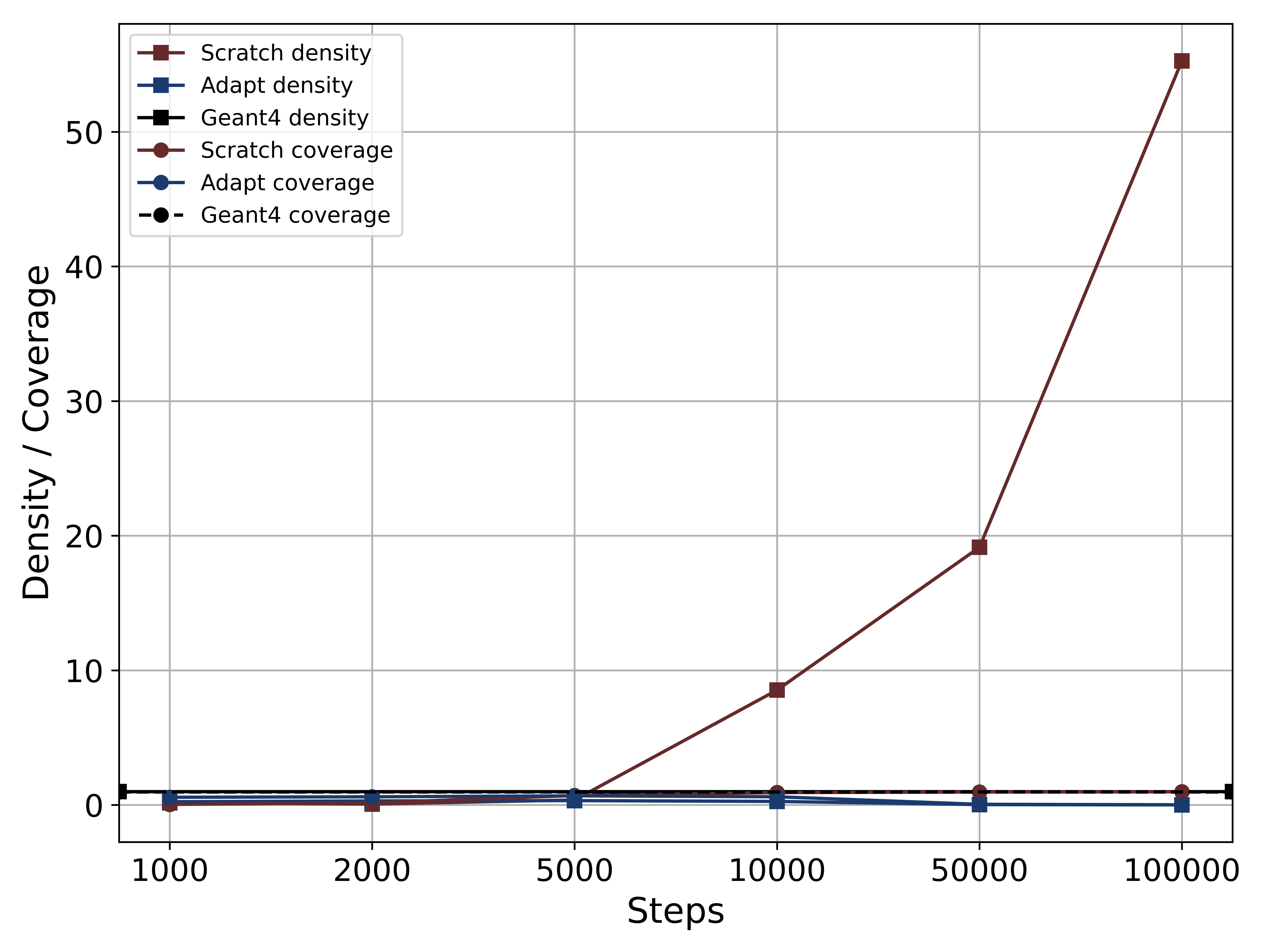}
    \caption{Density and Coverage for 1K data}
    \label{fig:dc_1K}
  \end{subfigure}
  \hfill
  \begin{subfigure}[b]{0.49\linewidth}
    \centering
    \includegraphics[height=\imgheight]{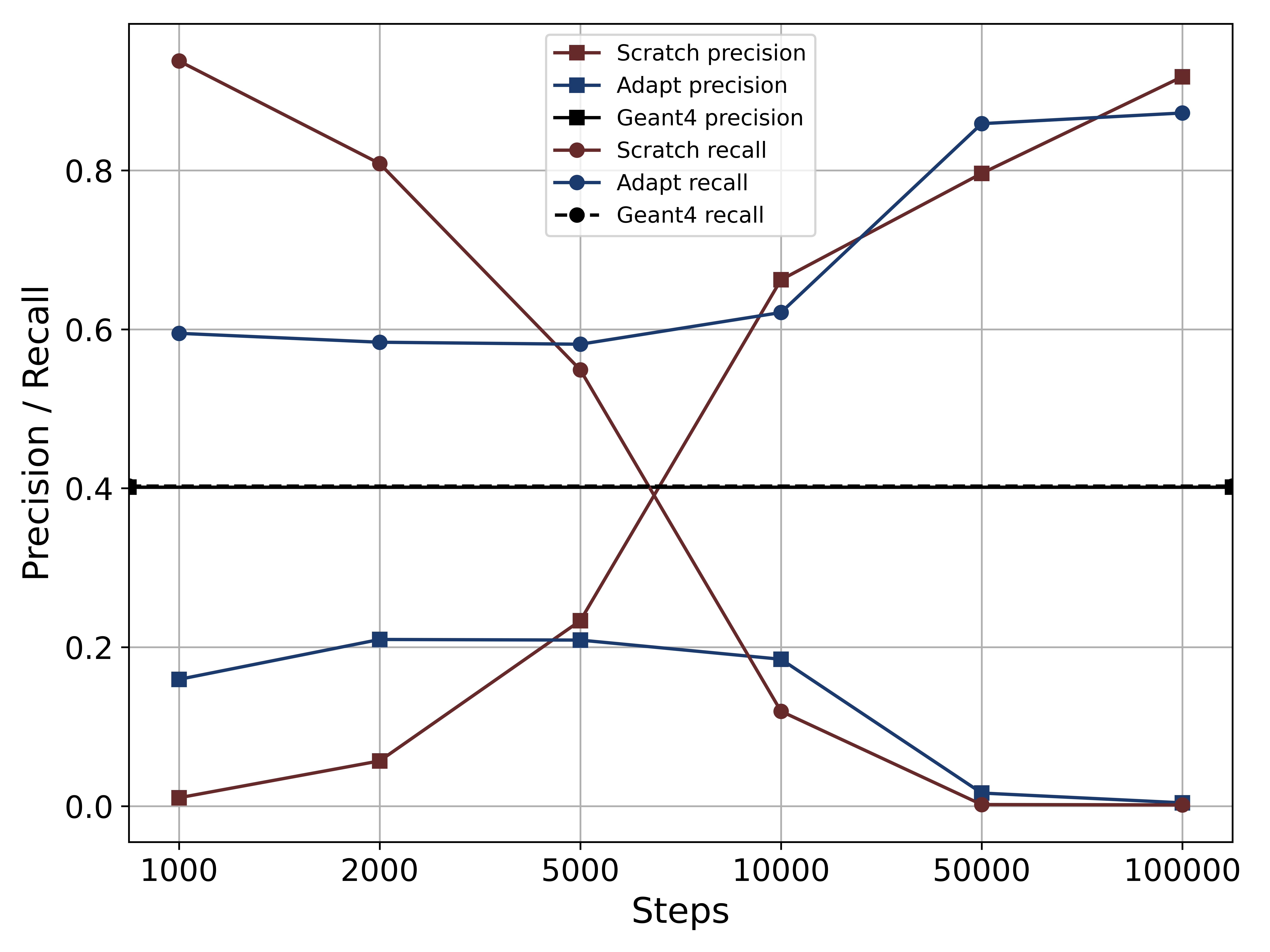}
    \caption{Precision and Recall for 1K data}
    \label{fig:pr_1K}
  \end{subfigure}
  \end{adjustwidth}
  \caption{(a) Density and coverage for when the amount of data is 1K. (b) Precision and recall for when the amount of data is 1K. Both of these shows memorization for a large number of training steps.}
  \label{fig:prdc_1K}
\end{figure}

Unexpectedly, training from scratch at 10,000 steps using 1K samples shows a better FPD than a similar training with more data. This is surprising because the amount of data is too low to train any kind of generative model with sufficient quality. This phenomenon is explained by using precision and density \cite{kynkaanniemi2019improved, naeem2020reliable} to assess the faithfulness of generated showers to \textsc{Geant4} showers. Figure \ref{fig:dc_1K} and Figure \ref{fig:pr_1K} show density and precision overshooting beyond \textsc{Geant4} v/s \textsc{Geant4}, implying that the model trained from scratch is memorizing a limited amount of data, and is therefore not sufficiently covering the actual data manifold. In contrast, adaptation is rather consistent for a reasonable number of steps and does not show the same memorizing behavior.  

These experiments indicate that adaptation can achieve the same FPD at 5,000 training steps using 1K samples of new detector data, as training for 50,000 steps from scratch using 25K samples. This is a significant reduction in required resources, consuming $20\times$ less training time (EDM and CD combined) and using $25\times$ less data. In practical terms, training a FastSim model takes a few hours instead of several days, not including the time taken to simulate the required training data. It should be noted that these results will vary based on the new detector, the architecture, and the hyperparameters. In our particular case, FCCeeALLEGRO can be considered very different compared to Par04-SiW, Par04-SciPb, ODD, and CLD from the physics perspective. Also, given sufficient data and sufficient training steps, the training from scratch may, at some stage, outperform adaptation. This is because the pre-trained model holds knowledge regarding several detectors and thus contains data biases, unlike the model trained from scratch. However, in the low data and low training steps regime, adaptation will significantly outperform training from scratch, as evident from the Figure \ref{fig:adaptation_fpd}. This allows for a computationally efficient way to train a model that requires less training compute and less data, i.e. a lower number of computationally extensive simulations from \textsc{Geant4}, taking a step towards a foundation model for calorimeter simulation.

\subsection{For Users}
\label{sec:for_users}
CaloDiT-2 is released in both ONNX and TorchScript formats, ensuring compatibility with C++ frameworks. Integration with \textsc{Geant4} is demonstrated in the Par04 extended example included with version 11.4.0-beta \cite{Par04}. In addition, PyTorch versions of the pre-trained diffusion model (EDM) and the distilled model (CD) are publicly available \cite{implementation}, together with scripts for adaptation and for conversion to ONNX or TorchScript.  

The workflow for applying CaloDiT-2 to a new detector geometry is as follows:  
\begin{enumerate}
    \item Generate training data using the virtual mesh demonstrated in the Par04 example \cite{Par04}.  
    \item Adapt the pre-trained EDM model \cite{implementation} to the new geometry by conditioning on the $(K+1)$ detector position.  
    \item Perform distillation by initializing the pre-trained CD model as the student and the adapted EDM model from step (2) as the teacher.  
    \item Convert the adapted CD model into ONNX or TorchScript format using the provided utilities.  
    \item Integrate the converted model into \textsc{Geant4} following the workflow of the Par04 example.  
\end{enumerate}

As we also release the scripts for training from scratch, users may also train the EDM and CD models directly, following the training pipelines in the repository \cite{implementation}. 

\section{Discussion}
\label{sec:discussion}
While CaloDiT-2 demonstrates strong performance and adaptability, several investigations remain.  

First, although the virtual mesh provides an elegant way to capture detector-agnostic data representation, it is affected by sparsity and fixed dimensionality. Point clouds provide a natural way to address sparsity, but flexibility regarding dimensionality remains a challenge. Different detectors may correspond to drastically different numbers of points, making uniform handling non-trivial.  

Second, while we investigated trivial preprocessing approaches towards encompassing multiple detectors, more sophisticated approaches are possible. In particular, the choice of preprocessing interacts differently with the two model families: the diffusion model (EDM) distributes capacity across the full diffusion trajectory and can often compensate for imperfect preprocessing through multi-step denoising. In contrast, the distilled model (CD) compresses the process into a single step (or a few), and is therefore more sensitive to preprocessing. This suggests that improvements in preprocessing, or explicitly accounting for detector-dependent quantities such as sampling fraction, could allow for a more uniform level of data representation, and thus allow the model to learn more robust representations.  

\iftoggle{oldarch}{}{\PR{Reconfirm this}}%
Third, our preliminary studies with multi-step CD did not show significant improvements over the single-step setup. This suggests that the current discrete-time distillation approach, while near-EDM performance, does not fully exploit the continuous generative dynamics of diffusion models. An interesting direction is to investigate continuous-time CD, leveraging the underlying ODE/SDE trajectory of the diffusion process to enable smoother interpolation and potentially better trade-offs between speed and accuracy.  

Finally, although suboptimal in expressiveness, adapting only the geometry condition layer has the benefit of extremely fast transfer. In principle, this could enable adaptation on limited compute resources, or even on a CPU, making it attractive for lightweight deployment scenarios.  

Despite these limitations, CaloDiT-2 demonstrates strong performance and, most importantly, provides the first clear evidence that pre-training with adaptation substantially reduces data and training requirements while preserving accuracy.

\section{Conclusion}
\label{sec:conclusion}
We presented CaloDiT-2, a fast and accurate transformer-based diffusion model for electromagnetic calorimeter shower simulation. This model can be used as a standalone model, as it is typically studied, but this study underlines the feasibility of going a step further, towards a more sustainable ML models. To the best of our knowledge, CaloDiT-2 is the first published model that demonstrates the strength of pre-training and adaptation in the context of FastSim, highlighting their potential benefits and paving the way toward a foundation model for calorimeter simulation. Our strategy enables up to $25\times$ reduction in training data requirements and $20\times$ faster training, while also allowing a model trained on a simplified version of a detector geometry to be efficiently adapted to its more complex counterpart, significantly reducing the computational cost of producing training datasets.  

Even without adaptation, CaloDiT-2 achieves competitive performance and maintains fast inference of $\sim 100$ ms per shower on a single-core CPU. Beyond the methodological contributions, we release pre-trained models, datasets, and implementation scripts to facilitate broad adoption within the community.  

\section{Acknowledgements}
This work benefited from support by the CERN Strategic R$\&$D Programme on Technologies for Future Experiments \cite{EPRD} and has received funding from the European Union’s Horizon 2020 Research and Innovation programme under Grant Agreement No. 101004761. 

\bibliography{example_paper}
\bibliographystyle{unsrt}


\newpage
\appendix
\section*{Appendix}

\section{Additional Results On Single Detector}

\begin{figure}[htbp]
\begin{adjustwidth}{-\figenvleftextend}{-\figenvrightextend}
  \centering
  \begin{subfigure}[b]{0.49\linewidth}
    \centering
    \includegraphics[height=\imgheight]{assets/plots_single_detector/LongTotalEnergy_Geo_Par04_E_50_Phi_0.00_Theta_1.57.pdf}
    \caption{Longitudinal profile}
  \end{subfigure}
  \hfill
  \begin{subfigure}[b]{0.49\linewidth}
    \centering
    \includegraphics[height=\imgheight]{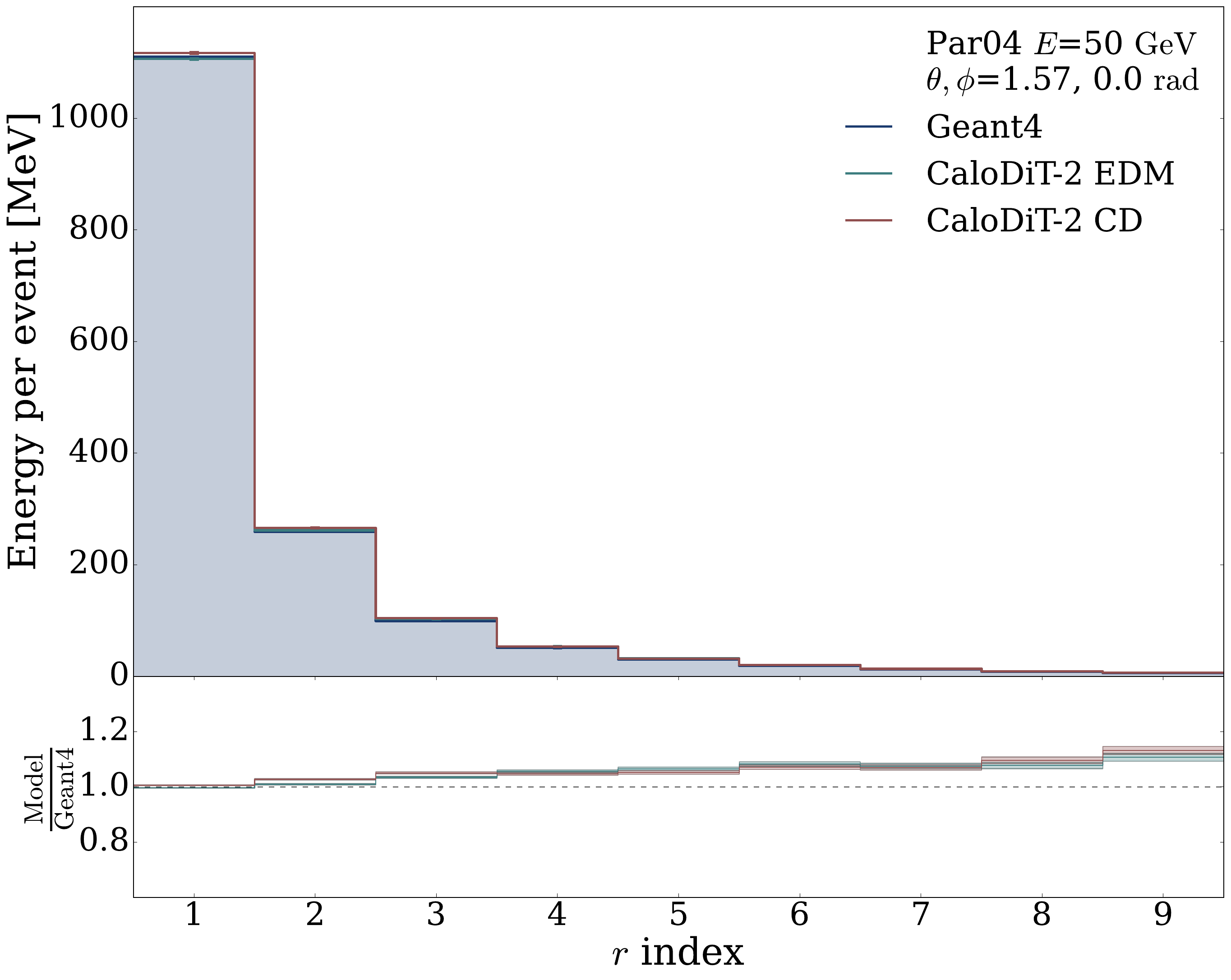}
    \caption{Transverse profile}
  \end{subfigure}
  \hfill
  \begin{subfigure}[b]{0.49\linewidth}
    \centering
    \includegraphics[height=\imgheight]{assets/plots_single_detector/LongFirstMoment_Geo_Par04_E_50_Phi_0.00_Theta_1.57.pdf}
    \caption{First moment of longitudinal profile}
  \end{subfigure}
  \hfill
  \begin{subfigure}[b]{0.49\linewidth}
    \centering
    \includegraphics[height=\imgheight]{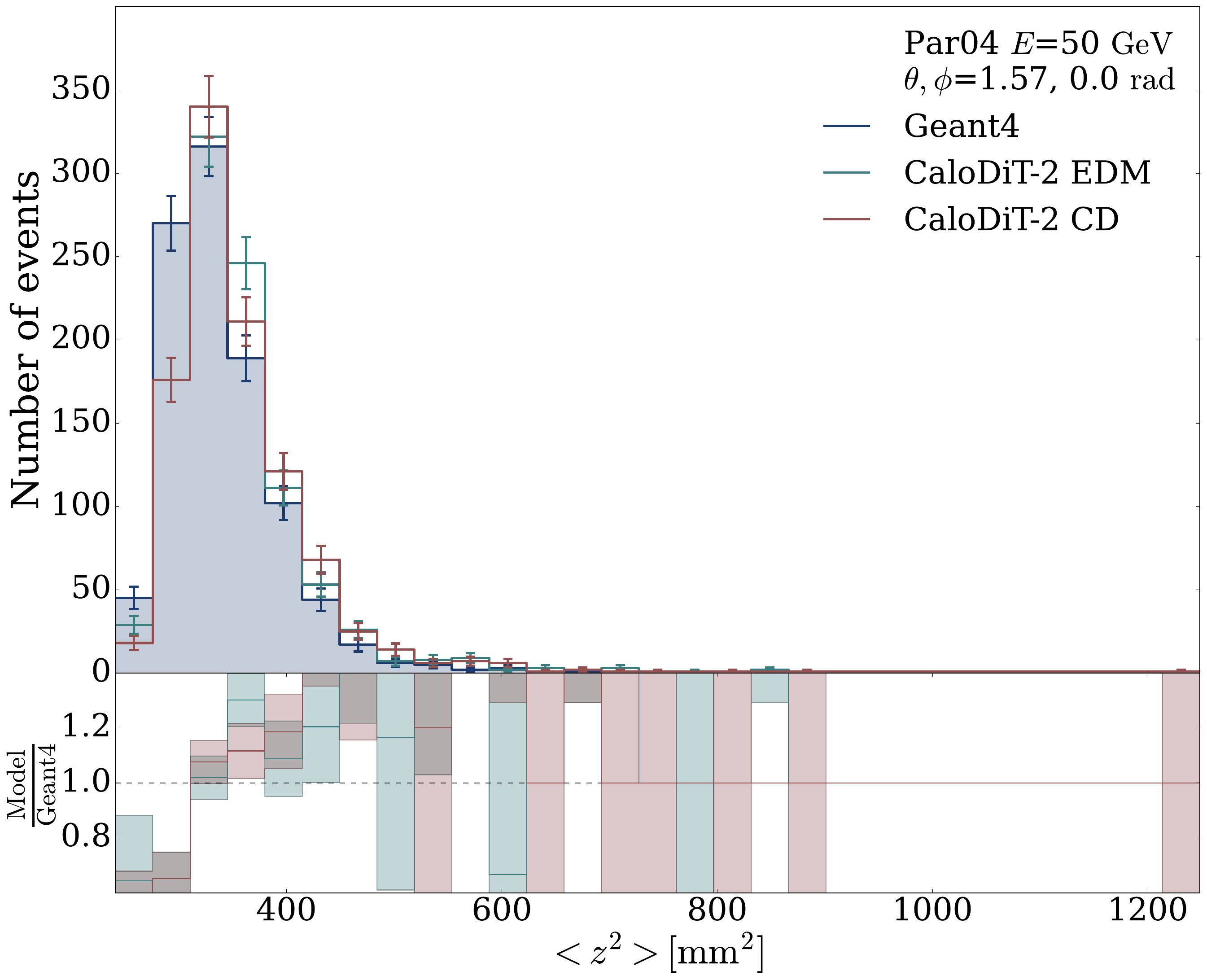}
    \caption{Second moment of longitudinal profile}
  \end{subfigure}
  \hfill
  \begin{subfigure}[b]{0.49\linewidth}
    \centering
    \includegraphics[height=\imgheight]{assets/plots_single_detector/CellLogEnergy_Geo_Par04_E_50_Phi_0.00_Theta_1.57.pdf}
    \caption{Cell energy distribution}
  \end{subfigure}
  \hfill
  \begin{subfigure}[b]{0.49\linewidth}
    \centering
    \includegraphics[height=\imgheight]{assets/plots_single_detector/TotalEventEnergy_Geo_Par04_E_50_Phi_0.00_Theta_1.57.pdf}
    \caption{Total visible energy distribution}
  \end{subfigure}
  \caption{Shower observables for 50 GeV $\gamma$ generated for Par04-SiW detector}
  \label{fig:appendix_single_detector_50GeV}
\end{adjustwidth}
\end{figure}

\begin{figure}[htbp]
\begin{adjustwidth}{-\figenvleftextend}{-\figenvrightextend}
  \centering
  \begin{subfigure}[b]{0.49\linewidth}
    \centering
    \includegraphics[height=\imgheight]{assets/plots_single_detector/LongTotalEnergy_Geo_Par04_E_500_Phi_0.00_Theta_1.57.pdf}
    \caption{Longitudinal profile}
  \end{subfigure}
  \hfill
  \begin{subfigure}[b]{0.49\linewidth}
    \centering
    \includegraphics[height=\imgheight]{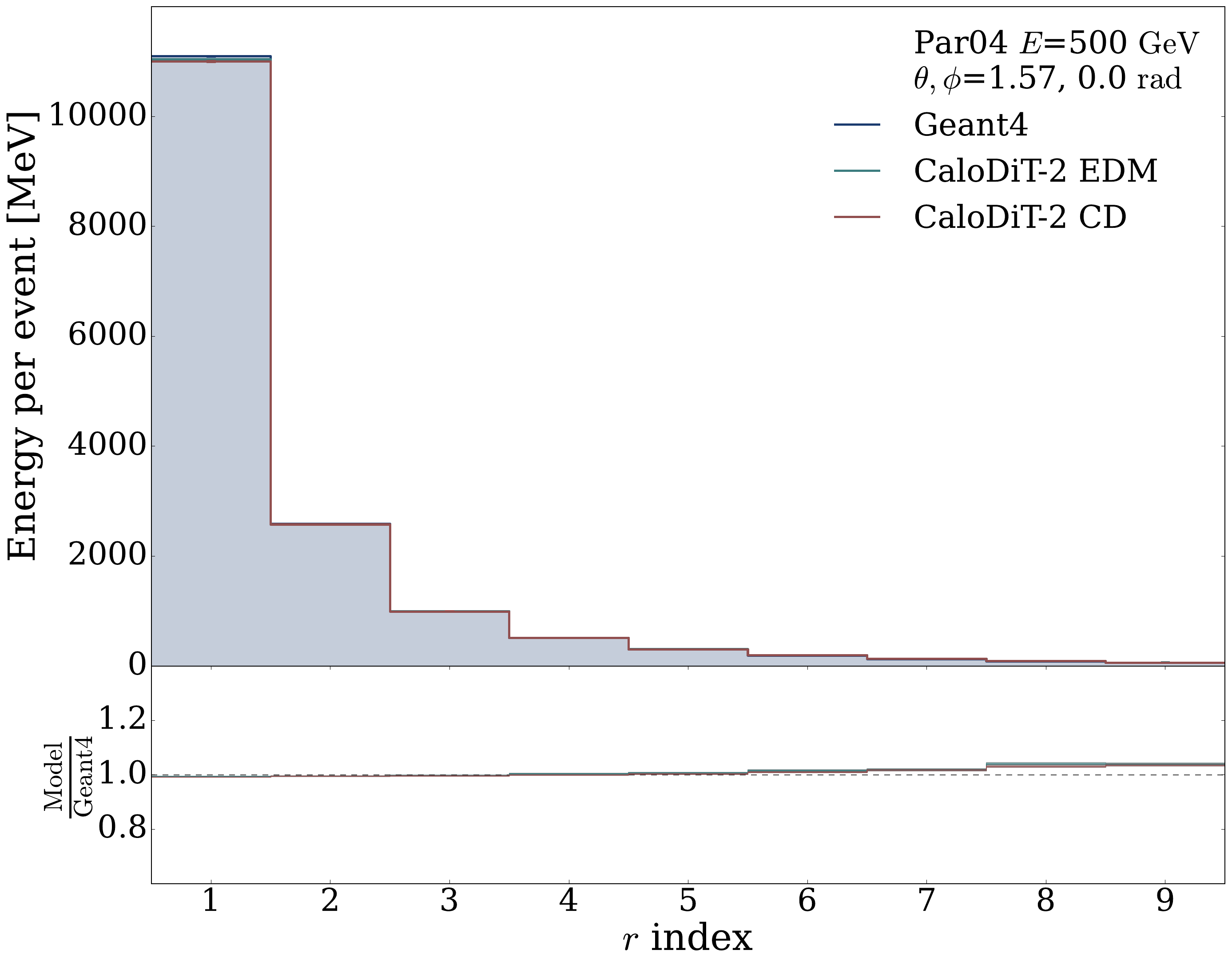}
    \caption{Transverse profile}
  \end{subfigure}
  \hfill
  \begin{subfigure}[b]{0.49\linewidth}
    \centering
    \includegraphics[height=\imgheight]{assets/plots_single_detector/LongFirstMoment_Geo_Par04_E_500_Phi_0.00_Theta_1.57.pdf}
    \caption{First moment of longitudinal profile}
  \end{subfigure}
  \hfill
  \begin{subfigure}[b]{0.49\linewidth}
    \centering
    \includegraphics[height=\imgheight]{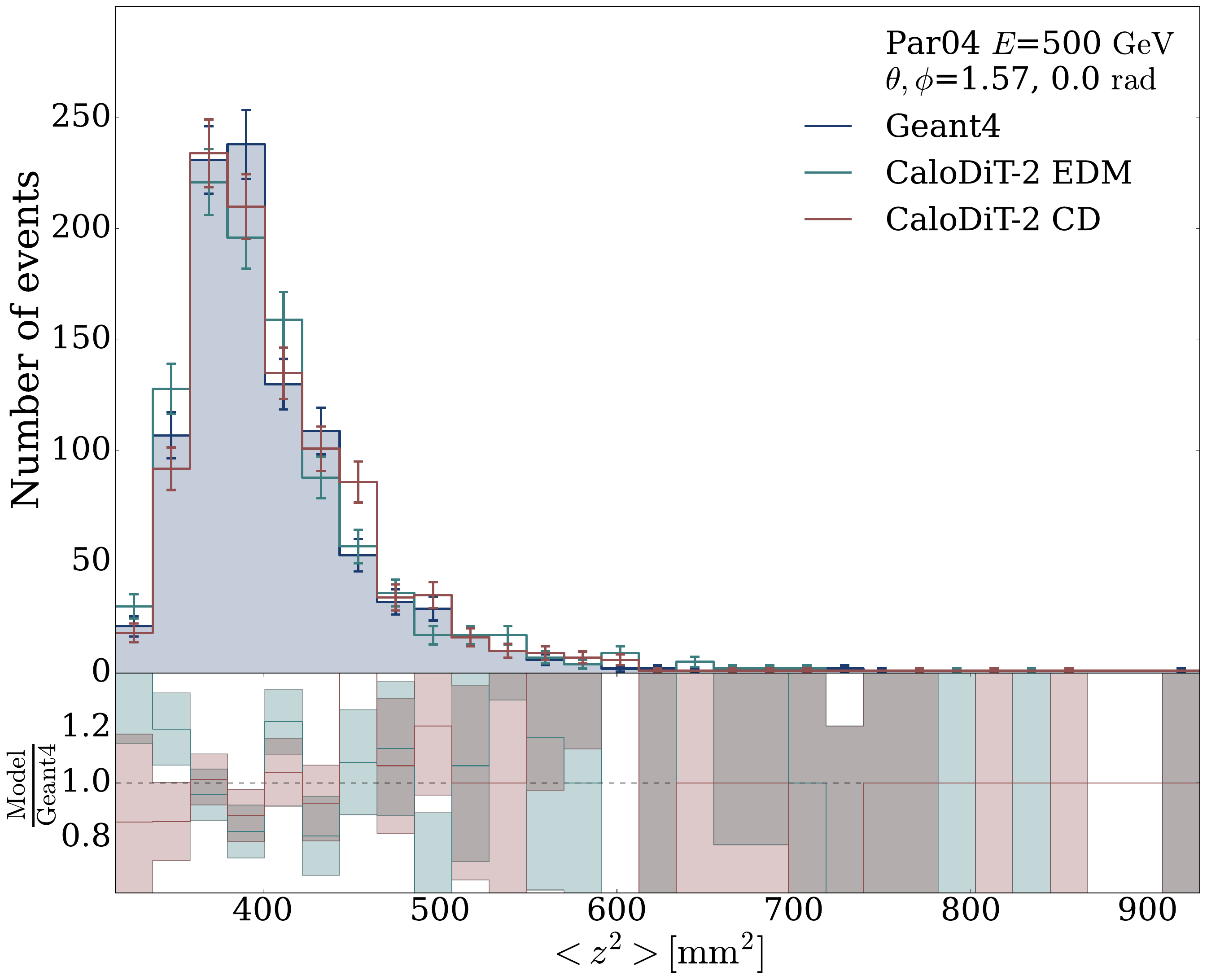}
    \caption{Second moment of longitudinal profile}
  \end{subfigure}
  \hfill
  \begin{subfigure}[b]{0.49\linewidth}
    \centering
    \includegraphics[height=\imgheight]{assets/plots_single_detector/CellLogEnergy_Geo_Par04_E_500_Phi_0.00_Theta_1.57.pdf}
    \caption{Cell energy distribution}
  \end{subfigure}
  \hfill
  \begin{subfigure}[b]{0.49\linewidth}
    \centering
    \includegraphics[height=\imgheight]{assets/plots_single_detector/TotalEventEnergy_Geo_Par04_E_500_Phi_0.00_Theta_1.57.pdf}
    \caption{Total visible energy distribution}
  \end{subfigure}
  \caption{Shower observables for 500 GeV $\gamma$ generated for Par04-SiW detector}
  \label{fig:appendix_single_detector_500GeV}
\end{adjustwidth}
\end{figure}

\begin{figure}[htbp]
\begin{adjustwidth}{-\figenvleftextend}{-\figenvrightextend}
  \centering
  \begin{subfigure}[b]{0.49\linewidth}
    \centering
    \includegraphics[height=\imgheight]{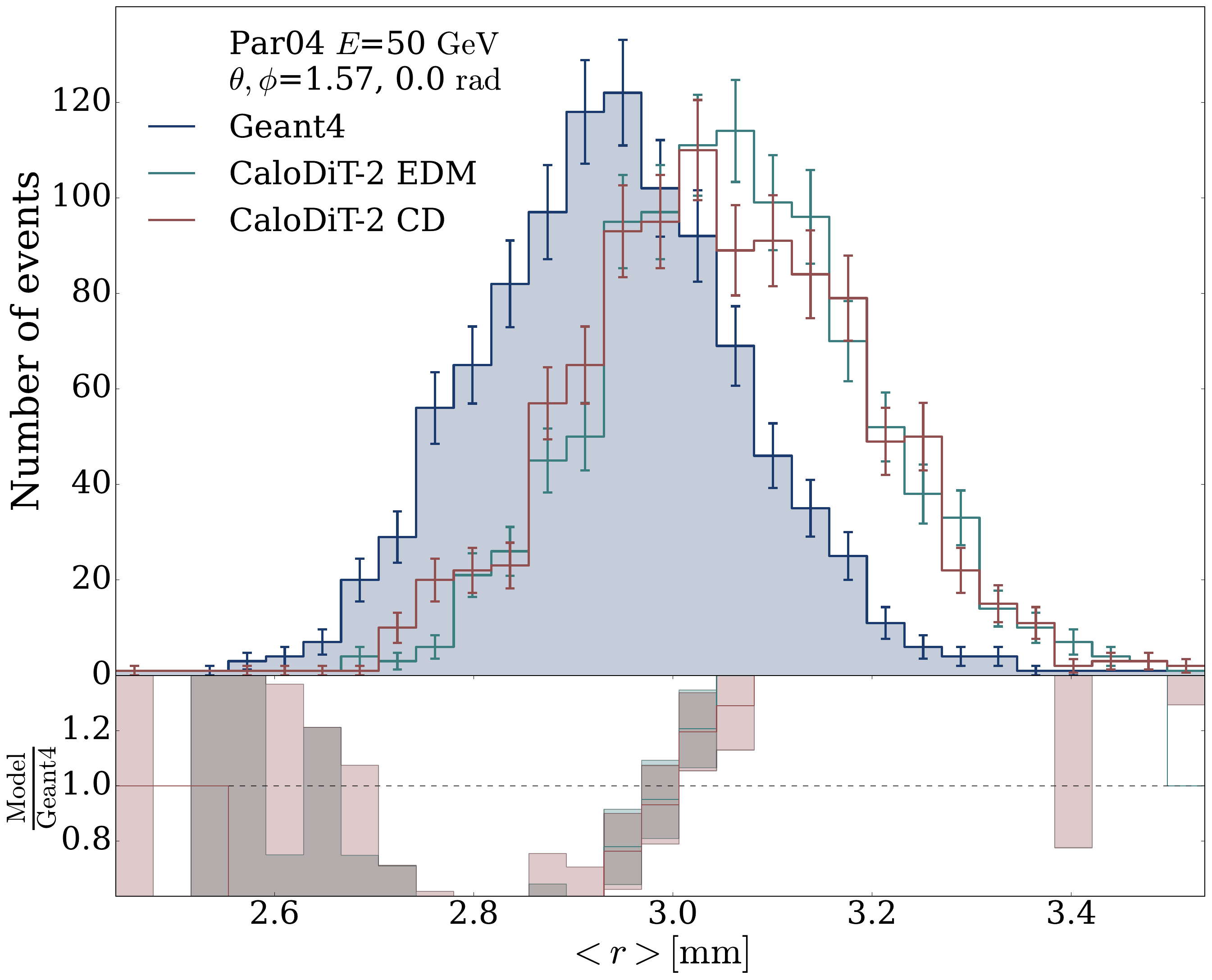}
    \caption{Transverse first moment 50 GeV}
  \end{subfigure}
  \hfill
  \begin{subfigure}[b]{0.49\linewidth}
    \centering
    \includegraphics[height=\imgheight]{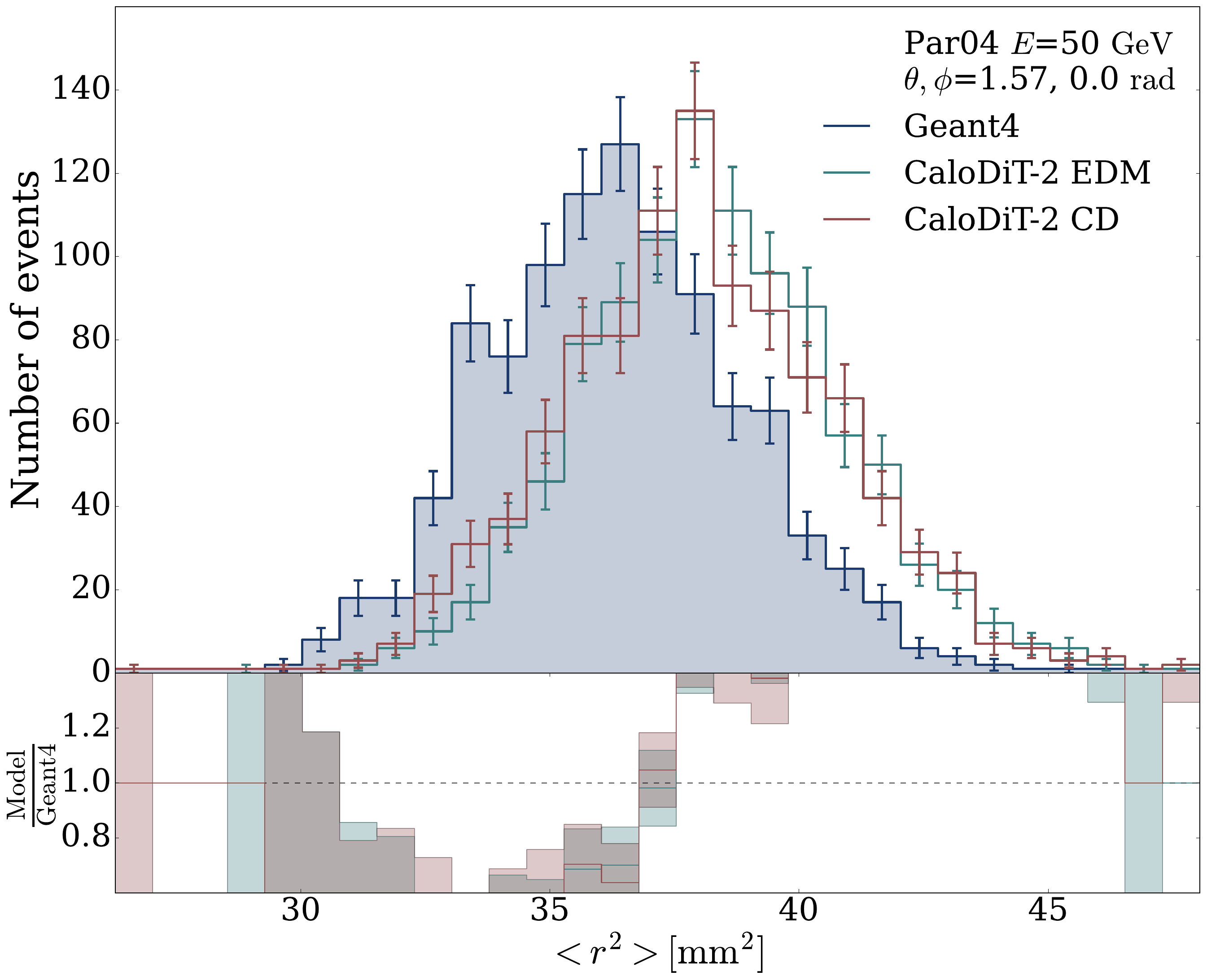}
    \caption{Transverse second moment 50 GeV}
  \end{subfigure}
  \hfill
  \begin{subfigure}[b]{0.49\linewidth}
    \centering
    \includegraphics[height=\imgheight]{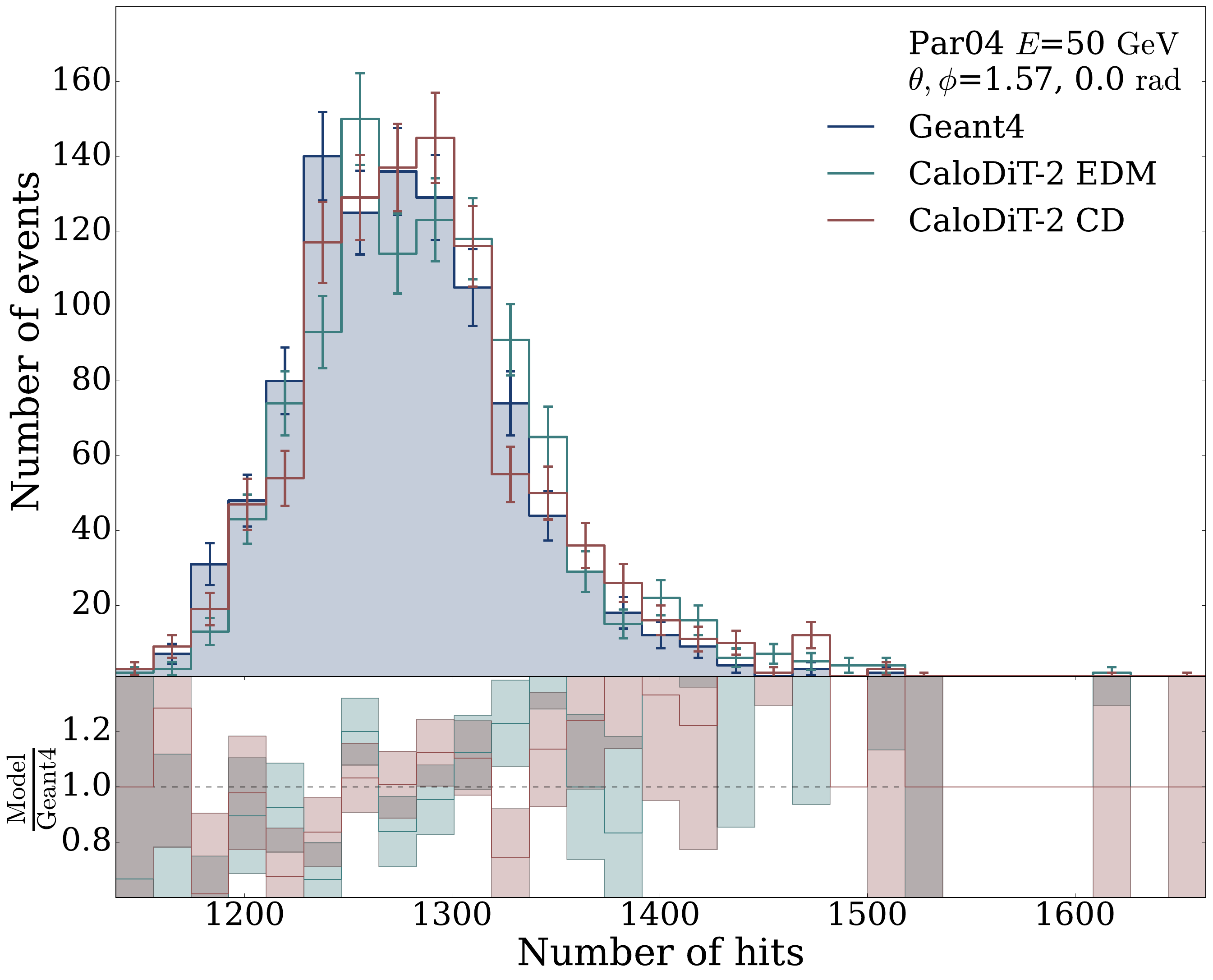}
    \caption{Total hits 50 GeV}
  \end{subfigure}
  \hfill
  \begin{subfigure}[b]{0.49\linewidth}
    \centering
    \includegraphics[height=\imgheight]{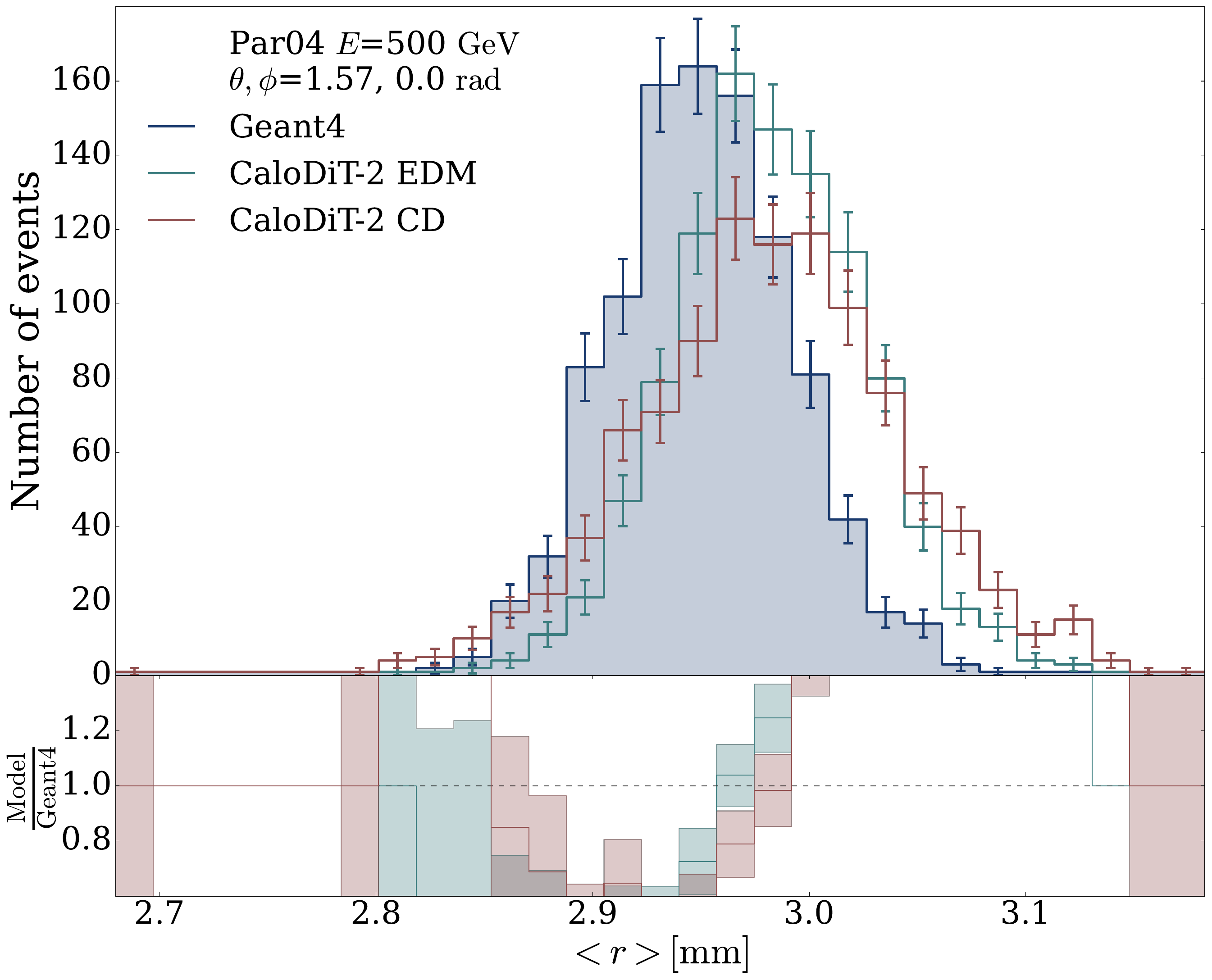}
    \caption{Transverse first moment 500 GeV}
  \end{subfigure}
  \hfill
  \begin{subfigure}[b]{0.49\linewidth}
    \centering
    \includegraphics[height=\imgheight]{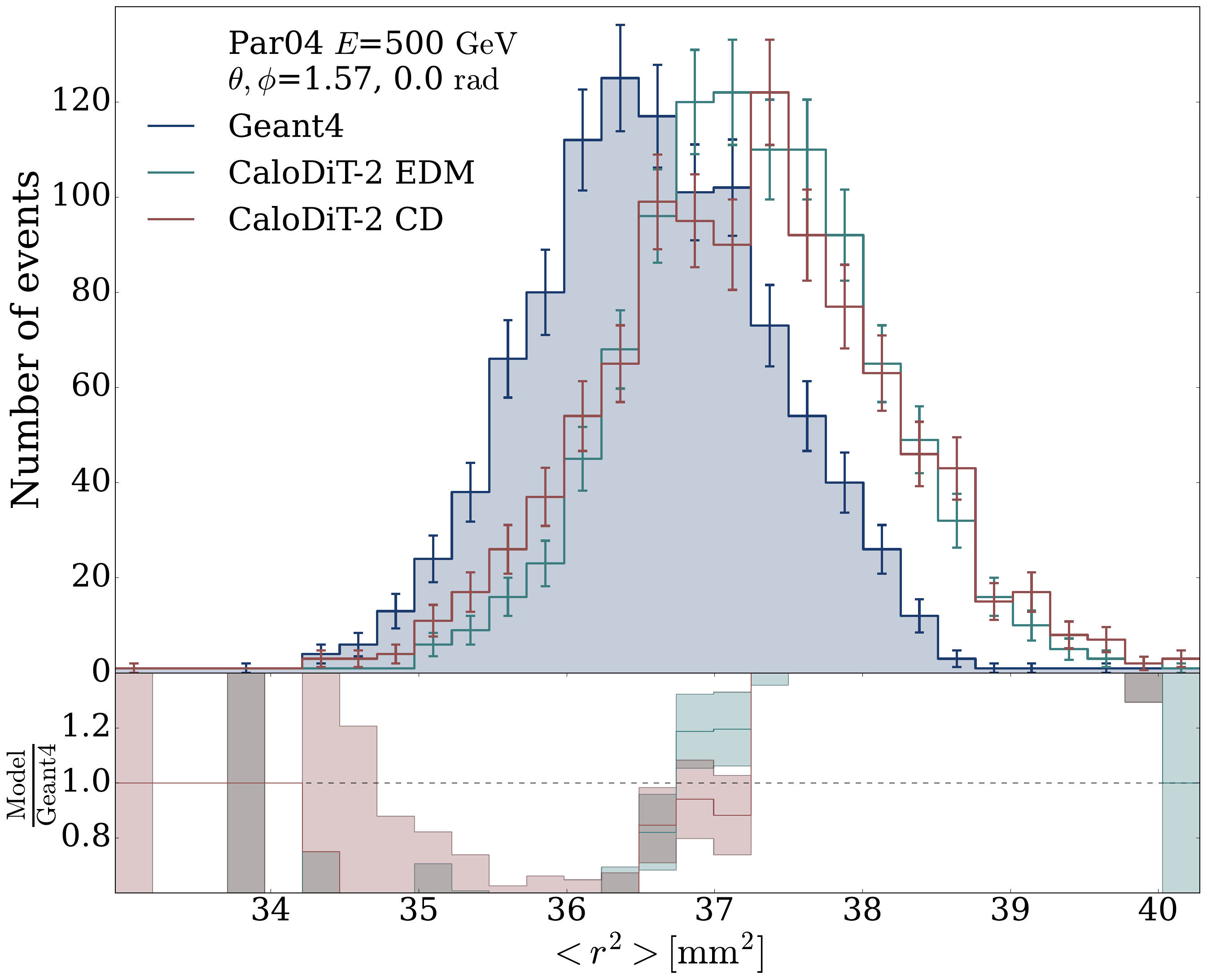}
    \caption{Transverse second moment 500 GeV}
  \end{subfigure}
  \hfill
  \begin{subfigure}[b]{0.49\linewidth}
    \centering
    \includegraphics[height=\imgheight]{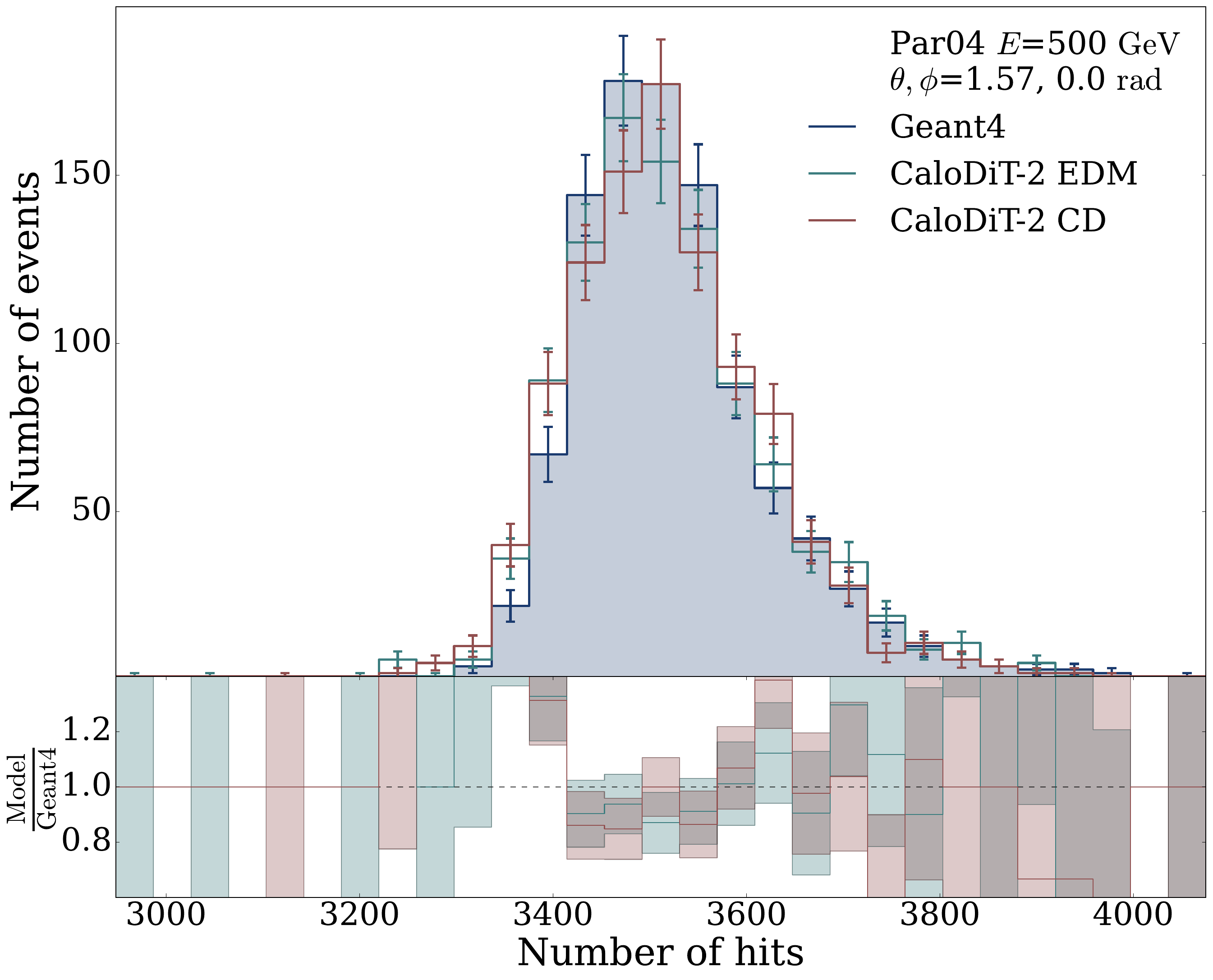}
    \caption{Total hits 500 GeV}
  \end{subfigure}
  \caption{More shower observables for 50 GeV and 500 GeV $\gamma$ generated for Par04-SiW detector}
  \label{fig:appendix_more_single_detector}
\end{adjustwidth}
\end{figure}

\section{Incident Energy Distribution}
\label{sec:appendix_flat_power}

\begin{figure}[htbp]
\begin{adjustwidth}{-\figenvleftextend}{-\figenvrightextend}
  \centering
  \begin{subfigure}[b]{0.49\linewidth}
    \centering
    \includegraphics[height=\imgheight]{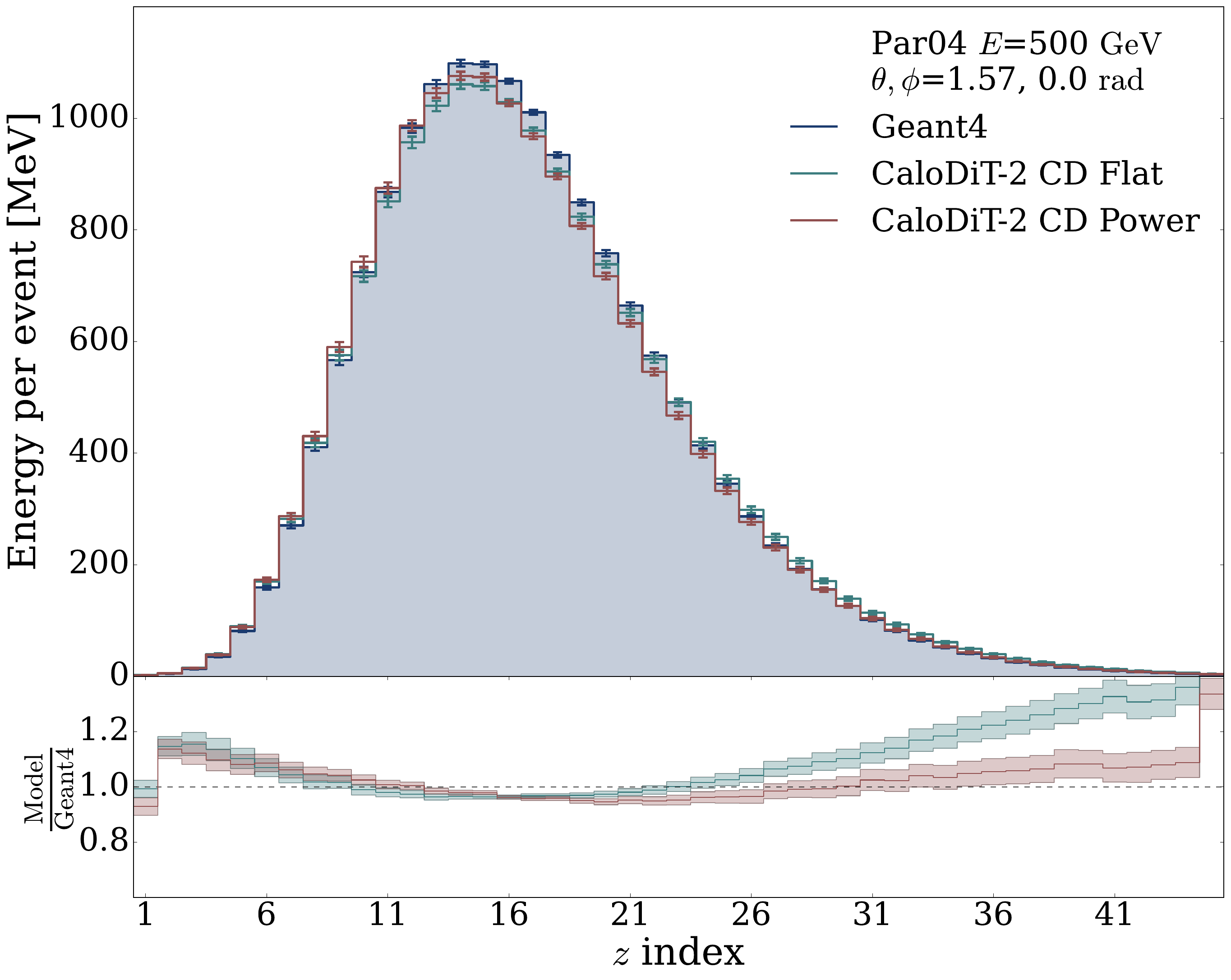}
    \caption{Longitudinal profile}
  \end{subfigure}
  \hfill
  \begin{subfigure}[b]{0.49\linewidth}
    \centering
    \includegraphics[height=\imgheight]{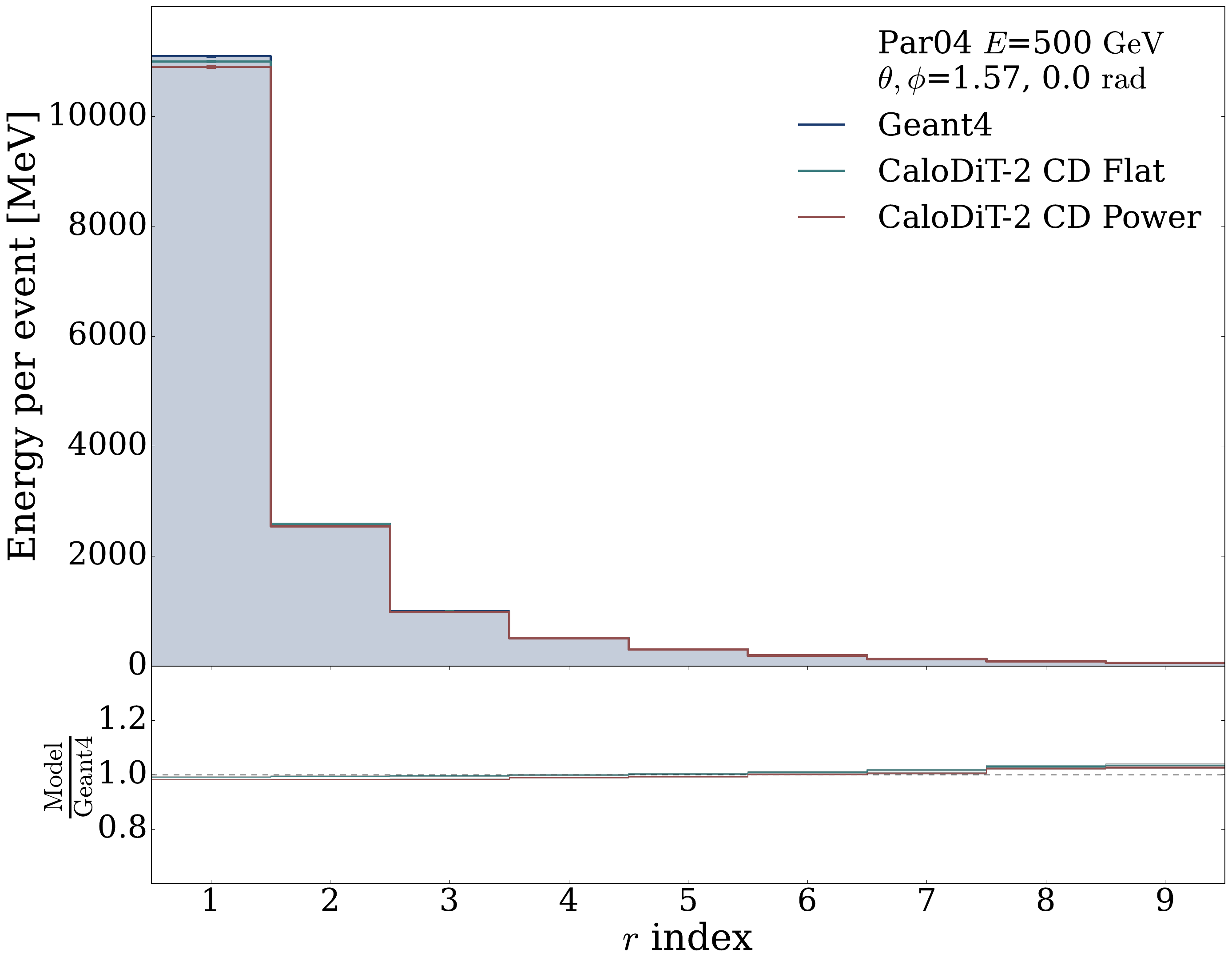}
    \caption{Transverse profile}
  \end{subfigure}
  \hfill
  \begin{subfigure}[b]{0.49\linewidth}
    \centering
    \includegraphics[height=\imgheight]{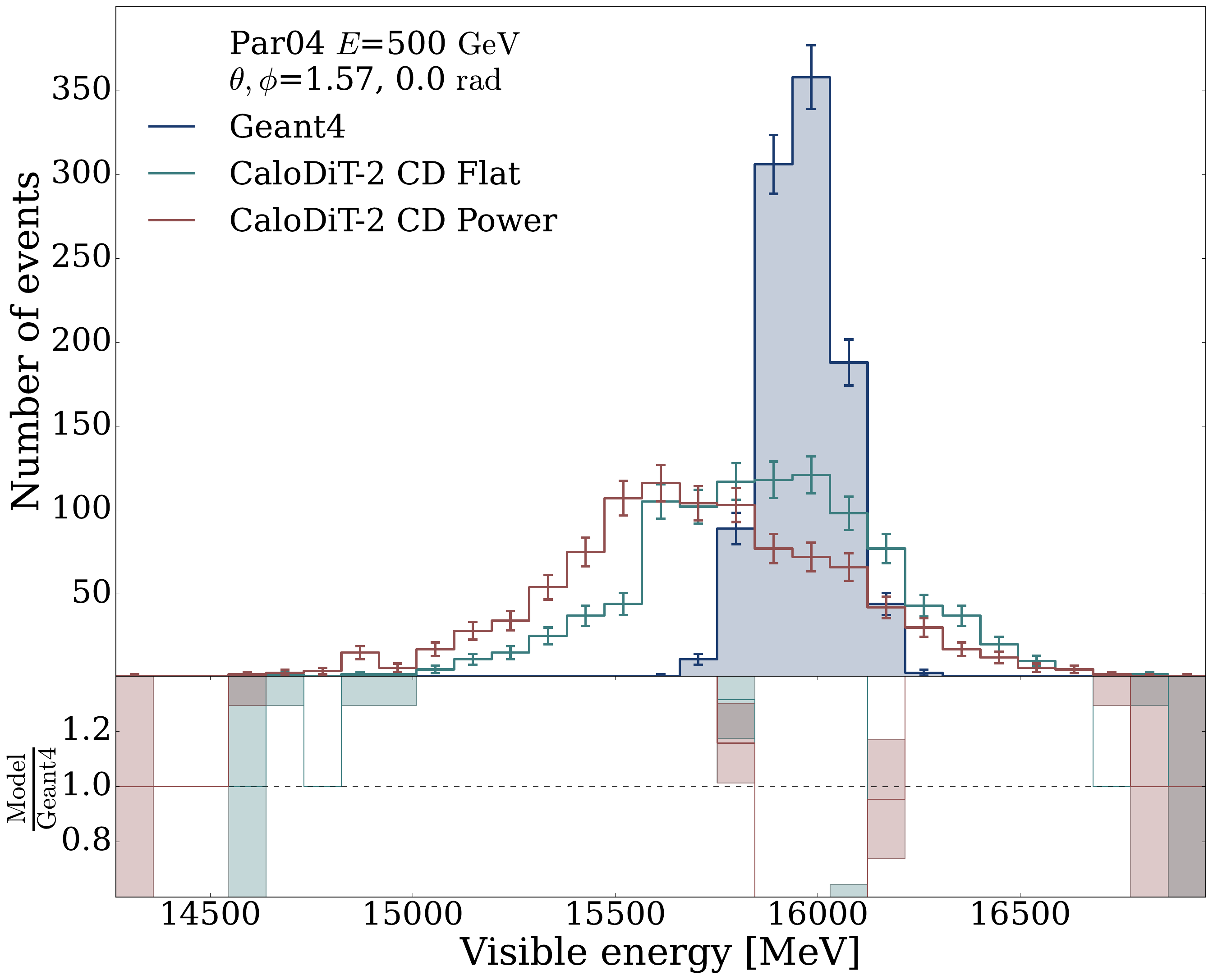}
    \caption{Total visible energy}
  \end{subfigure}
  \hfill
  \begin{subfigure}[b]{0.49\linewidth}
    \centering
    \includegraphics[height=\imgheight]{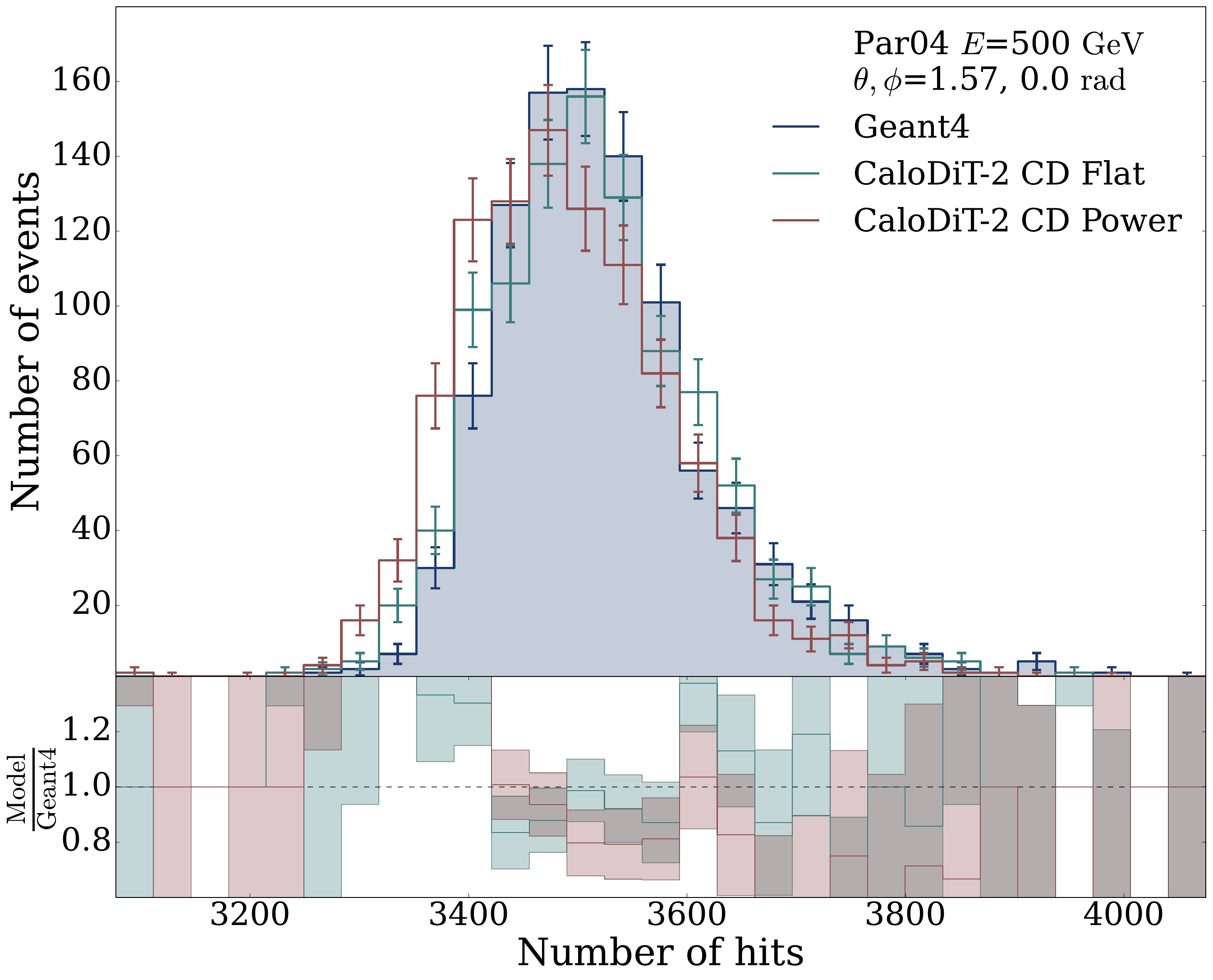}
    \caption{Total hits}
  \end{subfigure}
  \hfill
  \begin{subfigure}[b]{0.49\linewidth}
    \centering
    \includegraphics[height=\imgheight]{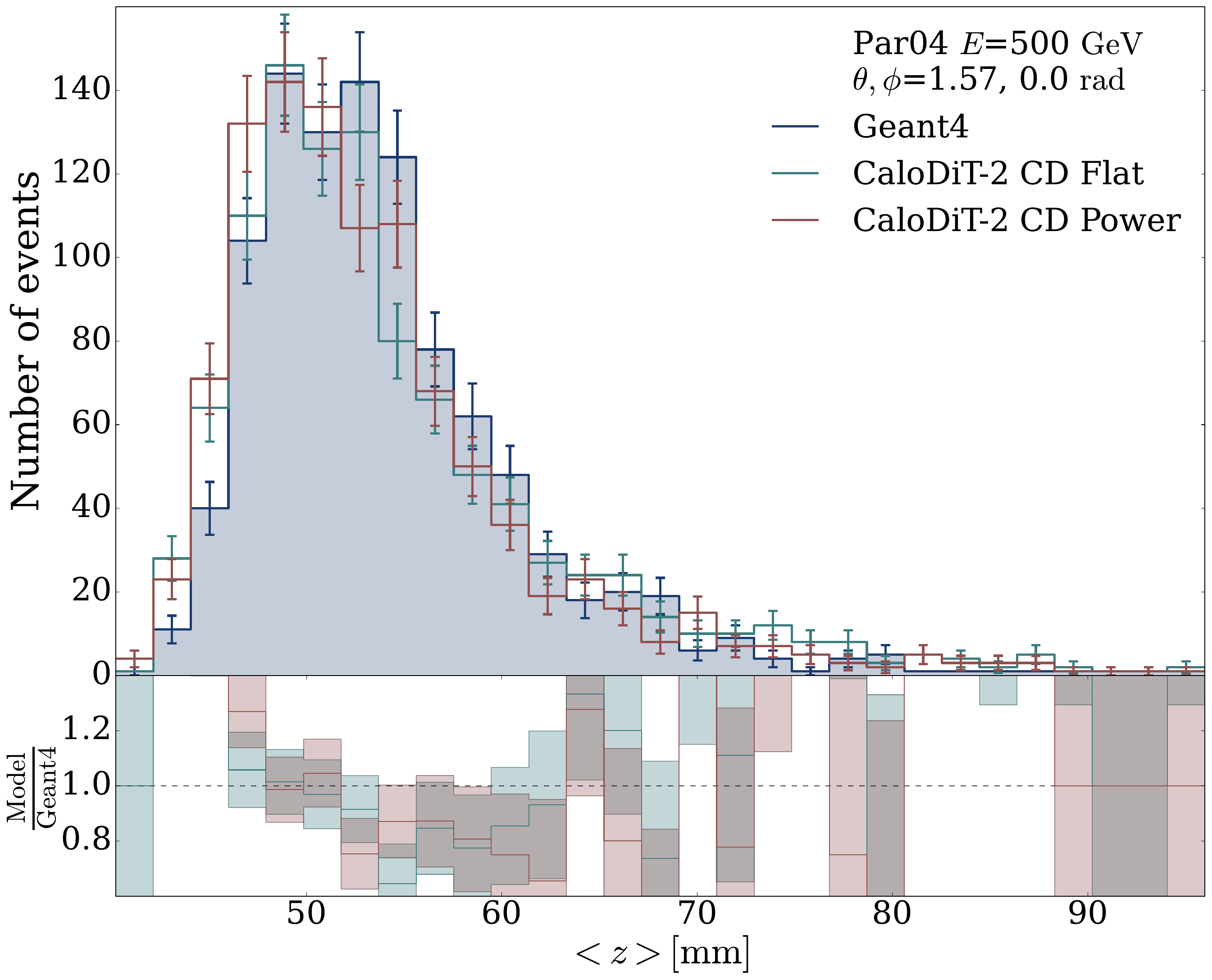}
    \caption{Longitudinal first moment}
  \end{subfigure}
  \hfill
  \begin{subfigure}[b]{0.49\linewidth}
    \centering
    \includegraphics[height=\imgheight]{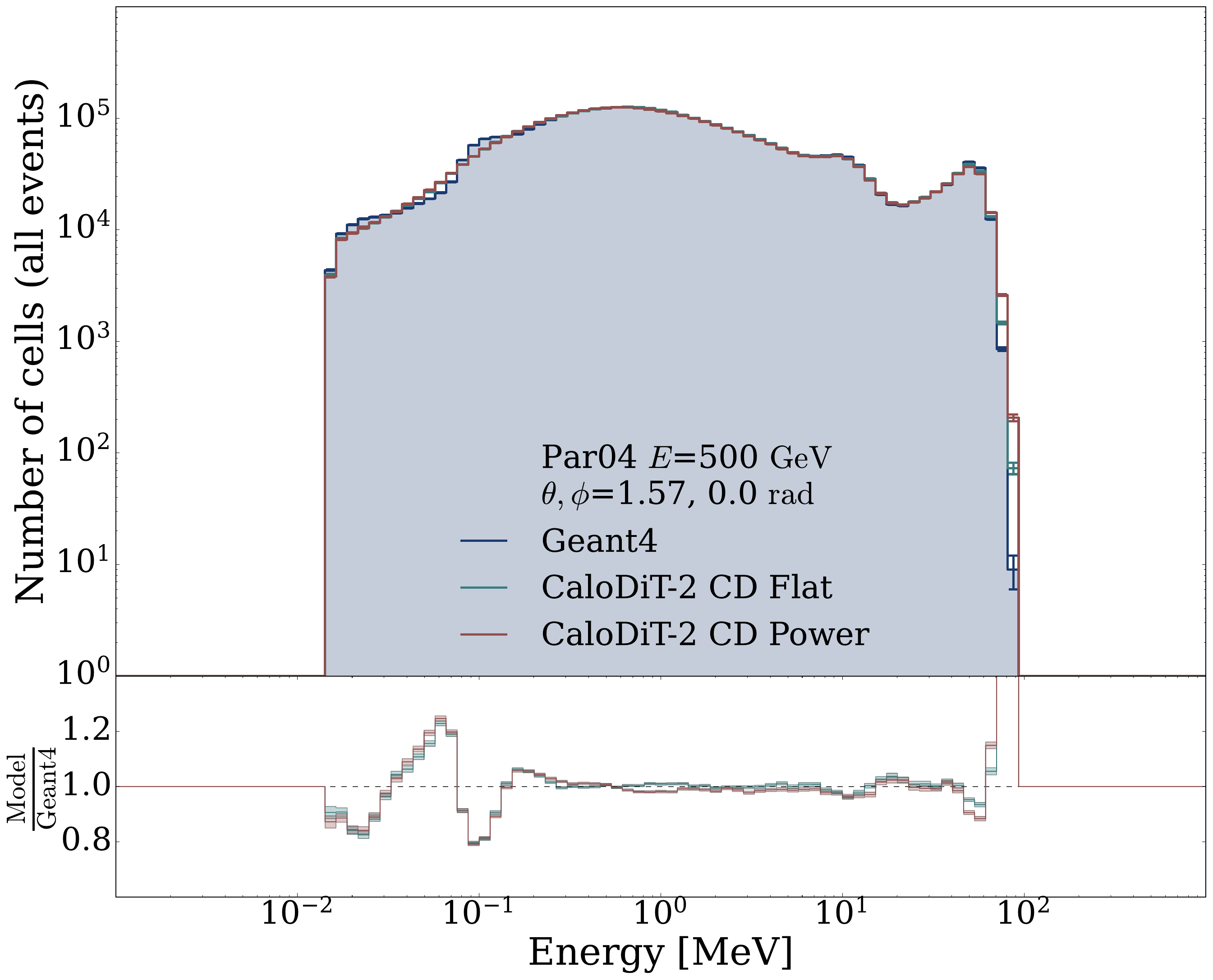}
    \caption{Cell energy}
  \end{subfigure}
  \caption{Shower observables for training energy spectrum. Comparing uniform vs power distribution of incident energy for 500 GeV $\gamma$. We observe that a model trained with uniform distribution of incident energy leads to better results.}
  \label{fig:appendix_flat_vs_power}
\end{adjustwidth}
\end{figure}

\section{Additional Results On Generalisable Model}
\label{sec:appendix_adaptation}

\begin{figure}[htbp]
  \begin{adjustwidth}{-\figenvleftextend}{-\figenvrightextend}
  \centering
  \begin{subfigure}[b]{0.49\linewidth}
    \centering
    \includegraphics[height=\imgheight]{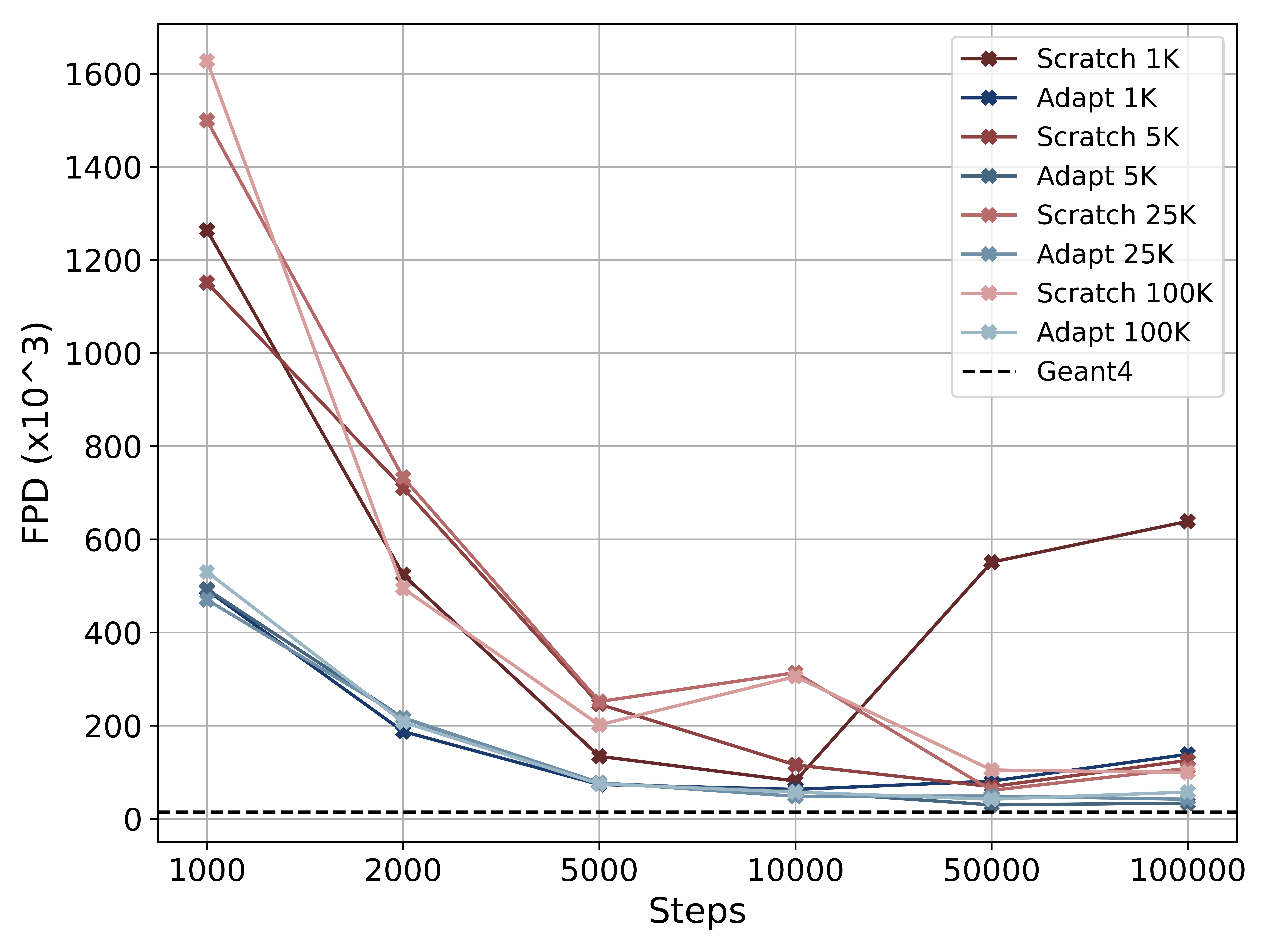}
    \caption{FPD vs steps with a linear y-axis across all configurations.}
  \end{subfigure}
  \hfill
  \begin{subfigure}[b]{0.49\linewidth}
    \centering
    \includegraphics[height=\imgheight]{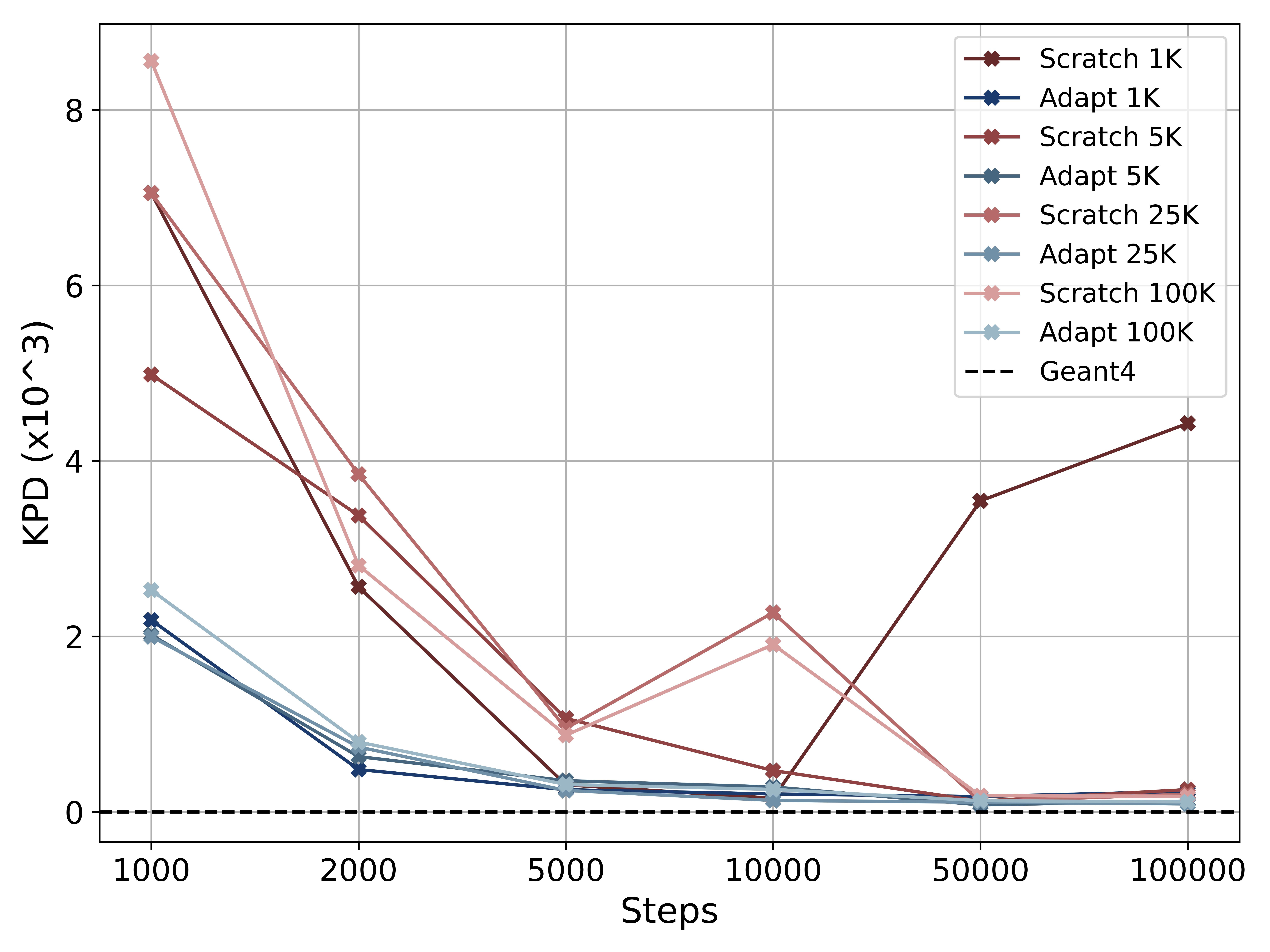}
    \caption{KPD vs steps with a linear y-axis across all configurations.}
  \end{subfigure}
  \end{adjustwidth}
  \caption{Comparing training from a pre-trained model with training from scratch. We show the FPD (a) and KPD (Kernel Physics Distance) (b) \cite{kansal2023evaluating} with a linear y-axis. In both cases, we see advantages of adaptation. KPD plots also follow similar trends as FPD plots.}
  \label{fig:appendix_adaptation_fpd_linear}
\end{figure}

\begin{figure}[htbp]
\begin{adjustwidth}{-\figenvleftextend}{-\figenvrightextend}
  \centering
  \begin{subfigure}[b]{0.49\linewidth}
    \centering
    \includegraphics[height=\imgheight]{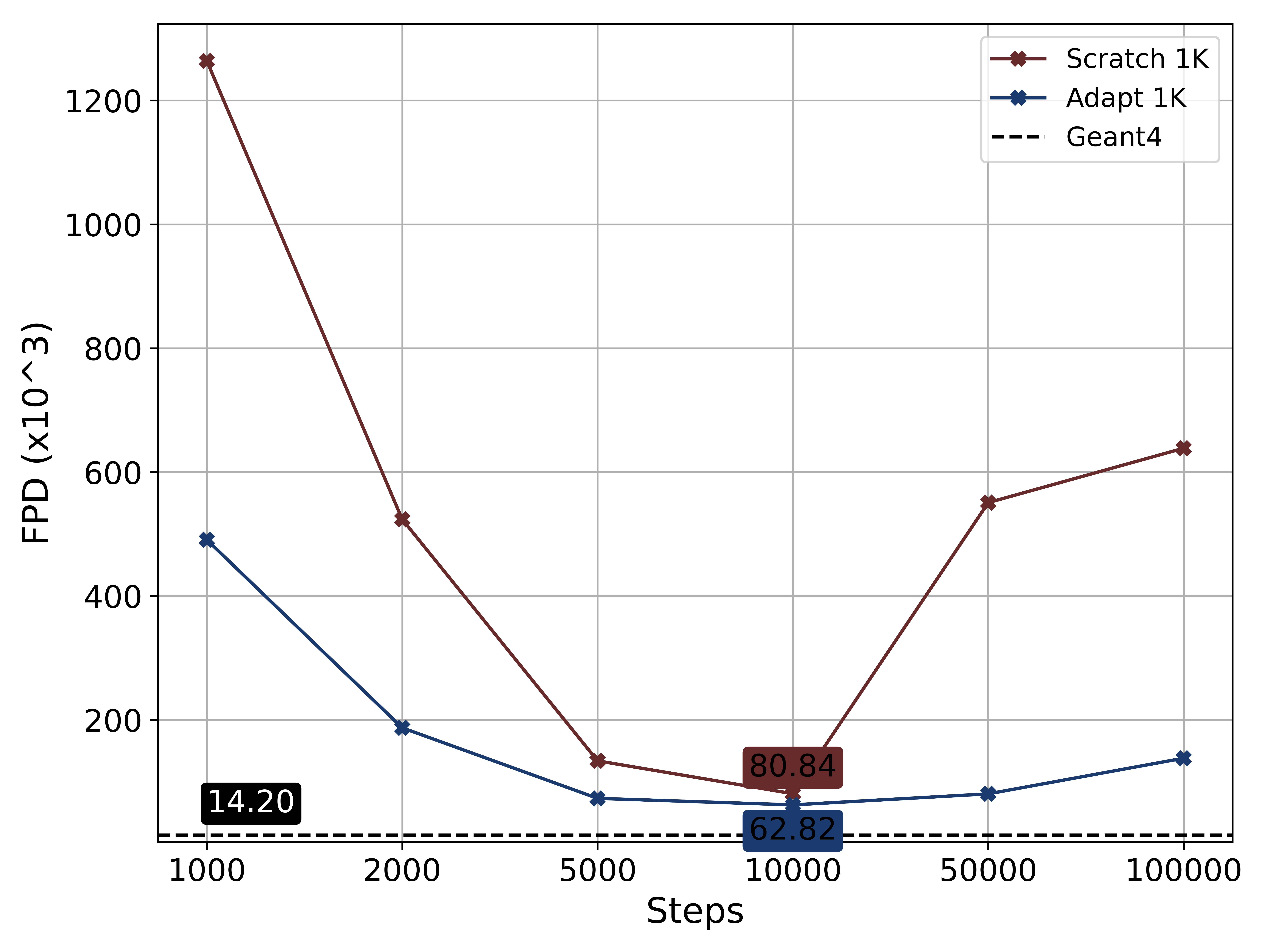}
    \caption{Training samples 1K}
  \end{subfigure}
  \hfill
  \begin{subfigure}[b]{0.49\linewidth}
    \centering
    \includegraphics[height=\imgheight]{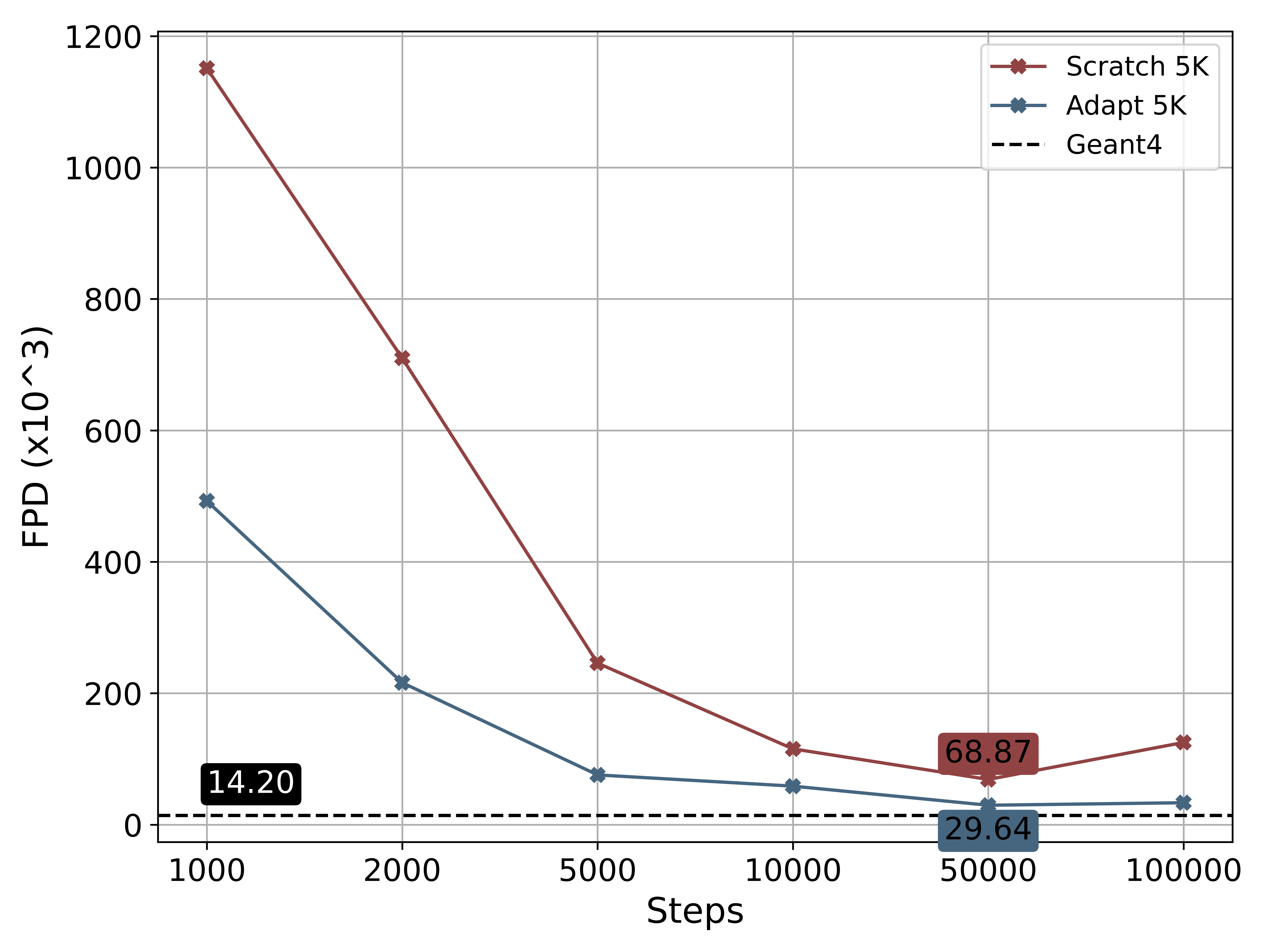}
    \caption{Training samples 5K}
  \end{subfigure}
  \hfill
  \begin{subfigure}[b]{0.49\linewidth}
    \centering
    \includegraphics[height=\imgheight]{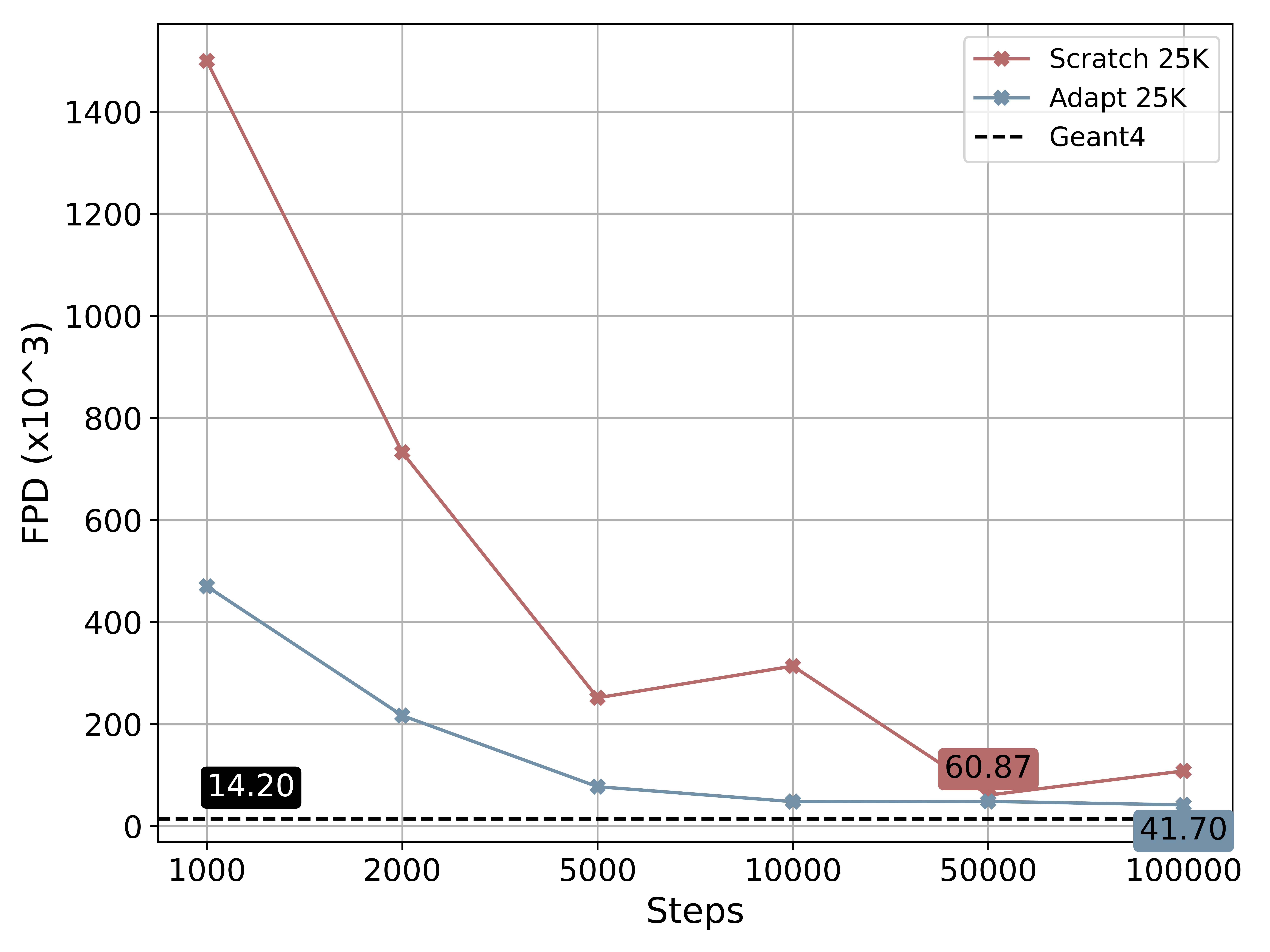}
    \caption{Training samples 25K}
  \end{subfigure}
  \hfill
  \begin{subfigure}[b]{0.49\linewidth}
    \centering
    \includegraphics[height=\imgheight]{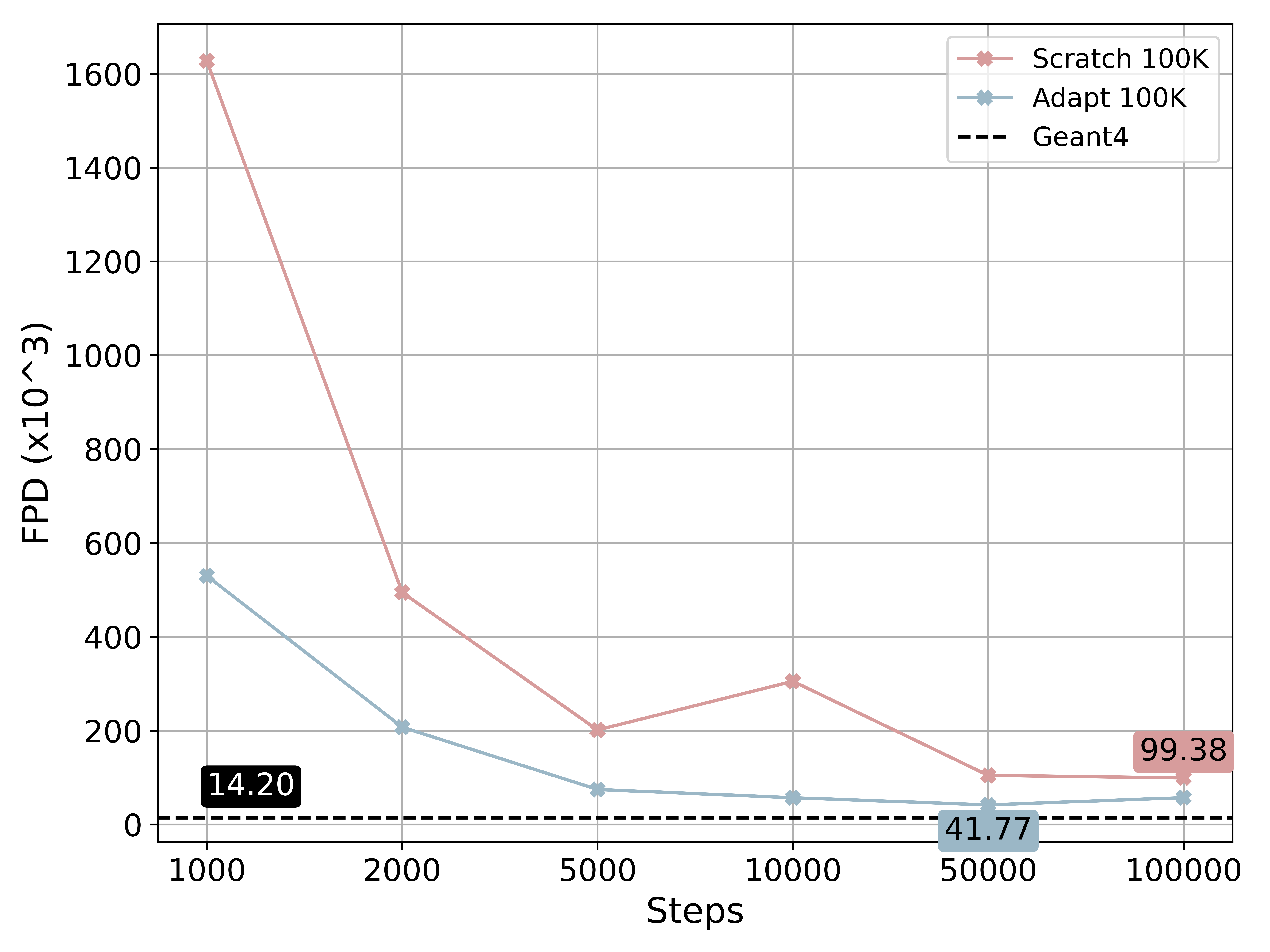}
    \caption{Training samples 100K}
  \end{subfigure}
  \caption{FPD vs steps for each configuration of the number of training data samples with a linear y-axis.}
  \label{fig:appendix_adaptation_linear}
\end{adjustwidth}
\end{figure}

\begin{figure}[htbp]
\begin{adjustwidth}{-\figenvleftextend}{-\figenvrightextend}
  \centering
  \begin{subfigure}[b]{0.49\linewidth}
    \centering
    \includegraphics[height=\imgheight]{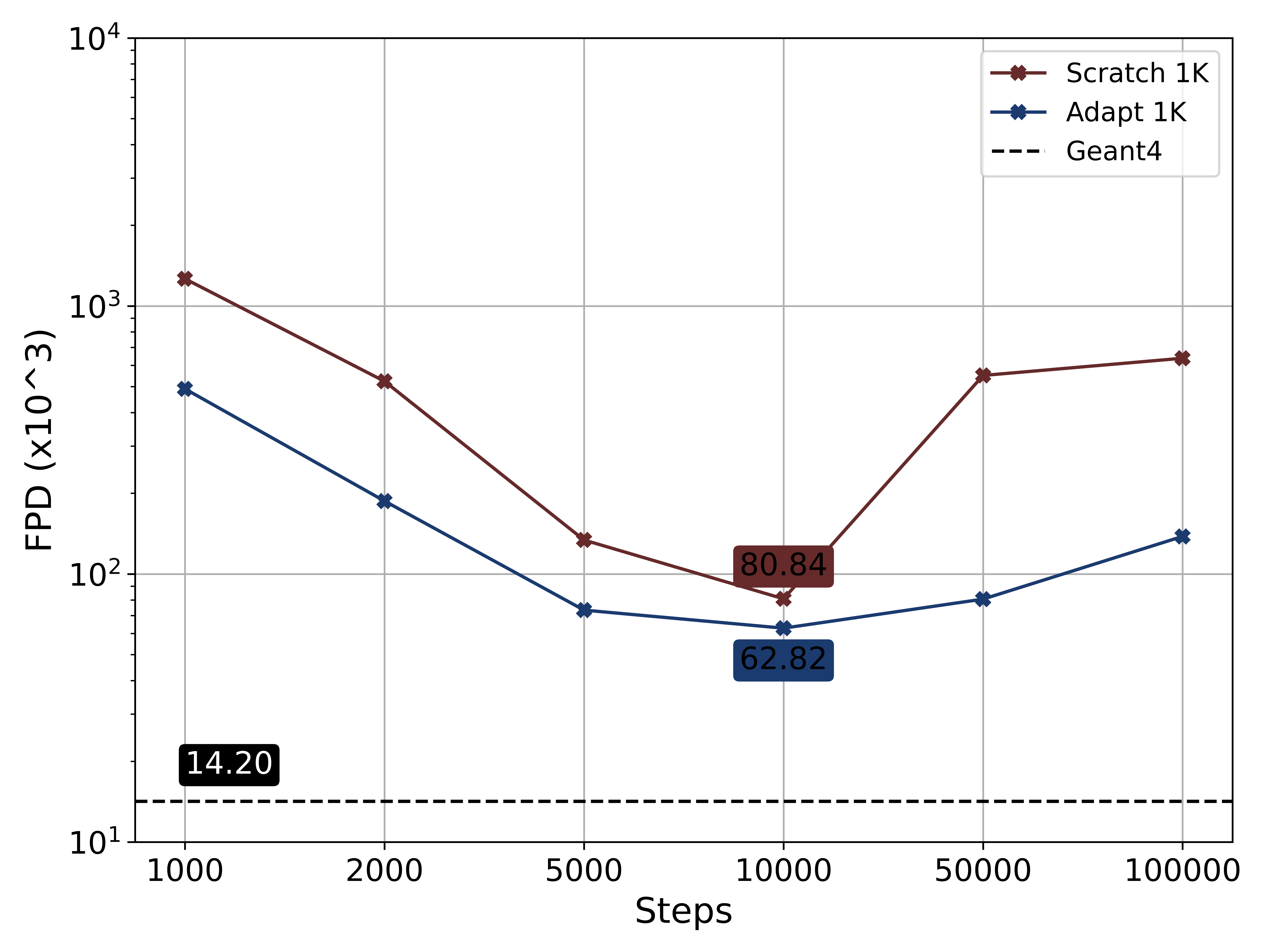}
    \caption{Training samples 1K}
  \end{subfigure}
  \hfill
  \begin{subfigure}[b]{0.49\linewidth}
    \centering
    \includegraphics[height=\imgheight]{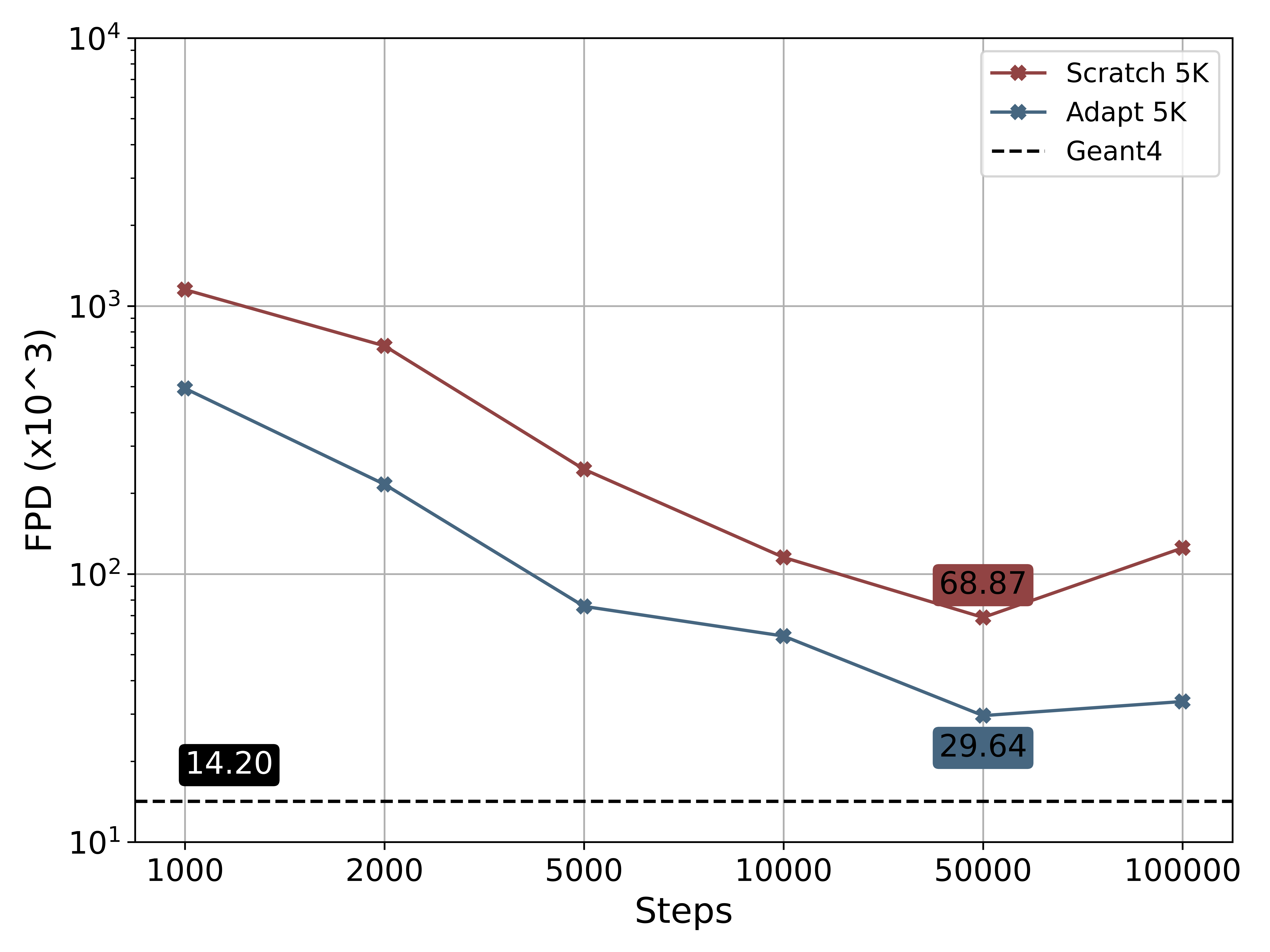}
    \caption{Training samples 5K}
  \end{subfigure}
  \hfill
  \begin{subfigure}[b]{0.49\linewidth}
    \centering
    \includegraphics[height=\imgheight]{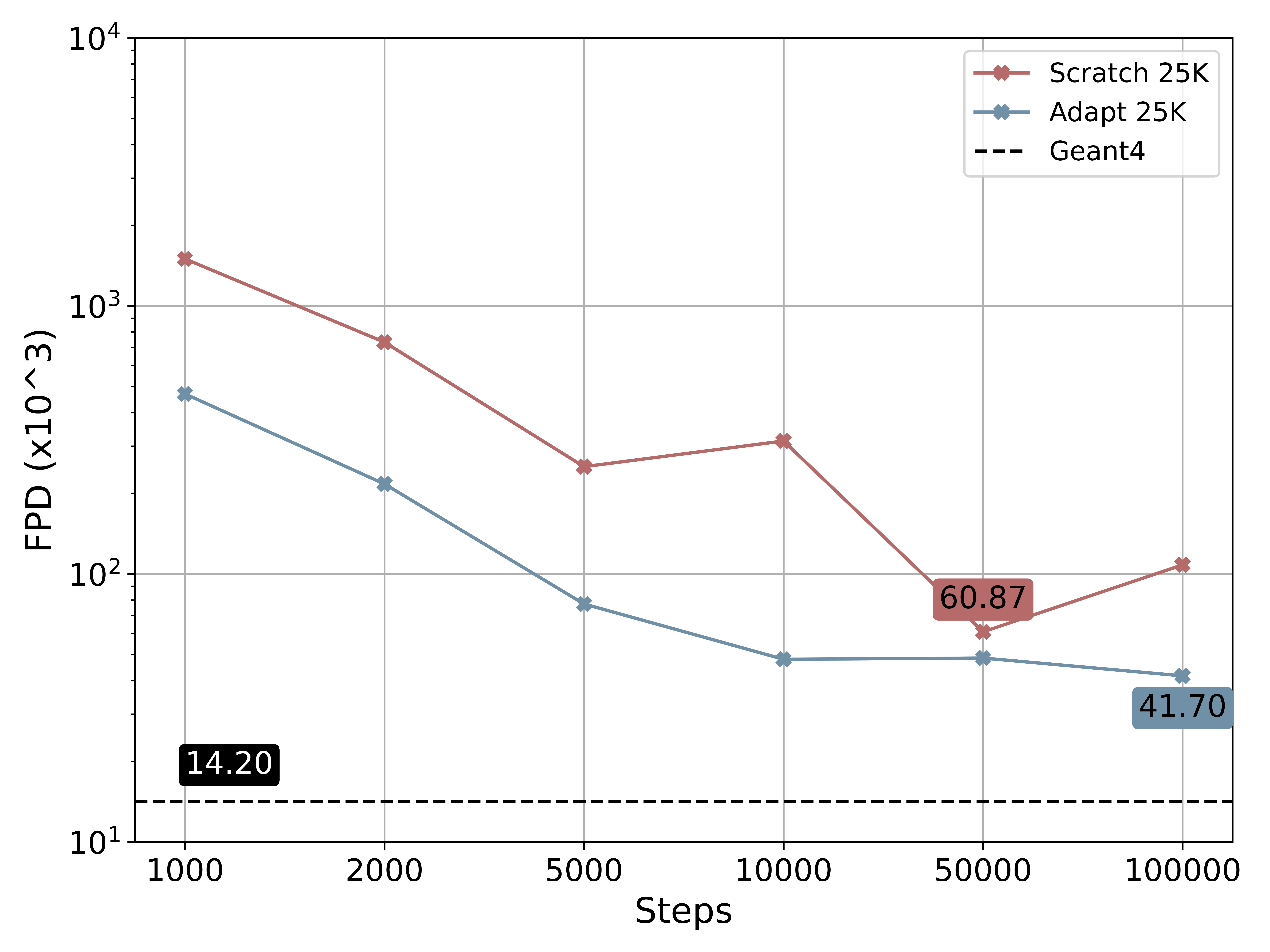}
    \caption{Training samples 25K}
  \end{subfigure}
  \hfill
  \begin{subfigure}[b]{0.49\linewidth}
    \centering
    \includegraphics[height=\imgheight]{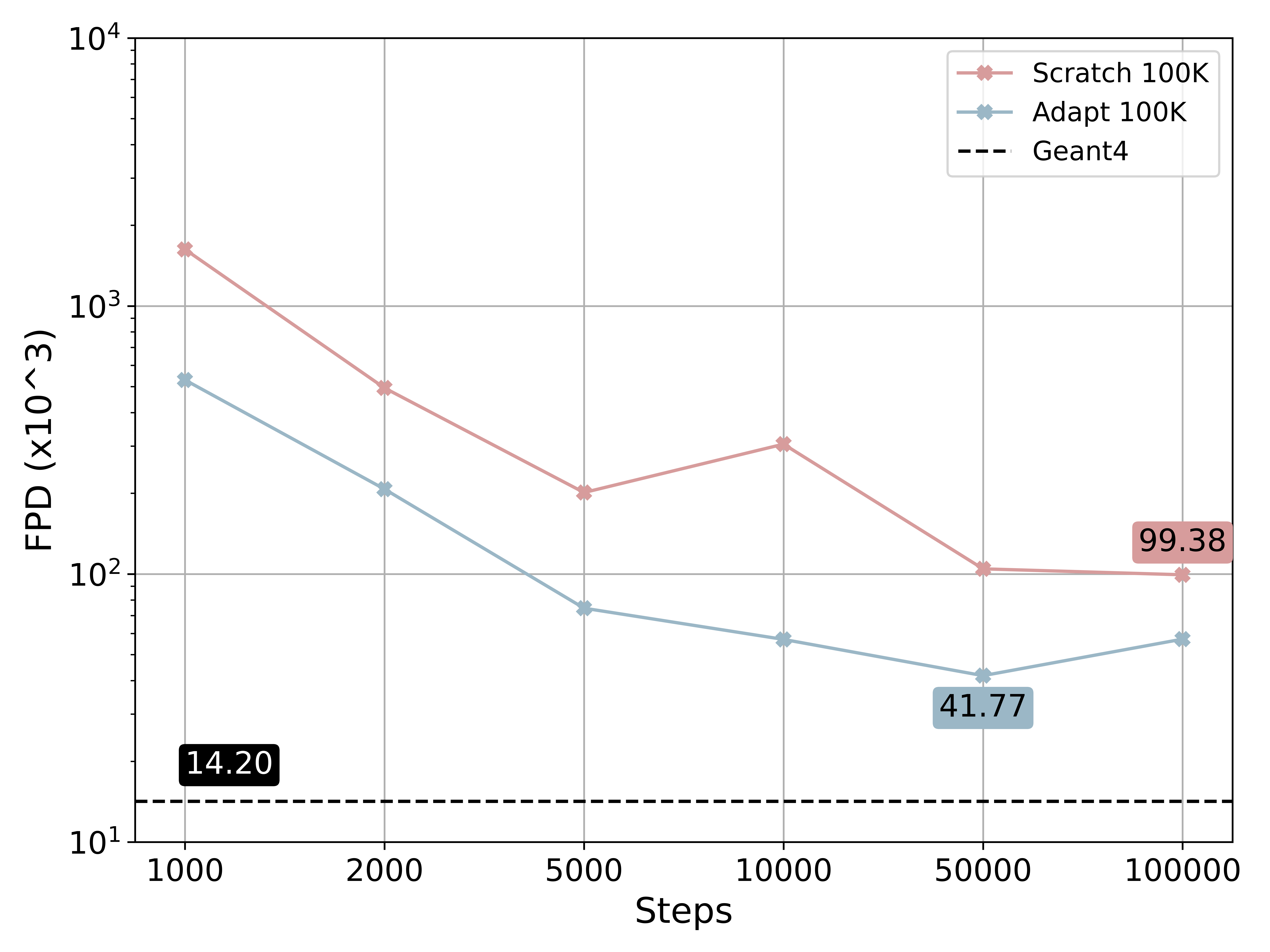}
    \caption{Training samples 100K}
  \end{subfigure}
  \caption{FPD vs steps for each configuration of the number of training data samples with a logarithmic y-axis.}
  \label{fig:appendix_adaptation_log}
\end{adjustwidth}
\end{figure}

\begin{figure}[htbp]
\begin{adjustwidth}{-\figenvleftextend}{-\figenvrightextend}
  \centering
  \begin{subfigure}[b]{0.49\linewidth}
    \centering
    \includegraphics[height=\imgheight]{assets/adaptation/pr_vs_steps_with_data_1K.png}
    \caption{Training samples 1K}
  \end{subfigure}
  \hfill
  \begin{subfigure}[b]{0.49\linewidth}
    \centering
    \includegraphics[height=\imgheight]{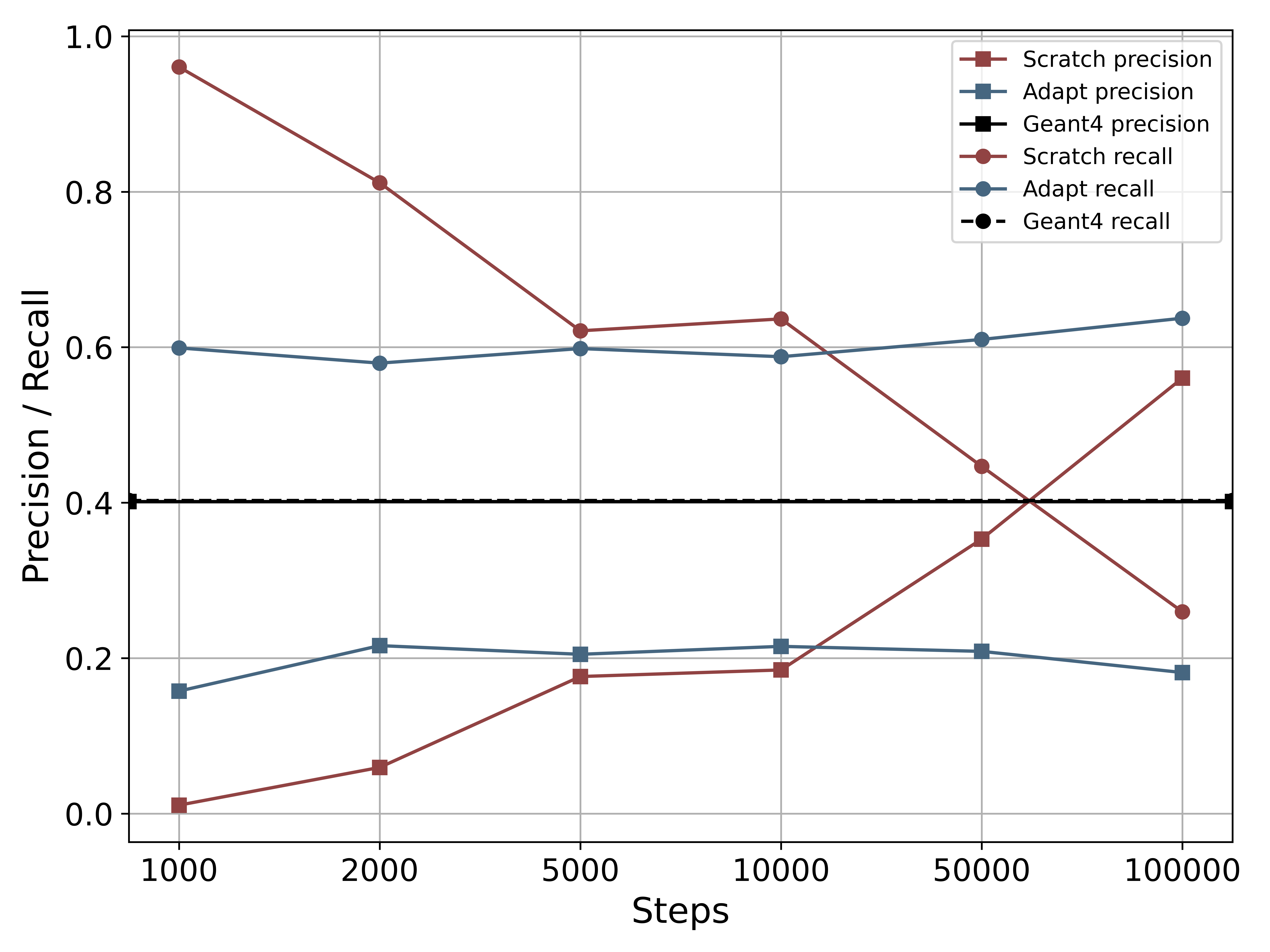}
    \caption{Training samples 5K}
  \end{subfigure}
  \hfill
  \begin{subfigure}[b]{0.49\linewidth}
    \centering
    \includegraphics[height=\imgheight]{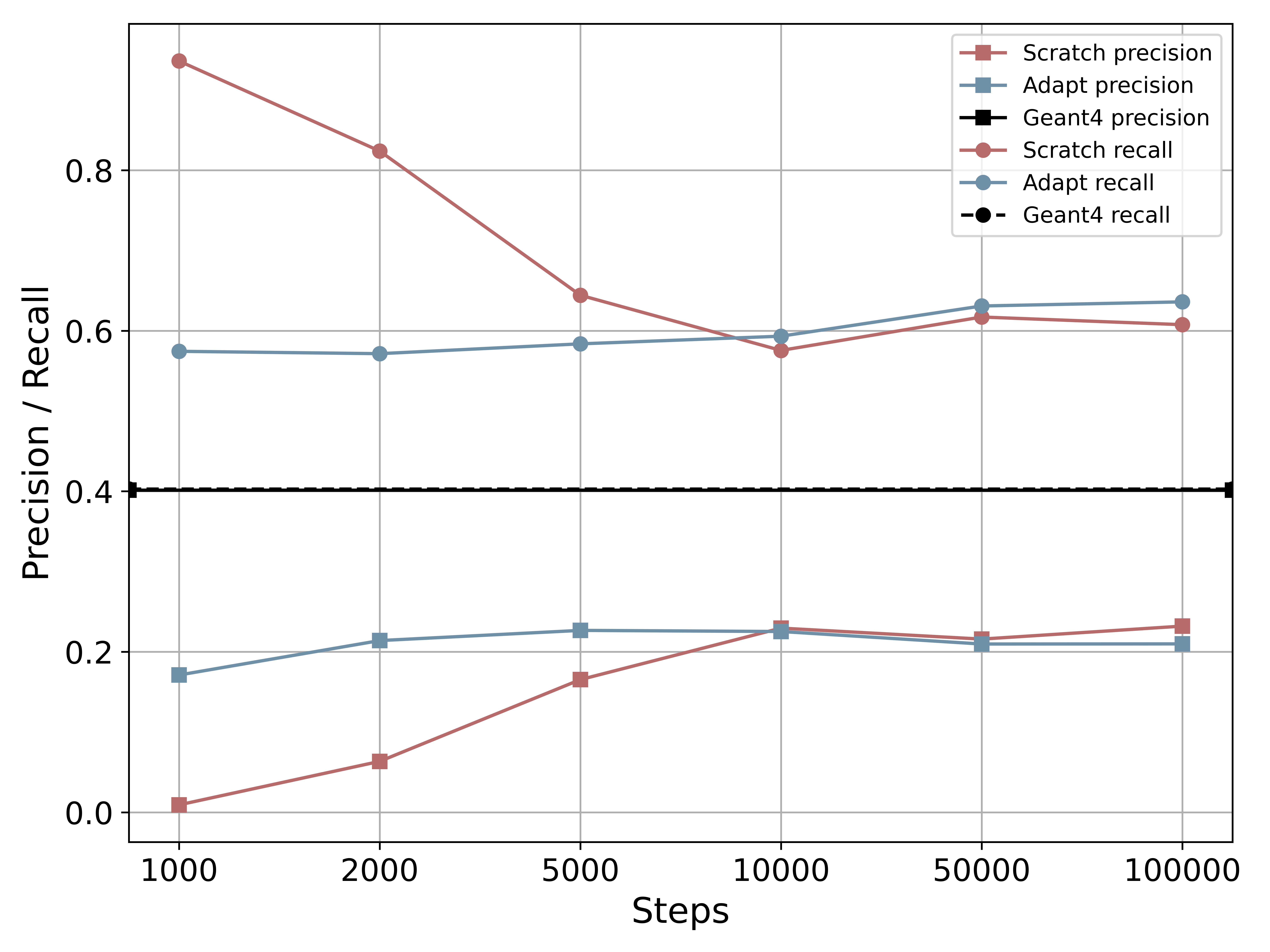}
    \caption{Training samples 25K}
  \end{subfigure}
  \hfill
  \begin{subfigure}[b]{0.49\linewidth}
    \centering
    \includegraphics[height=\imgheight]{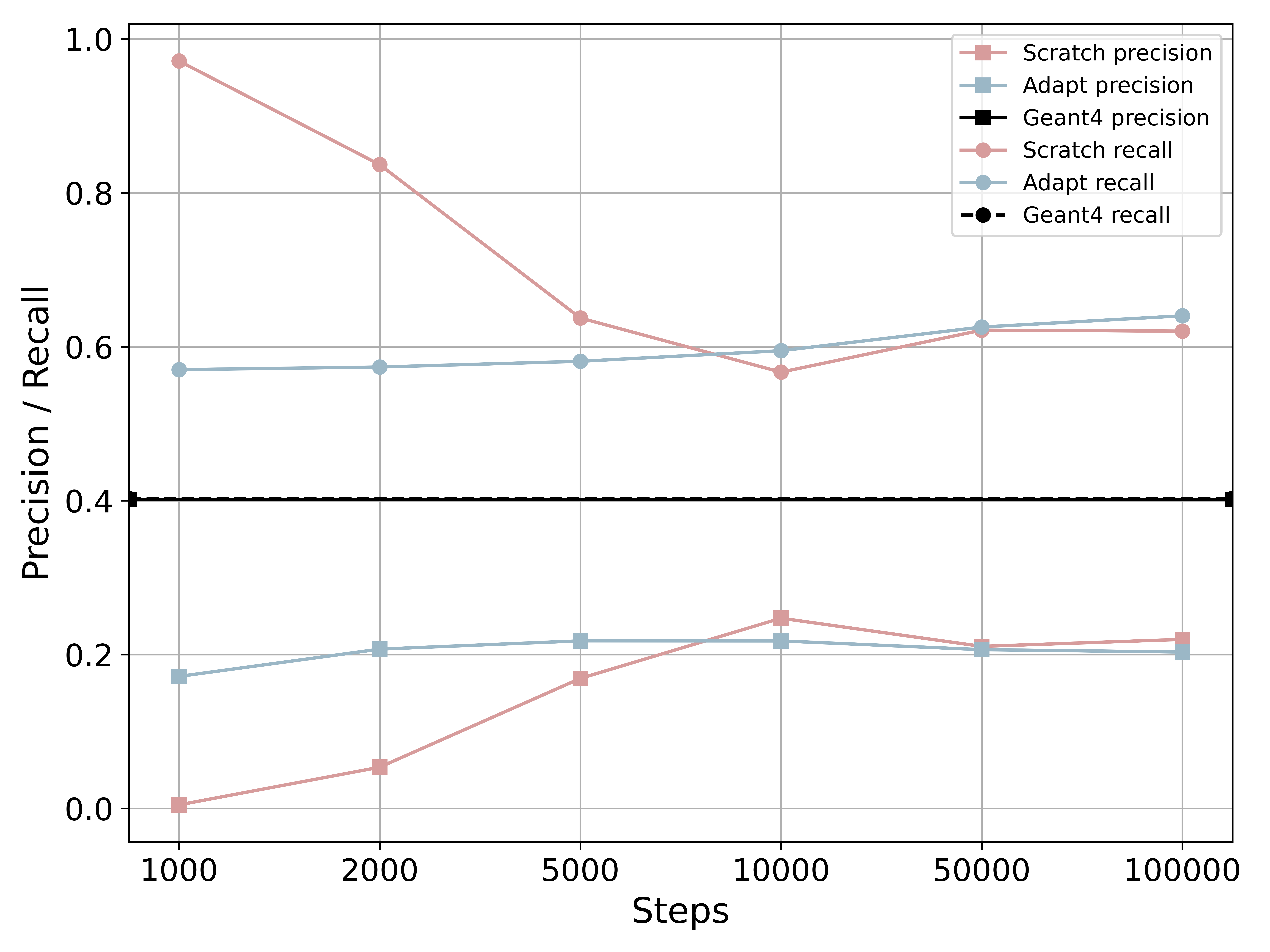}
    \caption{Training samples 100K}
  \end{subfigure}
  \caption{Precision and recall for each configuration of the number of training data samples.}
  \label{fig:appendix_adaptation_pr}
\end{adjustwidth}
\end{figure}

\begin{figure}[htbp]
\begin{adjustwidth}{-\figenvleftextend}{-\figenvrightextend}
  \centering
  \begin{subfigure}[b]{0.49\linewidth}
    \centering
    \includegraphics[height=\imgheight]{assets/adaptation/dc_vs_steps_with_data_1K.png}
    \caption{Training samples 1K}
  \end{subfigure}
  \hfill
  \begin{subfigure}[b]{0.49\linewidth}
    \centering
    \includegraphics[height=\imgheight]{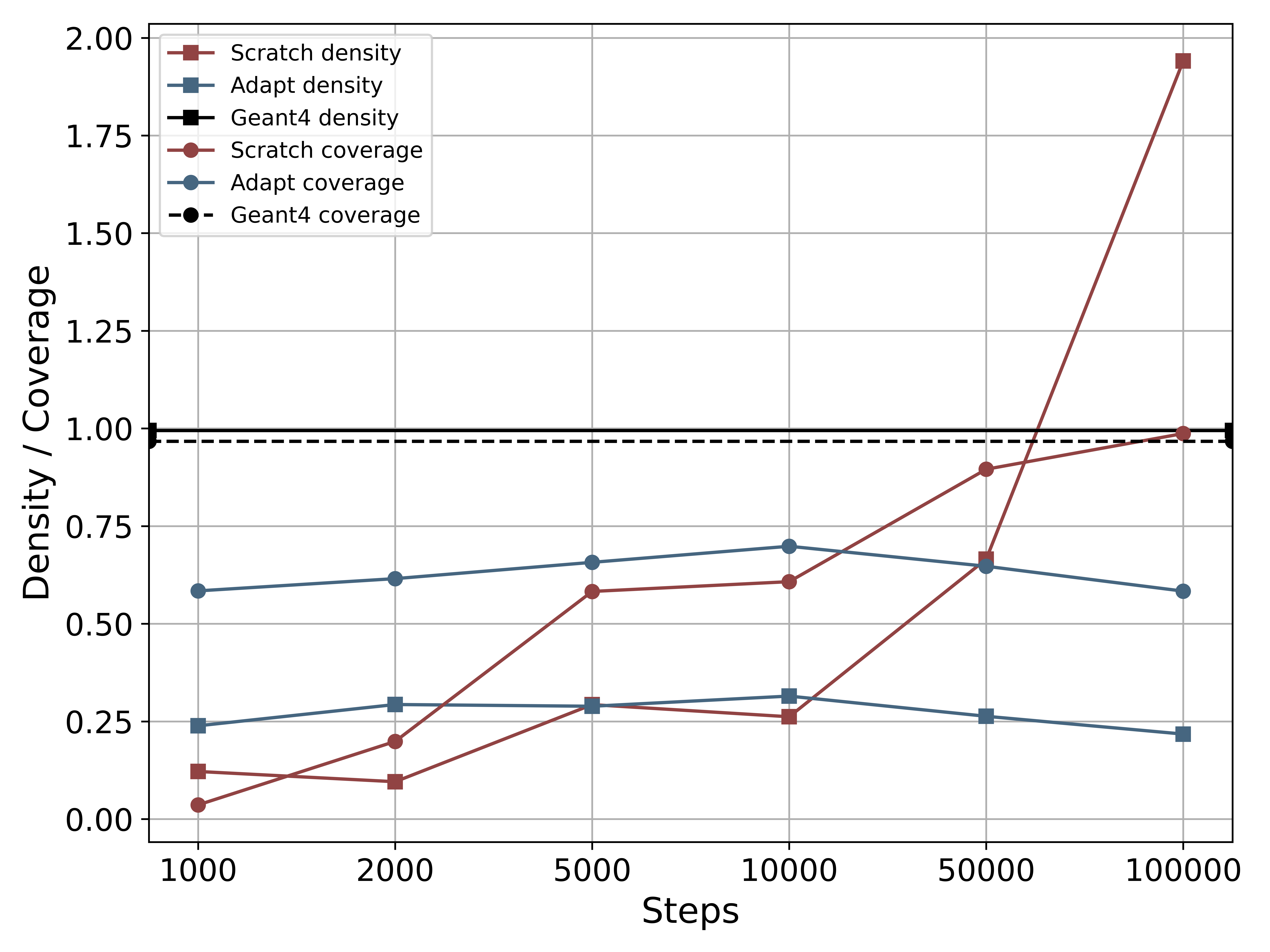}
    \caption{Training samples 5K}
  \end{subfigure}
  \hfill
  \begin{subfigure}[b]{0.49\linewidth}
    \centering
    \includegraphics[height=\imgheight]{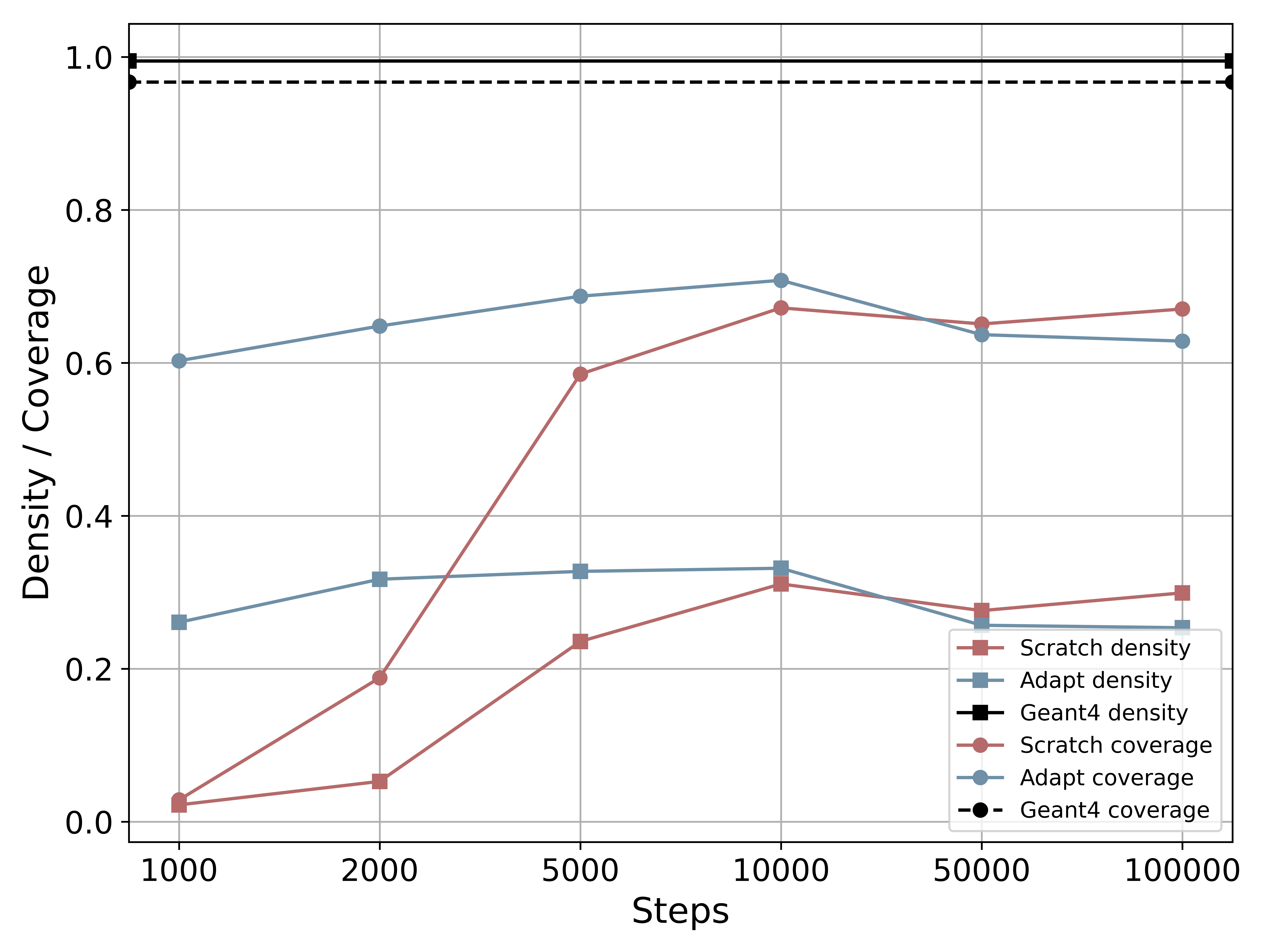}
    \caption{Training samples 25K}
  \end{subfigure}
  \hfill
  \begin{subfigure}[b]{0.49\linewidth}
    \centering
    \includegraphics[height=\imgheight]{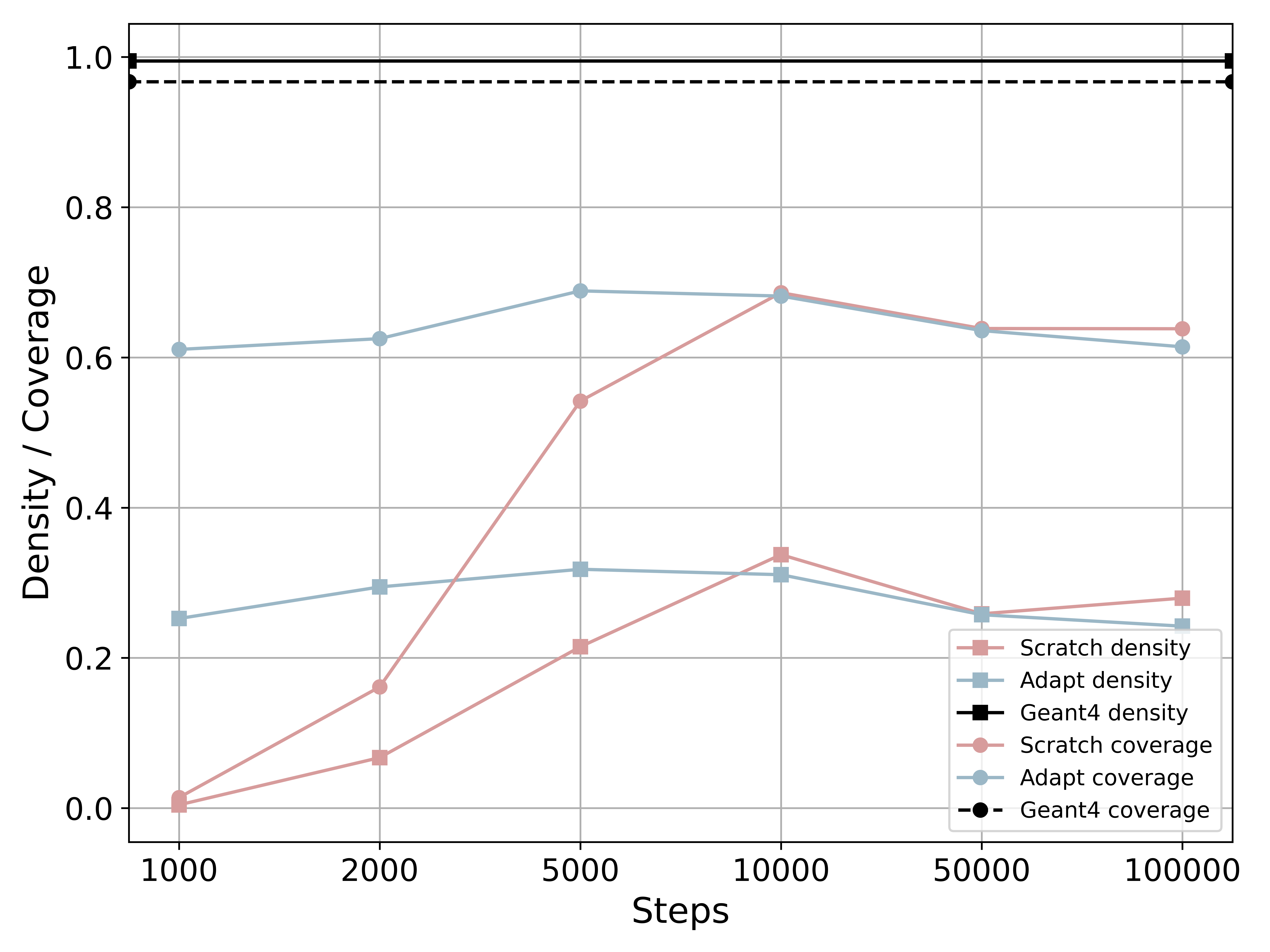}
    \caption{Training samples 100K}
  \end{subfigure}
  \caption{Density and coverage for each configuration of the number of training data samples.}
  \label{fig:appendix_adaptation_dc}
\end{adjustwidth}
\end{figure}

\end{document}